\documentclass[11pt]{article}
\usepackage{amsmath,amssymb,amsthm}
\usepackage[backref=page]{hyperref}
\usepackage[smalltableaux,centertableaux]{ytableau}
\usepackage{bbold}
\usepackage{authblk}
\usepackage{xcolor}
\usepackage{graphicx}
\usepackage{tikz}
\usepackage{caption}
\usepackage{subcaption}

\graphicspath{{./Figures/}}

\usetikzlibrary{decorations.pathmorphing}
\tikzset{grav/.style={decorate, decoration=snake}}

\numberwithin{equation}{section}

\addtocontents{toc}{\setcounter{tocdepth}{2}}

\newcommand{\RR}{\mathbb{R}}
\newcommand{\CC}{\mathbb{C}}

\newcommand{\NN}{\mathbb{N}}
\renewcommand{\Re}{\operatorname{Re}}
\renewcommand{\Im}{\operatorname{Im}}
\DeclareMathOperator*{\Res}{Res}

\DeclareMathOperator{\dDisc}{dDisc}

\newcommand{\op}{\mathcal{O}}

\title{Holographic solar systems and hydrogen atoms: \\non-relativistic physics in AdS and its CFT dual}
\author{Henry Maxfield\thanks{\href{mailto:henrym@stanford.edu}\texttt{henrym@stanford.edu}} }
\author{Zahra Zahraee\thanks{\href{mailto:zr.zahraee@physics.mcgill.ca}\texttt{zr.zahraee@physics.mcgill.ca}}}
\affil{$\ast$ Stanford Institute for Theoretical Physics,\\ Stanford University, Stanford, CA 94305, USA}
\affil{\dag Department of Physics, McGill University, \\Montreal, QC H3A 2T8, Canada}

\begin{document}

\maketitle

\vspace{-.8cm}

\begin{abstract}
	We study a non-relativistic limit of physics in AdS which retains the curvature through a harmonic Newtonian potential. This limit appears in a CFT dual through the spectrum of operators of large dimension and correlation functions of those operators with appropriate kinematics. In an additional flat spacetime limit, the spectrum is determined by scattering phase shifts (proportional to anomalous dimensions), and a CFT correlation function is proportional to the S-matrix. In particular, we describe the effect of resonances on the spectrum and correlation functions. As an example, we discuss the Coulomb potential (describing solar systems and hydrogen atoms in AdS) in detail.
\end{abstract}

\vspace{-.4cm}

\tableofcontents

\section{Introduction}

In our efforts to understand quantum mechanical theories of gravity, we suffer from a dearth of examples with complete non-perturbative descriptions. Perhaps the only such models use the AdS/CFT correspondence, where the definition is provided through `duality' by an ordinary conformally invariant quantum field theory (CFT). By now it is well-established that a CFT with appropriate properties has a low-energy description as a weakly-coupled local quantum field theory in AdS coupled to Einstein gravity \cite{Heemskerk:2009pn,El-Showk:2011yvt,Afkhami-Jeddi:2016ntf,Penedones:2016voo}.\footnote{Specifically, we require two properties. Firstly we have 't Hooft factorisation, so connected correlation functions of `single-trace' operators are suppressed by powers of the central charge $c_T\gg 1$, which controls the weak coupling. Secondly, we have a `large gap' $\Delta_\mathrm{gap}\gg 1$, defined by the minimal dimension of higher spin ($s>2$) single-trace primary operators, controlling the scale on which the QFT in AdS is local.} But the relation between CFT and AdS degrees of freedom is understood completely only in perturbation theory: we lack a complete independent definition of the theory in terms of gravitational variables, since we do not know how to translate the geometric language of fluctuating spacetime into CFT terms. This is true not only in regimes where we expect quantum gravitational effects to be very important (so spacetime itself may no longer meaningfully exist), but also in more controlled settings with spacetimes that are far from empty AdS but nonetheless weakly curved.

With this in mind, it is useful to have examples of gravitational systems in AdS which are non-perturbative but nonetheless under good control: more complicated than small fluctuations around empty spacetime, but simpler than a generic QFT in a strongly-interacting regime or black holes (whose microscopic description remains murky). In this paper we introduce a class of such examples that will be recognisable from any first course in quantum mechanics. Namely, we study non-relativistic interacting particles with the novelty that we include effects arising from placing them in AdS.

To take a standard model of interacting non-relativistic particles and place it in AdS, we need only make the small modification of adding a background Newtonian gravitational potential to the Hamiltonian.  For each particle of mass $m$, we add
\begin{equation}\label{eq:VAdS}
V_\mathrm{AdS}  = \tfrac{1}{2} m \omega^2 r^2 \,,
\end{equation}
where $r$ is the distance from the origin in appropriate coordinates. The frequency $\omega$ of this harmonic potential is the inverse of the usual AdS length $L$, with $\omega= \frac{c}{L}$. The resulting models are amenable to solution using a myriad of familiar analytic and numerical methods, including interesting cases with strong interactions that cannot be treated using perturbation theory, and have extremely well-understood non-perturbative properties. Accordingly, they provide an excellent setting to explore the CFT description of non-perturbative AdS physics.

\begin{figure}
	\centering
	\begin{tikzpicture}
		\draw[rounded corners] (-6,5) rectangle (-1.5,1.);
		\draw[rounded corners] (6,5) rectangle (1.5,1.);
		\draw[rounded corners] (-6,-5) rectangle (-1.5,-1.);
		\draw[rounded corners] (6,-5) rectangle (1.5,-1.);
		
		\begin{scope}[xshift=-105,yshift=111,scale=.5]
		\draw (0,0) ellipse (1.25 and 0.5);
		\draw (-1.25,0) -- (-1.25,-3.5);
		\draw (-1.25,-3.5) arc (180:360:1.25 and 0.5);
		\draw [dashed] (-1.25,-3.5) arc (180:360:1.25 and -0.5);
		\draw (1.25,-3.5) -- (1.25,0);
		\draw[ultra thick] [red] (-.1,-3.4) to [out=140, in=-120] (-.3,-2) to [out=60, in=-60] (.2,.1);
		\draw[ultra thick] [blue] (.4,-3.4) to [out=80, in=-50] (.3,-1.5) to [out=130, in=-100] (-.2,.1);
		\end{scope}
		\node at (-3.75,4.5) {Relativistic AdS$_{d+1}$};
  		\node at (-3.75,1.4) {$\mathfrak{so}(d,2)$};
  		
  		\begin{scope}[xshift=100,yshift=85,scale=1.1]
		\draw (0,-1) -- (1,0)--(0,1);
		\draw [dashed] (0,-1)--(0,1);
		\draw[ultra thick,red] (.2,-.5) to  (0,-.2) to (.4,.4);
		\draw[ultra thick,blue] (.4,-.5) to  (0,.3) to (.2,.6);
		\end{scope}
		\node at (3.75,4.6) {Relativistic $\RR^{d,1}$};
  		\node at (3.75,1.4) {Poincar\'e algebra};
  		
  		\draw[-{stealth},thick] (-1,3) -- (1,3);
  		\draw[-{stealth},thick] (-1,-3) -- (1,-3);
  		\draw[-{stealth},thick] (-3.75,.75) -- (-3.75,-.75);
  		\draw[-{stealth},thick] (3.75,.75) -- (3.75,-.75);
  		
  		\node at (0,3.3) {$L\to\infty$};
  		\node at (0,-2.7) {$\omega\to 0$};
  		
  		\node at (-3,.2) {$c\to\infty$};
  		\node at (-3,-.2) {$\omega$ fixed};
  		
  		\node at (4.5,0) {$c\to\infty$};
  		
  		\node at (-3.75,-1.4) {Non-relativistic AdS$_{d+1}$};
  		\node at (-3.75,-4.6) {\small Harmonic oscillator algebra};
  		\begin{scope}[xshift=-105,yshift=-110,scale=1.7]
		\draw[ domain=-1:1, smooth, variable=\x] plot ({\x}, {\x*\x});
		\node [circle,fill=blue,inner sep=2] at (.5,.3) {};
		\node [circle,fill=red,inner sep=2] at (-.6,.4) {};
		\draw [->,blue, ultra thick] (.5,.3) to (.2,.3);
		\draw [->, red, ultra thick] (-.6,.4) to (-.2,.4);
		\end{scope}

  		\node at (3.75,-1.4) {Non-relativistic $\RR^{d,1}$};
  		\node at (3.75,-4.6) {Galilean algebra};
  		\begin{scope}[xshift=105,yshift=-90,scale=1]
		\draw (-1.5,0)--(1.5,0);
		\node [circle,fill=blue,inner sep=2] at (.7,0.1) {};
		\node [circle,fill=red,inner sep=2] at (-.9,0.1) {};
		\draw [->,blue, ultra thick] (.7,0.1) to (.2,0.1);
		\draw [->, red, ultra thick] (-.8,0.1) to (-.3,0.1);
		\end{scope}

 	\end{tikzpicture}
	\caption{A sketch of relativistic and non-relativistic physics in AdS and flat spacetime, the symmetries in each case, and the relations between them. The top line represents relativistic physics by the worldlines (red and blue) of a pair of interacting particles moving in AdS (left) and Minkowski (right) spacetimes, which are related by the flat spacetime limit of large AdS length. The bottom right represents Galilean-invariant non-relativistic dynamics. This paper fills in the bottom left corner, namely non-relativistic dynamics with a quadratic Newtonian potential $\Phi = \frac{1}{2}\omega^2r^2$ arising from the $c\to\infty$ limit of AdS. Each arrow represents an \.In\"on\"u-Wigner contraction relating the respective symmetry algebras. In particular, the arrow marked $\omega\to 0$ is the non-relativistic flat spacetime limit discussed in section \ref{sec:flat}. 
	 \label{fig:symmTable}}
\end{figure}
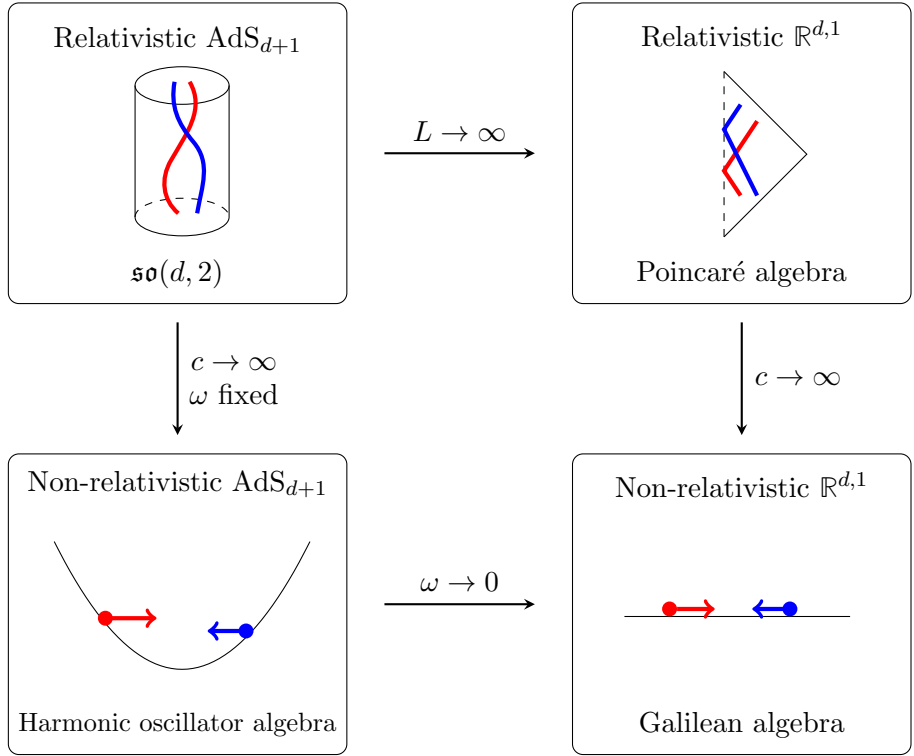

The non-relativistic AdS models we study have a nice characterisation in terms of their symmetries. If we take an ordinary non-relativistic theory with Galilean symmetry and add the AdS gravitational potential \eqref{eq:VAdS}, it would at first sight appear that we have broken the translation and boost symmetries. In fact, as might be anticipated from the maximal symmetry of AdS, the full symmetry group is retained in a modified form. This is a non-relativistic version of the statement that AdS acts a maximally symmetric `box' \cite{Hawking:1982dh}. In particular, the familiar decoupling of centre-of-mass and relative motion (a consequence of boost symmetry) is retained. The Galilean symmetries are deformed to become the creation/annihilation algebra of the harmonic oscillator acting on the centre-of-mass wavefunction.  More precisely,   an \.In\"on\"u-Wigner contraction of the harmonic oscillator algebra  (the $\omega\to 0$ limit) gives the Galilean symmetry algebra. This is the non-relativistic analogue of the contraction of the AdS symmetry algebra (in the $L\to\infty$ limit) which gives the Poincar\'e algebra.   In fact, the passage from these relativistic symmetry algebras (AdS and Poincar\'e) to the non-relativistic groups (harmonic oscillator and Galilean respectively) can similarly be described as contractions in a $c\to\infty$ limit. These four different symmetry groups, the associated physics and the relations between them are summarised in  figure \ref{fig:symmTable}. Our paper fills in the bottom-left corner of this diagram, and describes its connections to its neighbours.

Our goal is not only to describe non-relativistic systems in AdS, but also to relate this physics to observables in a dual CFT. A prerequisite is to identify which CFT operators can be dual to non-relativistic particles in AdS. The non-relativistic AdS limit is self-consistent only if the rest-energy $m c^2$ of the particles we study is large compared to the quantum $\hbar \omega$ of the harmonic oscillator. The ratio of these energies is (up to order one corrections) the dimension $\Delta$ of the dual operator, which must therefore be large:
\begin{equation}\label{eq:largeDelta}
\Delta = \frac{m c^2}{\hbar \omega} + \frac{d}{2}+\cdots \gg 1 \,.
\end{equation}
The correction $\frac{d}{2}$ (the first term from expanding the familiar AdS/CFT relation $L^2 m^2 = \Delta(\Delta-d)$) can be thought of as coming from the ground state energy $\frac{d}{2}\hbar \omega$ of the $d$-dimensional harmonic oscillator \eqref{eq:VAdS}. Another way to state the condition $\Delta\gg 1$ is that the Compton wavelength $\frac{\hbar}{m c}$ must be small compared to the AdS length $L= \frac{c}{\omega}$. More generally, if the operator is dual to some composite object in AdS which may be much bigger than its Compton wavelength (such as a black hole), then its size should also be much smaller than $L$.

The paper has three main aims. Firstly, we describe  properties and observables of non-relativistic physics in AdS, as defined by ordinary Galiliean-invariant quantum mechanics the addition of the harmonic potential \eqref{eq:VAdS}. This includes a discussion of the $\omega\to 0$ limit, which is pertinent for the flat spacetime limit of AdS/CFT. Secondly, we identify correlation functions of a dual CFT which are determined by this non-relativistic physics (and its flat limit), and compute the corresponding bulk observables in various examples and limits. The final aim is to relate this to the data of a CFT --- the spectrum of primary operators and their OPE coefficients --- via the conformal block decomposition of correlation functions. This final aspect splits into two distinct parts: the `S-channel' data corresponds to interacting multi-particle states, while the `T-channel' operators mediate the interaction potential. These two sets of data are related by the conformal bootstrap \cite{Simmons-Duffin:2016gjk,Poland:2018epd}. In this paper we discuss only the S-channel data, and T-channel exchanges will be considered in a companion paper \cite{companion}, along with a discussion of perturbation theory in non-relativistic AdS.

\subsection*{Summary}

In \textbf{\autoref{sec:NRAdS}} we explain the details of the generalities discussed above, including the definition of the non-relativistic AdS limit and the symmetries in both classical and quantum models. We introduce the observables that we study, which are directly related to correlation functions of a dual CFT. Here and throughout, we focus on the simplest interesting example of two particles interacting via a potential depending on their separation.

The centre-of-mass motion is determined by symmetry to be described by a simple harmonic oscillator. In particular, the special conformal generators $K_i$ of the CFT become the annihilation operators $A_i$ of the harmonic oscillator (see \eqref{eq:Adef}), so primary states of the CFT (annihilated by $K_i$) correspond to the centre-of-mass wavefunction in its ground state. The non-trivial dynamics is all in the relative motion, which for two particles is effectively a problem of a single particle moving in a potential coming from both interactions and the AdS gravitational potential. In particular, the masses $m_1,m_2$ of two particles appear only in the combination $\mu = \frac{m_1m_2}{m_1+m_2}$, the reduced mass.

To translate CFT correlation functions to non-relativistic bulk observables, the key observation is that inserting a CFT operator on the plane $\RR^d$ at a radius of order $\Delta^{-1/2}$ creates (via radial quantisation) a state on the unit ball with a simple bulk description, namely a coherent state for the AdS harmonic oscillator. By inserting several operators we can prepare multi-particle states, and the overlap of two such states with an intermediate time-evolution (in either real or imaginary time) gives a CFT correlation function which probes the interactions of these particles. For two particles, this gives us a four-point function, schematically
\begin{equation}\label{eq:Gintro}
	\langle \op_1\op_2\op_2\op_1\rangle \sim \langle \alpha'|e^{- \tau H}|\alpha\rangle\,,
\end{equation}
where $|\alpha\rangle$, $|\alpha'\rangle$ are coherent states and $\tau$ gives an evolution in Euclidean time; for real time evolution we give $\tau$ an imaginary part. The precise definition of the right hand side of \eqref{eq:Gintro} is given in \eqref{eq:scatterM}. The resulting four-point function depends on $\tau$ and an angle $\theta$ relating the initial and final coherent states, and has cross-ratios $z,\bar{z}$ of order $\Delta^{-1}$: the kinematics are defined in \eqref{eq:Gdef}.

In \textbf{\autoref{sec:QM}} we describe the evaluation of our four-point function by summing over a complete set of eigenstates for the Hamiltonian governing relative motion. This sum over states corresponds precisely to the conformal block decomposition into S-channel intermediate primary states. Descendant states are absent due to a simplifying feature of the non-relativistic limit: with judicious choice of kinematics, a state created by local insertions of two (or more) operators has centre-of-mass wavefunction in the ground state, which means it is a superposition of primary states.

The sum over states in the conformal block expansion is weighted by the S-channel OPE coefficients. In the non-relativistic bulk theory, these coefficients are determined by the asymptotic decay of the corresponding normalised eigenfunction of the Hamiltonian.

We check these results by solving the free problem, with no interaction potential. This is dual to `mean field theory' (MFT), sometimes called a generalised free theory. The S-channel intermediate states are double trace operators $[\op_1\op_2]_{l,n}$, with a spectrum precisely matching the $d$-dimensional harmonic oscillator. The non-relativistic limit of MFT OPE coefficients (large $\Delta_{1,2}$ with fixed $l,n$) gives the result expected from the decay of harmonic oscillator wavefunctions,  depending on the dimensions of the external operators only as a power of the reduced mass $\mu \sim\frac{\Delta_1\Delta_2}{\Delta_1+\Delta_2}$.

In \textbf{\autoref{sec:classical}} we discuss the classical limit. Firstly, we describe direct evaluation of the correlation function using Lagrangian methods, by finding a classical solution and computing its on-shell action. Secondly, we determine the spectrum of states and OPE coefficients in the classical limit by using the WKB approximation. Finally we discuss the relation between these. This is more subtle than one might expect, since the on-shell energy and angular momentum of the classical solutions does not typically correspond to a physical energy and angular momentum of any eigenstate.

In \textbf{\autoref{sec:flat}} we turn our attention to the flat spacetime limit, where we take $\omega$ to be small while fixing the interaction potential $V(r)$. This is interesting as an example of the flat spacetime limit of AdS/CFT, though with the advantage that we have a far better non-perturbative understanding of the bulk physics than is typical in QFT.

The spectrum in this limit divides into two regimes. There may be bound states $E<0$ for an attractive potential, which are essentially unaffected by the presence of a small AdS harmonic potential. More interestingly, we have states with $E>0$ that would be scattering states in the absence of the harmonic potential, but get resolved into a discrete spectrum separated by small energy gaps $\Delta E\sim 2\omega$. While this gap between states is always the same in the flat limit, the precise spectrum depends on the interactions, and is determined by the scattering phase shift $\delta$ at the given energy and angular momentum: $\gamma \sim -\frac{2}{\pi}\delta$, where $\gamma$ is the anomalous dimension, or the energy shift relative to non-interacting particles. Note that this result is leading order in the flat space (small $\omega$) limit, but not perturbative in the interactions so $\delta$ can be very large. It implies (but is stronger than) the phase shift formula of \cite{Paulos:2016fap}, and we expect it to remain true for relativistic elastic scattering. The same formula appeared for special kinematic regimes in \cite{Cornalba:2007zb,Kulaxizi:2019tkd}.

Flat spacetime scattering physics is encoded not only by the spectrum at small $\omega$, but is also directly accessible from a CFT correlation function. For the most interesting correlator we take \eqref{eq:Gintro} with $\Im\tau = \frac{\pi}{\omega}$. The result is essentially identical to a correlation function considered in \cite{Hijano:2019qmi,Komatsu:2020sag}, and has a nice physical interpretation shown in figure \ref{fig:scattering}. We prepare an initial coherent state of well-separated particles which are stationary, but elevated in the harmonic potential to give them some potential energy $E$. They then are accelerated by this potential until they meet with kinetic energy $E$ and scatter, before travelling back out up the potential until stationary once again. This takes total time $\frac{\pi}{\omega}$. The overlap with a similar rotated coherent state directly probes the scattering, and indeed the resulting correlation function is proportional to the usual scattering amplitude.

The advantage of the non-relativistic limit here is that we know precisely when the relationship between correlation function and scattering amplitude will be valid. The only obstruction is that the initial states prepared by operator insertions can sometimes fail to be the expected coherent states. This happens if there is a sufficiently deeply bound state, and we try to scatter at energies below an order one multiple of its binding energy: in that case, the initial state `tunnels' to be dominated instead by the bound state.

One interesting non-perturbative phenomenon that is easy for us to study is a long-lived resonance, which is of interest beyond the non-relativistic limit (and we expect our results to remain valid more generally). For decay rates $\Gamma$ slower than the AdS frequency $\omega$, these appear as `extra states' on top of the evenly spaced scattering states discussed above. As we vary the spin, the location of the resonance would typically cross the Regge trajectories of the scattering states, but the true Regge trajectories do not cross: instead, the scattering trajectories have a characteristic reconnection when they encounter the resonance, illustrated in figure \ref{fig:resonance}. Shorter-lived resonances $\Gamma\gtrsim \omega$ do not give rise to sharp reconnections, but show up as an exponential decay in our scattering correlation function if we add an additional real time evolution of order $\Gamma^{-1}$.

In \textbf{\autoref{sec:Coulomb}} we apply the general ideas to the interesting example of the Coulomb potential $V(r)= -\frac{g}{r}$ in $d=3$. This includes scattering physics, which requires slight modifications since the long-ranged nature of the interaction means the usual definition of scattering amplitude does not apply. The harmonic potential acts as a natural infrared regulator, so the required changes are minor and give us a nice regulated S-matrix \eqref{eq:SCoulomb} and scattering correlator \eqref{eq:GscatCoulomb}. We also discuss the spectrum and correlation function in the classical limit.

\begin{figure}
	\centering
	\begin{tikzpicture}
		\node at (0,0) {\includegraphics[width=10cm]{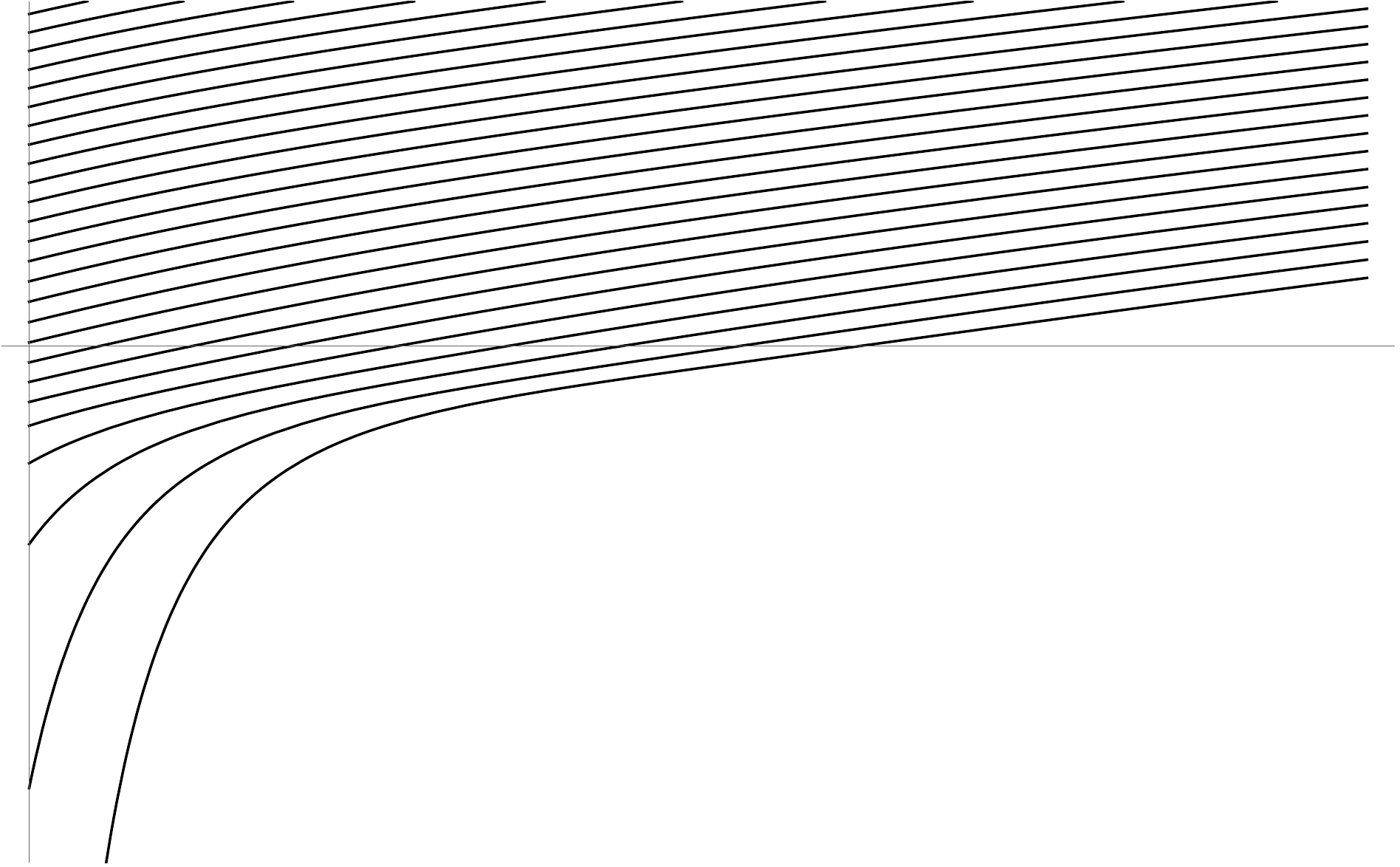}};
		\draw[ultra thick,->] (-4.8,.6) -- (5,.6);
		\draw[ultra thick,->] (-4.8,-3.4) -- (-4.8,3.6);
		\node at (-5.1,4.) {\huge $E$};
		\node at (5.1,.1) {\huge $l$};
		
		\node[red] at (-3.8,-1) { Hydrogen};
		\node[red] at (-3.8,-1.4) {atom};
		
		\node[red] at (-3.5,2) { Rutherford};
		\node[red] at (-3.5,1.6) {scattering};
		
		\node[red] at (-1.3,0.3) {Classical orbits};
		
		\node[red] at (2.6,2) {MFT};
		\node[red] at (2.6,1.6) {$+$ perturbations};
 	\end{tikzpicture}
	\caption{A sketch of the spectrum (Regge trajectories) of two non-relativistic particles in AdS$_4$ interacting via a Coulomb potential $V(r)=-\frac{g}{r}$ with $\mu g^2\gg \omega$. Energies $E_{l,n}$ depend on spin $l$ (plotted as a continuous variable, though physical states lie at integer $l$) and $n=0,1,2,\ldots$ labelling the states in order of increasing energy. Various (overlapping) regions of the diagram encode different interesting regimes. If $l$ is not too large, the spectrum is determined by flat spacetime physics: the phase shifts of Rutherford scattering for $E>0$, and the spectrum of the hydrogen atom for $E<0$. The leading $n=0$ Regge trajectory intersects the $l=0$ axis at $E_{0,0}\sim - \frac{1}{2}\mu g^2 $, the binding energy of the hydrogen ground state (not shown on the plot).	
	 These hydrogen bound states also describe classical Keplerian orbits for $l\gg 1$, circular for small $n$ and increasingly elliptical for larger $n$. At sufficiently large $l$ or $n$ the spectrum approaches the linear Regge trajectories of MFT, with perturbative corrections controlled by the exchange of the stress tensor and other conserved currents.
	 \label{fig:CoulombRegge}}
\end{figure}
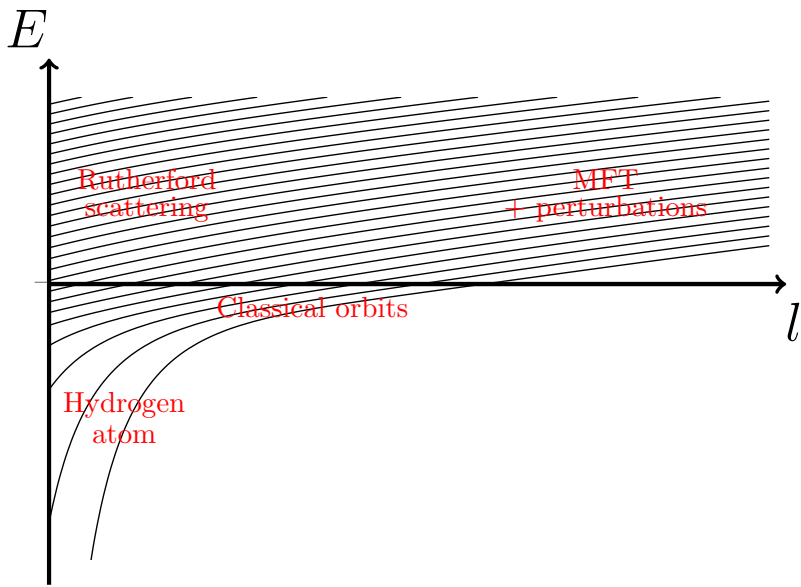

The bound states, scattering states and classical limit together give us a complete picture of the spectrum, illustrated in figure \ref{fig:CoulombRegge}. We have the familiar bound states of the hydrogen atom (or classical Keplerian orbits) at negative energy, and a spectrum encoding Rutherford scattering at positive energies and sufficiently small spin. Going to larger spin, the effect of the AdS potential becomes important, and the Regge trajectories interpolate from hydrogen bound states to the linear trajectories of MFT as $l\to\infty$, with the approach governed by perturbation theory (discussed in more detail in the companion paper \cite{companion}).

We conclude in \textbf{\autoref{sec:disc}} with a discussion of related ideas and open questions.

\section{Non-relativistic physics in AdS}\label{sec:NRAdS}

\subsection{The AdS gravitational potential}

The metric global of AdS$_{d+1}$ is given by
\begin{equation}\label{eq:AdSmetric}
	ds^2 = -c^2 f(r)dt^2+\frac{dr^2}{f(r)}+r^2 d\Omega_{d-1}^2,
\end{equation}
where $d\Omega_{d-1}^2$ is the metric on the unit sphere $S^{d-1}$, and $f(r) = \frac{r^2}{L^2}+1$ for AdS length $L$. In the limit we study, the interesting dynamics will take place in the region $r \ll L$ close to the origin, so the spatial curvature is negligible. Since we can disregard the spatial curvature, it will often be convenient to use Cartesian spatial coordinates $\vec{x} = r \vec{\Omega}$, where $r=|\vec{x}|$ and $\vec{\Omega}$ is a unit vector describing a point on $S^{d-1}$. The notation $\vec{\cdot}$ indicates $d$-dimensional vectors throughout.

 However, we will keep the effects of the nontrivial $g_{tt}$ component, which can be described by a quadratic Newtonian potential $\Phi \sim -\frac{1}{2}(g_{tt}+c^2)$:
\begin{equation}\label{eq:AdSpotential}
	\Phi = \frac{1}{2}\omega^2 r^2,
\end{equation}
where we have replaced the AdS length $L$ in favour of the AdS frequency $\omega = \frac{c}{L}$, which is the relevant scale for non-relativistic physics. This potential is sourced by the cosmological constant $\Lambda \propto -\omega^{2}$, which in Newtonian language we might think of as a constant contribution to the mass density $\rho$ in the Poisson equation $\nabla^2\Phi=4\pi G_N\rho$. If energies are non-relativistic ($E\ll mc^2$), this potential confines the dynamics to the flat space region $r\ll L$.

The curvature of AdS leaves its mark in the non-relativistic limit through the harmonic potential \eqref{eq:AdSpotential}. We will later consider an additional flat space\emph{time} limit $\omega \to 0$, relevant whenever the typical timescale for the dynamics (such as the orbital period for a bound state or the duration of a scattering process) is much shorter than the `cosmological time' $\omega^{-1}$.

\subsection{Symmetries of non-relativistic particle motion in AdS}\label{sec:classSym}

Since AdS introduces the gravitational potential \eqref{eq:AdSpotential}, it would appear that we have broken the translation and Galilean boost symmetries of Newtonian dynamics. However, these symmetries are in fact still present in a slightly modified form, along with the manifest symmetries under rotation and time-translation. This should be expected from the maximal symmetry of AdS. We give a more complete discussion of the symmetry algebra in the context of the quantum theory in section \ref{sec:symm}; here we illustrate its consequences for the main system we study.

Namely, we consider the dynamics of two particles with masses $m_{1,2}$ and positions $\vec{x}_1,\vec{x}_2$, moving under a central potential $V(|\vec{x}_1-\vec{x}_2|)$ along with the gravitational potential  \eqref{eq:AdSpotential}.   The non-relativistic Hamiltonian is
\begin{equation}
	H_\mathrm{NR} = \frac{p_1^2}{2m_1}+\frac{p_2^2}{2m_1} + \frac{1}{2}m_1\omega^2 r_1^2+ \frac{1}{2}m_2\omega^2 r_2^2 + V(|\vec{x}_1-\vec{x}_2|).
\end{equation}
 Given any solution $\vec{x}_{1,2}(t)$ to the equations of motion for the two particles we may construct a $2d$-dimensional family of solutions by shifting $\vec{x}_{1,2}(t) \rightarrow \vec{x}_{1,2}(t)+ \vec{X}(t)$, where $\vec{X}(t)$ is any solution to the harmonic oscillator equation of motion $\frac{d\vec{X}}{dt^2} = -\omega^2 \vec{X}$. Together with rotations and time-translations, this forms a $\frac{1}{2}(d+1)(d+2)$-dimensional group of symmetries, a deformation of the Galilean group to AdS.
 
 This symmetry implies that the decoupling between centre-of-mass and relative motion familiar from Galilean symmetry remains true with the harmonic potential. Write
\begin{equation}\label{eq:COM}
\begin{gathered}
	\vec{x}_1 = \vec{x}_\mathrm{COM} - \frac{m_2}{m_1+m_2} \vec{x}, \quad \vec{x}_2 = \vec{x}_\mathrm{COM} + \frac{m_1}{m_1+m_2} \vec{x},\\
	\text{with inverse}\qquad \vec{x} = \vec{x}_2-\vec{x}_1, \quad \vec{x}_\mathrm{COM} = \frac{m_1 \vec{x}_1+m_2 \vec{x}_2}{m_1+m_2}.
\end{gathered}
\end{equation}
Then the dynamics of $\vec{x}_\mathrm{COM}$ and $\vec{x}$ become independent. The centre-of-mass $\vec{x}_\mathrm{COM}$ moves freely under the harmonic potential only, while $\vec{x}$ behaves as a particle with `reduced mass' $\mu$ given by the harmonic mean of $m_{1,2}$,
\begin{equation}
	\mu = \frac{m_1m_2}{m_1+m_2},
\end{equation}
moving under the gravitational potential along with the central force $V(r)$. With this change of variables, the Hamiltonian becomes
\begin{gather}
	H_\mathrm{NR} = H_\mathrm{COM}+H \label{eq:COMrelH} \\
	H_\mathrm{COM} =  \frac{p_\mathrm{COM}^2}{2(m_1+m_2)} +\frac{1}{2}(m_1+m_2)\omega^2 r_\mathrm{COM}^2 \\
	H=	\frac{p^2}{2\mu} + \frac{1}{2}\mu\omega^2 r^2+  V(r).
\end{gather}
In particular, we may always use the symmetries to set $\vec{x}_\mathrm{COM}=0$. This is the classical non-relativistic version of choosing a primary state of the two particles, as will become clearer later. This leaves rotations and time translations as residual symmetries.

\subsection{Euclidean scattering}\label{sec:Escat}

Our ultimate aim is not just to describe non-relativistic physics in AdS, but to relate it to observables in a dual CFT. The AdS gravitational potential initially appears to pose a challenge to direct CFT probes of non-relativistic physics, since it confines particles with non-relativistic energies to lie close to the centre of AdS, while simple CFT observables are sources at the boundary. Here we describe how to circumvent this difficultly in the language of classical particle dynamics, by considering evolution in Euclidean (or imaginary) time $t_E = i t$. This dynamics directly determines a correlation function in the dual CFT, in an appropriate limit. Later in the quantum theory, we also describe this Euclidean time evolution as a mechanism to prepare initial and final states of interest, allowing us to study real time dynamics for the intermediate evolution.

 To obtain non-relativistic classical equations of motion in Euclidean time, we invert the potential: 
 \begin{equation}
 	\mu \frac{d^2 \vec{x}}{dt_E^2} = \mu \omega^2 \vec{x} + \vec{\nabla} V \,.
 \end{equation}
 The harmonic gravitational potential does not then confine the particles, but rather accelerates them away from one another. We are interested in `Euclidean scattering solutions' for which the particles become widely separated at large positive and negative Euclidean times, interacting while they are nearby; we take the interaction potential $V(r)$ to decay at large $r$ so it becomes negligible at early and late times.
 
 For such a solution, at sufficiently large $|t_E|$ the separation grows exponentially, so the particle motion becomes radial (at approximately constant angle), and we can write
\begin{equation}\label{eq:expr}
	r \sim \frac{1}{\sqrt{\omega\mu}}\exp\left(\omega|t_E|-\tfrac{1}{2}\omega \tau\right)
\end{equation}
for some $\tau$. We have chosen the origin of time so the solution is symmetric under $t_E \mapsto -t_E$ (along with a reflection of space).  The prefactor is a length scale in the non-relativistic quantum theory (restoring $\hbar$, it is $\sqrt{\frac{\hbar}{\omega\mu}}$, the width of the ground state wavefunction of the harmonic oscillator). The constant $\tau$ is the length of time the particles spend in the scattering region; more precisely, they spend a time $\tau+\omega^{-1}\log(\mu \omega R^2)$ within some distance $R$ which is large compared to a typical length-scale of interactions, but small compared to the AdS length.

Now, at late times when the radius $r$ in \eqref{eq:expr} becomes of order $L$, our non-relativistic approximation will break down: the particle will be accelerated to relativistic velocities by the inverted harmonic potential, and the spatial curvature will become important. This time is large in the non-relativistic limit, since a slowly moving particle takes much longer than an AdS time to move an AdS distance away from the scattering, but only logarithmically large since the particles are exponentially accelerated by the AdS potential. But in this regime, the particles will be sufficiently well-separated that their interactions are negligible. They subsequently move independently along radial geodesics in AdS, which we can determine separately for each particle. Parameterising each by proper length $s_{1,2}$, these geodesics are given (now using units where $c=1$ and $\omega=1$) by 
\begin{equation}\label{eq:geodesics}
	r_{1,2} = e^{s_{1,2}},\quad (t_E)_{1,2} = s_{1,2}  - \tfrac{1}{2}\log\left(1+e^{2s_{1,2}}\right) + \tfrac{1}{2}\tau +\tfrac{1}{2}\log\tfrac{m_{1,2}^2}{\mu},
\end{equation}
and constant angle on $S^{d-1}$. These solutions match \eqref{eq:expr} when $s$ is large and negative, after passing to the centre-of-mass and separation coordinates \eqref{eq:COM}.

Taking $s_{1,2}\to\infty$, these geodesics meet the boundary $r\to \infty$ at Euclidean times
\begin{equation}
	(t_E)_{1,2} =  \tfrac{1}{2}\tau + \tfrac{1}{2} \log\tfrac{m_{1,2}^2}{\mu}.
\end{equation}
These points correspond to the location of operator insertions in the dual CFT, relating our two-particle dynamics to a CFT four-point function, which we now introduce.

\subsection{The CFT correlation function}\label{sec:correlator}

The above non-relativistic dynamics (with coordinates chosen to trivialise centre-of-mass motion) motivates us to study the following family of four-point functions of operators $\op_{1,2}$ (and their conjugates  $\op^\dag_{1,2}$ if they are not Hermitian):
\begin{equation}\label{eq:Gdef}
\begin{gathered}
	G(\tau,\theta) = \frac{1}{\mathcal{N}} \Big\langle \op^\dag_1\left(\tau_1, \Omega'\right)\op^\dag_2( \tau_2,- \Omega') \op_1(-\tau_1,\Omega)	\op_2(-\tau_2,-\Omega) \Big\rangle_{\RR\times S^{d-1}} \\
		\tau_{1,2} =  \tfrac{1}{2}\tau + \tfrac{1}{2} \log\tfrac{m_{1,2}^2}{\mu}, \quad \cos\theta = \Omega\cdot \Omega', \\
		\mathcal{N} = e^{-2(\Delta_1\tau_1+\Delta_2\tau_2)}. 
\end{gathered}
\end{equation}
This is a correlator on the Euclidean cylinder $\RR\times S^{d-1}$, with coordinates labelling the Euclidean time and an angle on $S^{d-1}$ ($-\Omega$ is the antipodal point to $\Omega$; $\theta$ is the angle between incoming and outgoing directions). The normalisation constant $\mathcal{N}$ is a convenient convention which cancels a contribution from the free relativistic part of the dynamics.
The operator dimensions are related to the masses of the dual particles as $\Delta\sim \frac{m c^2}{\hbar \omega}+O(1)$ (with the order one quantum correction to be discussed in the next section).

Any four-point function can be written as in \eqref{eq:Gdef} by an appropriate conformal transformation, but it describes non-relativistic physics in the kinematic regime of large operator dimensions $\Delta_{1,2} \gg 1$, with with $\tau$, $\theta$ held fixed in the limit.

We can also write this as a correlation function on the plane $\RR^d$ with coordinates $\vec{y}$, using the conformal map $\vec{y} = e^{\tau}\vec{\Omega}$. Due to the symmetric insertion of the operators, there is no overall conformal factor passing from the cylinder to the plane, and we have
\begin{gather}
	G(\tau,\theta) = \frac{1}{\mathcal{N}}\left\langle \op_1(\vec{y}_1)\op_2(\vec{y}_2) \op^\dag_2\left(\vec{y}_3\right)\op^\dag_1\left(\vec{y}_4\right) \right\rangle_{\RR^d} \\
	\begin{aligned}
		\vec{y}_1&=\tfrac{\sqrt{\mu}}{m_1}e^{-\frac{\tau}{2}} \vec{\Omega},\quad \vec{y}_2 = -\tfrac{\sqrt{\mu}}{m_2}e^{-\frac{\tau}{2}} \vec{\Omega},\\
	 \vec{y}_3 &=-\tfrac{m_2}{\sqrt{\mu}}e^{\frac{\tau}{2}} \vec{\Omega}' , \quad \vec{y}_4=\tfrac{m_1}{\sqrt{\mu}}e^{\frac{\tau}{2}} \vec{\Omega}'.
	\end{aligned}
\end{gather}
 The conformally invariant cross-ratios $z,\bar{z}$ for the correlator of interest are given by
\begin{equation}
	z 
	=  \tfrac{1}{\mu}e^{-(\tau+i\theta)}\left(1+\tfrac{e^{-(\tau+i\theta)}}{m_1+m_2}\right)^{-2}\sim  \tfrac{1}{\mu}e^{-(\tau+i\theta)},
\end{equation}
with $\bar{z}=z^*$ (the complex conjugate of $z$). We have written an expansion in the relevant limit of large $m_{1,2}$, from which we see that the non-relativistic limit corresponds to large $\Delta$ with cross-ratio scaling as $\Delta^{-1}$. For equal masses $m_1=m_2$, we have a simple relation $\rho = \frac{1}{2m}e^{-(\tau+i\theta)}$ to the `radial' coordinate of \cite{Hogervorst:2013sma} (and indeed we arrive at this from a very similar configuration of operator insertions).

To compare our results with the CFT literature, it is also useful to write the correlator in terms of a different function $g$ of cross-ratios $z,\bar{z}$, by stripping off some conventional kinematic factors:
\begin{align}
	G(\tau,\theta) &= \frac{1}{\mathcal{N}}\frac{1}{(y_{12}^2)^\frac{\Delta_1+\Delta_2}{2}(y_{34}^2)^\frac{\Delta_3+\Delta_4}{2}}\left(\frac{y_{24}^2}{y_{14}^2}\right)^\frac{\Delta_1-\Delta_2}{2} \left(\frac{y_{14}^2}{y_{13}^2}\right)^\frac{\Delta_3-\Delta_4}{2} g\left(z,\bar{z}\right) \nonumber \\
	&\sim \left(\mu e^{\tau}\right)^{\Delta_1+\Delta_2}\exp\left(2\tfrac{\Delta_1-\Delta_2}{\Delta_1} e^{-\tau}\cos\theta\right) g(z,\bar{z}) \, .\label{eq:Gg}
\end{align}
Here, $y_{ij}^2 = |\vec{y}_i-\vec{y}_j|^2$, and in the second line we have written the kinematic factors in the relevant limit of large masses with $\tau,\theta$ fixed.

A useful equivalent representation of the correlator $G(\tau,\theta)$ is obtained by decomposing into partial waves $G_l(\theta)$ on $S^{d-1}$. This means expanding in terms of Gegenbauer polynomials $C_l(\cos\theta)$, which are the spherical harmonics depending only on latitude $\theta$ (generalising the familiar Legendre polynomials from the case $d=3$):
\begin{equation}\label{eq:Gl}
\begin{aligned}
	G(\tau,\theta) = \frac{1}{\Omega_{d-1}}\sum_{l=0}^\infty \frac{2l+d-2}{d-2} C_l(\cos\theta) G_{l}(\tau),
\end{aligned}
\end{equation}
where $\Omega_{d-1} = \frac{2\pi^{\frac{d}{2}}}{\Gamma(\frac{d}{2})}$ is the volume of the unit $S^{d-1}$.
See appendix \ref{app:SH} for a brief summary of these harmonics including our conventions.

An important example is the MFT or generalised free correlation function, the product of two-point functions $\langle \op_1\op_2 \op_2^\dag \op_1^\dag\rangle=\langle \op_1 \op_1^\dag\rangle\langle \op_2 \op_2^\dag\rangle $. 
The two-point function of operators on the cylinder separated by Euclidean time $t_E$ and angle $\theta$ is
\begin{equation}\label{eq:2pt}
\Big\langle \op^\dag \op\Big\rangle = \frac{1}{\left(2\cosh(t_E)-2\cos(\theta)\right)^{\Delta}} \sim e^{-\Delta t_E} e^{2\Delta e^{-t_E}\cos\theta},
\end{equation}
where the asymptotic formula is valid in the limit of large $t_E$ and $\Delta$, with $\Delta e^{-t_E}$ held fixed. From this and the definition \eqref{eq:Gdef}, we find the result
\begin{equation}\label{eq:Gfree}
	G_\mathrm{free}(\tau,\theta) = e^{2 e^{-\tau}\cos\theta}.
\end{equation}
The corresponding partial wave decomposition is
\begin{equation}\label{eq:Glfree}
	G_l(\tau) = 2\pi^\frac{d}{2} e^{\frac{d-2}{2} \tau} \,  I_{l+\frac{d}{2}-1}(2e^{-\tau}),
\end{equation}
as shown in appendix \ref{app:SH}.

\subsection{Symmetries in the quantum theory}\label{sec:symm}

We now begin our discussion of quantum non-relativistic dynamics in AdS with a more thorough discussion of the symmetries. We have already seen that AdS becomes a harmonic gravitational potential in the non-relativistic limit; here, we will see that fact emerge in a rather abstract way, with the harmonic oscillator algebra appearing as a contraction of the AdS symmetry algebra.

The symmetry algebra of AdS$_{d+1}$ is $\mathfrak{so}(d,2)$, with complexification spanned by the dilatation $D$, momentum $P_i$, special conformal transformations $K_i$, and rotations $M_{ij}=-M_{ji}$.
We have
\begin{equation}
	D^\dag = D,\quad M_{ij}^\dag = M_{ij}, \quad K_i^\dag = P_i \,,
\end{equation}
with the real algebra of Killing vectors in AdS$_{d+1}$ spanned by the anti-Hermitian generators $iD$, $iM_{ij}$, $i(P_i+K_i)$ and $(P_i^\dag-K_i^\dag)$. The names of the generators come from their action as the conformal symmetries on the boundary after passing to Euclidean AdS and mapping to the plane, which we describe in a moment. The algebra is given by
\begin{equation}\label{eq:confAlg}
	[D,P_i]=P_i,\quad [D,K_i]=-K_i, \quad [K_i,P_j]=2\delta_{ij}D - 2i M_{ij},
\end{equation}
along with the $\mathfrak{so}(d)$ algebra of rotations for $M_{ij}$, and relations telling us that $P_i$ and $K_i$ transform like vectors and $D$ as a scalar under those rotations.

The simplest of these generators is the dilatation $D$, which is the generator of time-translation in global coordinates: $D=i\partial_t$. In other words, it is our Hamiltonian. In the non-relativistic limit for particles of total mass $M$, this will become the rest-energy $Mc^2$ plus the non-relativistic Hamiltonian:
\begin{equation}
	D \sim M + H_\mathrm{NR}.
\end{equation}
This split is already enough to see the relevant reduction of the symmetry algebra. If we neglect $H_\mathrm{NR}$ and $M_{ij}$ relative to the mass $M$, the algebra \eqref{eq:confAlg} reduces to
\begin{equation}\label{eq:NRalgebra}
	[H_\mathrm{NR},P_i]=P_i,\quad [H_\mathrm{NR},K_i]=-K_i, \quad [K_i,P_j]=2\delta_{ij} M.
\end{equation}
We can regard the mass $M$ as a central element of the algebra (commuting with all other elements), so the non-relativistic Hilbert space decomposes into superselection sectors for different masses (eigenvalues of $M$). The resulting algebra is the AdS analogue of the Galilean algebra (or more precisely, the Bargmann algebra including the central extension by mass $M$), which is obtained by an analogous contraction of the Poincar\'e algebra: see figure \ref{fig:symmTable}.\footnote{The Bargmann algebra is itself a contraction of \eqref{eq:NRalgebra}: the arrow  from bottom left to bottom right in figure \ref{fig:symmTable}. To obtain this, first restore units in \eqref{eq:NRalgebra} by including a factor of $\omega$ on the right hand side in in each commutator (giving $P_i,K_i$ units of momentum). Then define $P_i=p_i+i \omega c_i$ and $K_i=p_i-i \omega c_i$, so $p_i$ and $c_i$ generate translations and Galilean boosts respectively. With these definitions \eqref{eq:NRalgebra} is
\begin{equation}
	[c_i,H_\mathrm{NR}] = i p_i, \quad [p_i,H_\mathrm{NR}] = -i\omega^2 c_i, \quad [c_i,p_j]=i \delta_{ij}M,
\end{equation}
and if we set $\omega=0$ this becomes the Bargmann algebra.} This is the Newton algebra $N^-$ in the classification of \cite{Bacry:1968zf}, studied further in \cite{derome1972hooke,dubois1973hooke} under the name of the Hooke algebra $H_-$ (compare figure 1 of \cite{derome1972hooke} to our figure \ref{fig:symmTable}!).

 The algebra \eqref{eq:NRalgebra} is in fact extremely familiar.\footnote{Looking at the commutators of $P_i$, perhaps one might ask ``Is this $a^\dag$ which I see before me?'' \cite{shakespeare}.} To see this, we define rescaled generators $A_i,A^\dag_i$ (for a fixed value of $M$) by
\begin{equation}\label{eq:Adef}
	P_i = i\sqrt{2M} \, A^\dag_i,\quad K_i = -i\sqrt{2M} \, A_i,
\end{equation}
so \eqref{eq:NRalgebra} becomes the algebra of creation and annihilation operators in the $d-$dimensional harmonic oscillator:
\begin{equation}\label{eq:SHOalgebra}
	[H_\mathrm{NR},A_i^\dag] = A_i^\dag,\quad [H_\mathrm{NR},A_i] = -A_i,\quad [A_i,A_j^\dag]=\delta_{ij} \,.
\end{equation}
Thus we see the harmonic potential arise abstractly from the reduction of the symmetry algebra.

We can make this a little more concrete by examining the Killing vector corresponding to $P_i$ in the region $r \ll L$ where we have non-relativistic dynamics. This is not so simple in global coordinates, but is easy to write if we pass to  Poincar\'e coordinates $(\vec{y},z)$ in Euclidean signature:
\begin{equation}\label{eq:Poincare}
	z= \frac{e^{t_E}}{\sqrt{1+r^2}}, \quad \vec{y} = \frac{r}{\sqrt{1+r^2}} e^{t_E} \vec{\Omega} \implies ds^2 = \frac{dz^2+d\vec{y}^2}{z^2}.
\end{equation}
This change of coordinates extends into AdS the conformal map $\vec{y} =  e^{t_E} \vec{\Omega}$ on the boundary ($z\to 0$ or $r\to\infty$) from the plane to the cylinder. These coordinates make manifest the symmetry under translations of $\vec{y}$, generated by the momenta $P_i = -i \frac{\partial}{\partial y^i}$. To describe the action of $P_i$ on the non-relativistic Hilbert space, we now  re-express the Killing vector $\frac{\partial}{\partial y^i}$ in real-time global coordinates, and approximate it in the vicinity of the origin $r\ll 1$:
\begin{align}\label{eq:Pi}
	P_i  &\sim -i e^{-i t}(\partial_i - i x_i\partial_t) \\
	&\sim  i e^{-i t}(m x_i(t)  - i p_i(t)).
\end{align}
For the second line we have written the Killing vector in terms of its action on the wavefunction of a single particle of mass $m$, replacing the time-translation by its leading-order action in the non-relativistic limit $\partial_t \sim -i m$ coming from the rest energy. 
 We have explicitly indicated $t$-dependence in $x$ and $p$ to emphasise that these are Heisenberg picture operators. In fact, this Heisenberg evolution cancels the explicit $e^{-i t}$ factor, so $P_i$ is time-independent when written in terms of Schr\"odinger picture operators $x_i=x_i(0)$, $p_i=p_i(0)$. If we have many particles, $P_i$ acts simultaneously on all of them, so its action on the Hilbert space will be a sum of terms \eqref{eq:Pi} for each particle. But since it acts linearly in $ m\vec{x}$ and $\vec{p}$, this is equivalent to saying that it acts on the centre-of-mass wavefunction.  From this, $A_i^\dag$ in \eqref{eq:Adef} indeed becomes precisely the familiar creation operator of the harmonic oscillator, for the centre-of-mass wavefunction.

From this, we can write the non-relativistic Hamiltonian as
\begin{equation}
	H_\mathrm{NR} = (A^\dag \cdot A + \tfrac{d}{2}) + H,
\end{equation}
where $H$ commutes with the whole symmetry algebra (with a similar split for rotations $M_{ij}$). We interpret the first term as the Harmonic oscillator Hamiltonian for the centre of mass (including ground-state energy $\frac{d}{2}\omega$), and $H$ the Hamiltonian for the relative motion, as in \eqref{eq:COMrelH}.

 The Hilbert space factorises as $\mathcal{H}_\mathrm{NR} = \mathcal{H}_\mathrm{relative}\otimes \mathcal{H}_\mathrm{COM}$, where $\mathcal{H}_\mathrm{COM}$ is the familiar irreducible representation of the harmonic oscillator algebra. That is, it is spanned by a ground state $|0\rangle_\mathrm{COM}$ annihilated by the $A_i$, along with excited states with energies separated by integers (in units of the `AdS energy' $\hbar\omega$) created by acting with $A_i^\dag$ on $|0\rangle_\mathrm{COM}$. Since $A_i$ is proportional to the special conformal generator $K_i$, putting the centre-of-mass wavefunction in the ground state corresponds precisely to a primary state of the CFT (defined as a state annihilated by $K_i$). Acting with $A^\dag_i$ produces a tower of descendants.
 
 We may therefore take the symmetries into account by restricting our considerations to $\mathcal{H}_\mathrm{relative}$, placing the centre-of-mass motion in its ground state. In CFT language, we study dynamics in the subspace spanned by primary states only. These fall into representations of the residual symmetry group $SO(d)$, which commutes with the relative Hamiltonian $H$. In the non-relativistic limit this turns out to be rather convenient, since (as we will see in the next section) we can create such superpositions of multi-particle primary states from simple operator insertions at appropriate locations. This does not hold more generally: beyond the non-relativistic limit, one would have to smear the operators to remove descendants.

The conformal dimension $\Delta$ of a CFT primary state will be given by the eigenvalue of the dilatation operator $D$ in the oscillator vacuum. In the non-relativistic limit this becomes $\Delta \sim  M+\frac{d}{2}+E$, where $M$ is the total mass (in units of $\hbar \omega c^2$ to make it dimensionless), $\frac{d}{2}$ the ground-state energy of the centre-of-mass, and $E$ the eigenstate of the relative Hamiltonian $H$ (in units of $\hbar \omega$). For a single particle, $\mathcal{H}_\mathrm{relative}$ describes only internal degrees of freedom with degenerate energy (for example, a representation of $SO(d)$ corresponding to the particle's spin) so $H=0$, and we have $\Delta \sim m+\tfrac{d}{2}$. The same result can be recovered from expanding the familiar AdS/CFT relation between mass and conformal dimension for a field of any spin $s$ (for the formula at general spin see \cite{Sleight:2017krf}) at large mass,
\begin{equation}
	m^2 = \Delta(\Delta-d)-s \implies \Delta = \tfrac{d}{2} + \sqrt{m^2+s+\left(\tfrac{d}{2}\right)^2} \sim m + \tfrac{d}{2} + O(m^{-1}).
\end{equation}

\subsection{Correlation functions as coherent state amplitudes}\label{sec:Gamp}

We would now like to make contact with CFT correlation functions such as $G(\tau,\theta)$ described in section \ref{sec:correlator}. To do that, we should first understand the quantum states created by operator insertions in the Euclidean cylinder, beginning with single-particle states.

As explained above, a primary state $|\mathcal{O}\rangle$ corresponds to the ground state $|0\rangle$ of the harmonic oscillator. This state is created in the CFT by inserting an operator $\mathcal{O}$ at the origin in the $y$-plane, or in the distant Euclidean past $t_E=-\infty$ on the cylinder. To produce a state created by an operator inserted at another point at Eulidean time $t_E$ at angle $\Omega$, we can simply apply the translation operator $e^{-i \vec{y}\cdot \vec{P}}$ with an appropriate choice of $\vec{y}$, up to a conformal factor $|\vec{y}|^\Delta$ accounting for the conformal map from plane to cylinder:
\begin{align}
	|\mathcal{O}(-t_E,\Omega)\rangle  &= |\vec{y}|^{\Delta}e^{-i \vec{y}\cdot \vec{P}}|\mathcal{O}\rangle \quad (\vec{y} = e^{-t_E}\vec{\Omega}) \\
	&\sim e^{\Delta t_E}e^{\sqrt{2m} \vec{y}\cdot \vec{A}^\dag}|0\rangle.\label{eq:opWavefunc}
\end{align}
In the second line, we have approximated the momentum operators as creation operators acting on the oscillator ground state. This is appropriate as long as $|\vec{y}|\ll 1$ or $|t_E|\gg 1$, so that we remain within the non-relativistic limit. This exponential of a creation operator acting on the ground state gives a coherent state: a Gaussian wavefunction with the same width as the ground state, but centre offset from the origin. It is an eigenstate of the annihilation operators $\vec{A}$, with eigenvalues $\sqrt{2m}\vec{y}$.

As an explicit check of this, we may take the overlap of two such states using
\begin{equation}
	\langle 0| e^{\vec{\beta}\cdot \vec{A}} e^{\vec{\alpha}\cdot \vec{A}^\dag}|0\rangle = e^{\vec{\beta}\cdot\vec{\alpha}}.
\end{equation}
The result is the non-relativistic limit of the conformal two-point function given in \eqref{eq:2pt}.

Now we extend this to two particles, starting with the case when they do not interact. For non-interacting particles, the states created by two operator insertions are simply the tensor product of two coherent states \eqref{eq:opWavefunc}, with a factorised wavefunction. We would like to split this into dependence on the centre-of-mass and separation of the particles. Fortunately, with a judicious choice of operator insertions this is extremely simple. Without interactions, the centre-of-mass special conformal operators $\vec{K}$ split as a sum $\vec{K}_1+\vec{K}_2 = -i\sqrt{2m_1} \vec{a}_1-i \sqrt{2m_2} \vec{a}_2$, where $\vec{a}_{1,2}$ are $d$-tuples of annihilation operators acting only on a single particle. This means that any tensor product of coherent states for each particle (created by acting with $e^{\sqrt{2m_1} \vec{y}_1\cdot \vec{a}_{1}^\dag + \sqrt{2m_2} \vec{y}_2\cdot \vec{a}_{2}^\dag }$ 
on the two-particle ground state) is an eigenstate of $\vec{K}$, with eigenvalues $-2i(m_1\vec{y}_1+m_2 \vec{y}_2)$. Hence, by choosing $m_1\vec{y}_1+m_2\vec{y}_2=0$ the product state is annihilated by $\vec{K}$, so it must be a superposition of primary states, with centre-of-mass wavefunction in the ground state. And indeed, this choice is precisely the relation we found between the location of operator insertions when we studied classical solutions with $\vec{x}_\mathrm{COM}=0$ in \eqref{eq:Gdef}.

This motivates us the consider states $|\psi(\tau_\mathrm{in},\Omega_\mathrm{in})\rangle$ of the relative Hilbert space defined by
\begin{equation}\label{eq:scatDef}
\begin{aligned}
	\left|\op_1\left(-\tau_\mathrm{in} - \tfrac{1}{2}\log\tfrac{m_1^2}{\mu},\Omega_\mathrm{in}\right)\op_2\left(-\tau_\mathrm{in} - \tfrac{1}{2}\log\tfrac{m_2^2}{\mu},-\Omega_\mathrm{in}\right)\right\rangle &\\
	 = \left(\tfrac{\mu}{m_1^2}\right)^{\frac{\Delta_1}{2}} \left(\tfrac{\mu}{m_2^2}\right)^{\frac{\Delta_2}{2}} e^{-(\Delta_1+\Delta_2)\tau_\mathrm{in}} |\psi(\tau_\mathrm{in},\Omega_\mathrm{in})\rangle& \otimes |0\rangle_\mathrm{COM}.
\end{aligned}
\end{equation}
 The first line defines the state in CFT by operator insertions, with arguments denoting Euclidean times and angles on the cylinder $\RR\times S^{d-1}$, and $-\Omega$ means the antipodal point to $\Omega$. The second line defines a state in the non-relativistic Hilbert space, with the centre-of-mass wavefunction in the ground state $|0\rangle_\mathrm{COM}$, and $|\psi(\tau_\mathrm{in},\Omega_\mathrm{in})\rangle$ giving the relative wavefunction. We will explain in a moment why the wavefunction must take this form, even in the presence of interactions.

For non-interacting particles, we may find the wavefunction $|\psi(\tau_\mathrm{in},\Omega_\mathrm{in})\rangle$ from the product of two copies of \eqref{eq:opWavefunc}, after passing to the centre-of-mass and separation coordinates. We find Gaussian coherent states in terms of the separation variables:
\begin{equation}\label{eq:scatCoherent}
	|\psi(\tau_\mathrm{in},\Omega_\mathrm{in})\rangle \sim e^{\vec{\alpha}_\mathrm{in}\cdot \vec{a}^\dag}  |0\rangle, \quad \vec{\alpha}_\mathrm{in} = \sqrt{2} e^{-\tau_\mathrm{in}}\vec{\Omega}_\mathrm{in} \quad (\text{non-interacting}).
\end{equation}
Here, $\vec{a}^\dag$ are creation operators built from the separation variables ($\vec{x}=\vec{x}_1-\vec{x}_2$, etc), written as $\vec{a}$ to distinguish from the centre-of-mass versions $\vec{A}$ in section \ref{sec:symm}. Similarly, $|0\rangle$ is the harmonic oscillator ground state for the separation wavefunction.

We continue to use the same definition \eqref{eq:scatDef} of $|\psi(\tau_\mathrm{in},\Omega_\mathrm{in})\rangle$ when we include interactions between our particles. But for this definition to make sense, we must justify why the COM wavefunction must be in the ground state. In other words, we must justify why the state in the left of \eqref{eq:scatDef}  is annihilated by the centre-of-mass annihilation operators $\vec{a}$. First, note that this will hold for large negative $\tau_\mathrm{in}$ as long as the potential decays with distance: in that regime the particles will be well-separated so we are justified in neglecting interactions. To extend this to finite values of $\tau_\mathrm{in}$ where interactions become important, consider acting on the definition \eqref{eq:scatDef} by a  Euclidean time evolution $e^{-\tau D}$. On the left-hand-side (defining a state in terms of CFT operator insertions), this simply increases $\tau_\mathrm{in}$, mapping $\tau_\mathrm{in}\mapsto \tau_\mathrm{in}+\tau$. On the right-hand-side (defining a state in the non-relativistic Hilbert space), we use the non-relativistic decomposition of the dilatation operator,  $D=\Delta_1+\Delta_2-\frac{d}{2} + \vec{A}^\dag\cdot \vec{A} +H $ (with $\Delta_{1,2} = m_{1,2}+\frac{d}{2}$). The term $e^{-\tau\vec{A}^\dag\cdot \vec{A}}$ acting on the centre-of-mass wavefunction leaves the ground state $|0\rangle_\mathrm{COM}$ invariant (and if there were any overlap with excited states it would be suppressed, so any small corrections from interactions for large negative $\tau_\mathrm{in}$ do not get amplified). 

A corollary of this argument is an expression for the relative wavefunction $|\psi(\tau_\mathrm{in},\Omega_\mathrm{in})\rangle$ including interactions. First, the expression as a coherent state \eqref{eq:scatCoherent} holds in the $\tau_\mathrm{in}\to -\infty$ limit, where interactions are negligible. Secondly, from the action of $e^{-\tau D}$ we have the relation
\begin{equation}\label{eq:psiscatevolution}
	e^{-\tau H}|\psi(\tau_\mathrm{in},\Omega_\mathrm{in})\rangle =  e^{-\frac{d}{2}\tau} |\psi(\tau_\mathrm{in}+\tau,\Omega_\mathrm{in})\rangle.
\end{equation}
Combining these, we have the following expression:
\begin{equation}\label{eq:psiscat}
	|\psi(\tau_\mathrm{in},\Omega_\mathrm{in})\rangle = \lim_{\tau\to \infty} e^{-(H-\frac{d}{2})(\tau+\tau_\mathrm{in})} e^{\sqrt{2}e^{\tau}\Omega_\mathrm{in}\cdot a^\dag}|0\rangle.
\end{equation}
If we took $H$ to be the free Hamiltonian $\vec{a}^\dag\cdot\vec{a}+\frac{d}{2}$, the expression on the right would be independent of $\tau$. Our definition is self-consistent if this limit exists for all $\tau_\mathrm{in}$ and approaches the free result \eqref{eq:scatCoherent} for $\tau_\mathrm{in}\to-\infty$.

 The result \eqref{eq:psiscat} is similar to the definition of a scattering state in flat spacetime by evolving backward with the free Hamiltonian and forward with the interacting Hamiltonian. The difference here is that we use Euclidean time, since the particles do not become well-separated in real time in the presence of the harmonic AdS potential.

Now we may write our Euclidean correlation function \eqref{eq:Gdef} simply as the overlap of two of these states:
\begin{equation}\label{eq:scatterM}
\begin{gathered}
	G(\tau,\theta) =\langle \psi(\tau_\mathrm{out},\Omega_\mathrm{out})|\psi(\tau_\mathrm{in},\Omega_\mathrm{in})\rangle, \\
	\qquad \tau= \tau_\mathrm{in}+\tau_\mathrm{out},\quad \cos\theta = \Omega_\mathrm{in}\cdot \Omega_\mathrm{out}.
\end{gathered}
\end{equation}
More generally, we can write the same correlator including a time-evolution operator as $\langle \psi(\tau_\mathrm{out},\Omega_\mathrm{out})|e^{-H\tilde{\tau}}|\psi(\tau_\mathrm{in},\Omega_\mathrm{in})\rangle$, with $\tau= \tau_\mathrm{in}+\tau_\mathrm{out}+\tilde{\tau}$. We can also include evolution in real time by giving $\tau$ an imaginary part. In particular, we can write a correlator as a matrix element of the time-evolution operator $e^{-i H t}$:
\begin{equation}
	G(\tau+i t,\theta) =\langle \psi(\tfrac{\tau}{2},\Omega_\mathrm{out})|e^{-i H t}|\psi(\tfrac{\tau}{2},\Omega_\mathrm{in})\rangle
\end{equation}
with $\cos\theta = \Omega_\mathrm{in}\cdot \Omega_\mathrm{out}$. We will later use such a correlator with $t=\frac{\pi}{\omega}$ to extract flat spacetime scattering amplitudes.

\section{Quantum dynamics}\label{sec:QM}

In this section we describe how to evaluate the amplitude \eqref{eq:scatterM} as a sum over intermediate states, and relate this to the S-channel conformal block decomposition of the dual CFT correlation function \eqref{eq:Gdef}.

\subsection{The correlator from the spectrum}

We find the non-relativistic spectrum by diagonalising the Hamiltonian
\begin{equation}
	H =	\frac{p^2}{2\mu} + \frac{1}{2}\mu r^2+  V(r)
\end{equation}
for some interaction potential satisfying $V(r)\to 0$  as $r\to\infty$. Due to the confining quadratic potential, the spectrum of the Hamiltonian will be discrete, with states $|l,m,n\rangle$, satisfying  $H|l,m,n\rangle = E_{l,n}|l,m,n\rangle$. Using rotational invariance, they are labelled by angular momentum $l=0,1,2,\ldots$ (telling us that the states transform in the $l$-fold symmetric traceless representation of $SO(d)$) and the `magnetic quantum number' $m$ (labelling states within a representation of $SO(d)$). Finally, $n=0,1,2\ldots$ labels the states in order of increasing energy for given $l$ (that is, $n$ labels the Regge trajectories).

We can write these eigenstates as
\begin{equation}
\label{eq:psilnm}
	\psi_{l,m,n}(x) = \langle x|l,m,n\rangle = \frac{1}{r^\frac{d-1}{2}} \phi_{l,n}(r) Y_{l,m}(\Omega),
\end{equation}
where $Y_{l,m}$ are harmonics on $S^{d-1}$ satisfying $\nabla^2_{S^{d-1}} Y_{l,m}=-l(l+d-2)Y_{l,m}$, and $\phi_{l,n}$ is an eigenfunction of the radial Hamiltonian, solving
\begin{equation}\label{eq:radialSchr}
	-\frac{1}{2\mu}\phi''(r) + \left[\frac{\left(l+\tfrac{d-3}{2}\right)\left(l+\tfrac{d-1}{2}\right)}{2\mu r^2}+ \tfrac{1}{2}\mu r^2 + V(r)\right] \phi(r) = E \phi(r).
\end{equation}
The states are normalised as $\int |\psi_{l,m,n}|^2 d^dx=1$, or in terms of the radial wavefunction $\int_0^\infty|\phi_{l,n}(r)|^2dr=1$ (using normalisation of the harmonics, $\int_{S^{d-1}}|Y_{l,m}|^2=1$). 

At the origin, the we demand that $\phi\propto r^{l+\frac{d-1}{2}}$ (or $\psi\propto r^l$), so the coefficient of the independent solution to \eqref{eq:radialSchr} ($\phi\propto r^{-l-\frac{d-3}{2}}$) vanishes. This assumes that $V$ is less singular than $r^{-2}$ as $r\to 0$. At large $r$, the solution will decay as
\begin{equation}\label{eq:phiasymp}
	\phi_{l,n}(r)\sim \frac{A_{l,n}}{\sqrt{r}} (\mu r^2)^{\tfrac{E_{l,n}}{2}} e^{-\frac{1}{2}\mu r^2}, \quad r\to\infty
\end{equation}
for some dimensionless coefficient $A_{l,n}$ which we take to be real and positive to fix the phase, determined by normalisation of the wavefunction.

We can express the matrix element \eqref{eq:scatterM} as a sum over a complete set of these eigenstates. For this we need the overlap between the energy eigenstates and the states $|\psi(\tau_\mathrm{in},\Omega_\mathrm{in})\rangle$. For this, using the definition \eqref{eq:psiscat} we may use the wavefunction at large negative imaginary time. This is a coherent state $e^{\vec{\alpha}\cdot \vec{a}^\dag}|0\rangle$ with large $|\vec{\alpha}|$, a Gaussian of fixed width supported at very large radius $r\sim \sqrt{\frac{2}{\mu}}|\vec{\alpha}|$. In terms of angles on $S^{d-1}$, the wavefunction is supported in a small sector in the direction of $\vec{\alpha}$, so (for the purposes of integrating against a fixed function in the $|\vec{\alpha}|\to\infty$ limit) its angular dependence can be approximated by a $\delta$-function:
\begin{align}
	\langle \vec{x}|e^{\alpha \cdot a^\dag}|0\rangle &= \left(\frac{\mu}{\pi}\right)^{\frac{d}{4}} e^{-\tfrac{\mu}{2} \vec{x}^2+ \sqrt{2\mu}\,\vec{\alpha}\cdot \vec{x} - \frac{1}{2}\vec{\alpha}^2} \\
	&\sim \left(\frac{\mu}{\pi}\right)^{\frac{d}{4}}\left(\frac{2\pi^2}{\mu\alpha^2 r^2}\right)^{\frac{d-1}{4}}  e^{-\tfrac{\mu}{2} r^2+ \sqrt{2\mu}\,|\alpha| r - \frac{1}{2}|\alpha|^2} \delta_{\Omega}.
\end{align}
In the second line, $\delta_\Omega$ denotes a delta-function on $S^{d-1}$ supported in the direction of $\alpha$.

 We can choose coordinates on $S^{d-1}$ so that $\Omega_\mathrm{in}$ is aligned with the North Pole, so that the only intermediate state for given $n,l$ is proportional to the $m=0$ `zonal' harmonic $Y_{l,m=0}=Y_l$. This depends only on latitude $\theta$ (angle from the North pole), and is given by a Gegenbauer polynomial:
\begin{equation}
	Y_{l}(\theta) = \frac{1}{\sqrt{\mathcal{N}_l}} C_l(\cos\theta)
\end{equation}
for an appropriate normalisation constant $\mathcal{N}_l$ (see appendix \ref{app:SH}).  Using this and the asymptotics of the radial wavefunction \eqref{eq:phiasymp}, we find
\begin{equation}
	\langle l,m=0,n|e^{\vec{\alpha}\cdot \vec{a}^\dag}|0\rangle\sim A_{l,n}\frac{C_l(\cos\theta)}{\sqrt{\mathcal{N}_l}}  \pi^{\frac{d}{4}}  \left(\frac{|\vec{\alpha}|}{\sqrt{2}}\right)^{E_{l,n}-\frac{d}{2}} \quad\text{as } |\vec{\alpha}|\to\infty.
\end{equation}
From this we find the overlap between our Euclidean scattering states $|\psi(\tau,\Omega)\rangle$ and an eigenstate in terms of the energy $E_{l,n}$ and the decay coefficient $A_{l,n}$:
\begin{equation}\label{eq:psiscatnlm}
	\left\langle n,l,m=0\middle|\psi(\tau,\Omega)\right\rangle = \pi^{\frac{d}{4}}   A_{l,n}\frac{C_l(\cos\theta)}{\sqrt{\mathcal{N}_l}}   e^{-(E_{l,n}-\frac{d}{2})\tau},
\end{equation}
where $\theta$ is the angle between $\Omega$ and the North pole.

Now by summing over a complete set of states at given $l$, we can extract the partial waves of our correlation function \eqref{eq:Gl}:
\begin{equation}\label{eq:Glsum}
	G_l(\tau)\sim \pi^{\frac{d}{2}}  \sum_{n=0}^\infty   A_{l,n}^2 e^{-(E_{l,n}-\frac{d}{2})\tau}.
\end{equation}
We made use of the normalisation constant $N_l$ and the value of $C_l(1)$ (see appendix \ref{app:SH}). These partial waves can be assembled into the correlation function $G(\tau,\theta)$ using \eqref{eq:Gl}, as
\begin{equation}\label{eq:Schannel}
	G(\tau,\theta) = \frac{ \pi^{\frac{d}{2}}}{\Omega_{d-1}} \sum_{n=0}^\infty\sum_{l=0}^\infty \frac{2l+d-2}{d-2} C_l(\cos\theta) A_{l,n}^2 e^{-(E_{l,n}-\frac{d}{2})\tau}.
\end{equation}

\subsection{S-channel conformal block decomposition}

Unsurprisingly, the sum over states we found in \eqref{eq:Glsum} can be precisely matched with the decomposition of the correlation function into S-channel conformal blocks. Each term corresponds to a single conformal primary state in the CFT, with an absence of descendant contributions as expected from the discussion of section \ref{sec:Gamp}.

The S-channel conformal block decomposition is conventionally written in terms of the correlation function $g(z,\bar{z})$ defined in \eqref{eq:Gg}:
\begin{equation}\label{eq:OPES}
	g(z,\bar{z}) = \sum_{l=0}^\infty\sum_{n=0}^\infty f_{l,n}^2\; g_{\Delta_{l,n},l}(z,\bar{z}).
\end{equation}
The sum runs over conformal primary operators $\op_{l,n}$ of dimensions $\Delta_{l,n}$, weighted by OPE coefficients $ f_{l,n}= f_{\op_1\op_2\op_{l,n}}$. We have organised them into Regge trajectories labelled by $n=0,1,2,\ldots$, defined so that $\op_{l,n}$ is the $(n+1)$th lightest primary operator of spin $l$ (in the $l$-index symmetric traceless representation of $SO(d)$) appearing in the OPE of $\op_1$ and $\op_2$.

 In appendix \ref{app:Sblocks}, we determine the S-channel blocks for external scalars in the relevant limit of large $\Delta_{1,2}$, with internal dimension $\Delta=\Delta_1+\Delta_2+O(1)$ and cross-ratios $z,\bar{z}$ of order $\Delta^{-1}$. We do this first in $d=2,4$ by taking limits of closed form expressions in those cases, before verifying that the result obeys the conformal Casimir equation in general dimension in the appropriate limit. We find
 \begin{equation}\label{eq:Sblocks}
	g_{\Delta,l}(z,\bar{z}) \sim \mathcal{N}_{d,l} \left(z\bar{z}\right)^\frac{\Delta}{2} C_l\left(\tfrac{z+\bar{z}}{2\sqrt{z \bar{z}}}\right) \exp\left(\tfrac{\Delta_2^2}{\Delta_1+\Delta_2} (z+\bar{z})\right).
\end{equation}
In our conventions, the blocks are normalised by $g_{\Delta,l}(z,\bar{z})\sim  z^{\frac{\Delta-l}{2}}\bar{z}^{\frac{\Delta+l}{2}}$ in the limit $0<z\ll \bar{z}\ll 1$ (as in \cite{Caron-Huot:2017vep}, for example). This means that the coefficients $\mathcal{N}_{d,l}$ are given by\footnote{For comparison to other conventions, see \cite{Poland:2018epd}. Note that $\mathcal{N}_{d,l}$ here differs from $\mathcal{N}_{d,l}$ in that paper by a factor of $(-1)^l$, because we define the coefficient of the blocks as $f_{12p}^2$ rather than $f_{12p}f_{p21}$, and OPE coefficients change sign under odd permutation of indices.}
\begin{equation}
	\mathcal{N}_{d,l} = \frac{\Gamma(\frac{d}{2}-1)\Gamma(l+1)}{\Gamma(\frac{d}{2}+l-1)}.
\end{equation}

Translating the expansion \eqref{eq:OPES} to the correlator $G$ in terms of variables $\tau,\theta$ using \eqref{eq:Gg} and the appropriate limit of the blocks \eqref{eq:Sblocks}, the S-channel conformal block expansion becomes %
\begin{equation}
	G(\tau,\theta) \sim \sum_{l,n=0}^\infty f_{l,n}^2 \,\mathcal{N}_{d,l}\, C_l(\cos\theta) \left(\mu e^\tau\right)^{-(\Delta-\Delta_1-\Delta_2)}.
\end{equation}
The kinematic dependence of the terms precisely matches that of \eqref{eq:Schannel}, so we can identify the eigenstates of the nonrelativistic Hamiltonian with intermediate primary states of conformal dimension
\begin{equation}\label{eq:deltanl}
	\Delta_{l,n} = \Delta_1+\Delta_2+E_{l,n}-\tfrac{d}{2}.
\end{equation}
By matching coefficients, we read off the OPE coefficients in terms of the fall-off \eqref{eq:phiasymp} of the eigenstate wavefunctions:
\begin{equation}\label{eq:OPEA}
	f_{l,n}^2  
	= \frac{\Gamma\left(\tfrac{d}{2}+l\right)}{2l!}\mu^{E_{l,n}-\frac{d}{2}} A_{l,n}^2 \,.
\end{equation}
These squared OPE coefficients are manifestly positive, consistent with unitarity.

\subsection{Free correlation function}

We can check this against known results for the free correlation function, where the intermediate states and OPE coefficients are those of MFT.

The solution to the nonrelativistic Schr\"odinger equation with only the harmonic potential ($V=0$) which decays at infinity can be written in terms of a confluent hypergeometric function, as
\begin{equation}\label{eq:philn}
	\phi_{l,n}(r) = \sqrt{\frac{2\Gamma(\frac{d}{2}+l+n)}{n!\Gamma(\frac{d}{2}+l)^2}} \frac{e^{-\frac{1}{2}\mu r^2}}{\sqrt{r}}(\mu r^2)^{\frac{l}{2}+\frac{d}{4}}{}_1F_1(-n;l+\tfrac{d}{2};\mu r^2),
\end{equation}
with $E_{l,n}=\frac{d}{2}+2n+l$. This obeys the appropriate boundary conditions at $r=0$ for nonnegative integer $n$, and we have chosen the coefficient to normalise the state in that case. The primary states thus have energies
\begin{equation}
	E_{n,l} = \frac{d}{2}+2n+l\implies \Delta_{n,l} =\Delta_1+\Delta_2 + 2n +l,
\end{equation}
so the corresponding dimensions are precisely those of double-trace operators in MFT as expected. These wavefunctions $\phi_{l,n}$ in \eqref{eq:philn} can also be obtained by taking an appropriate large dimension, small radius limit of general expressions for relativistic wavefunctions in AdS (eigenvalues of the AdS Laplacian), for example from equation (2.19) of \cite{Fitzpatrick:2010zm}.

 From the asymptotics of the wavefunction we can extract the OPE coefficients:
\begin{equation}\label{eq:freeAnl}
	A_{n,l}^2 = \frac{2}{n!\Gamma(\frac{d}{2}+l+n)}\implies f_{l,n}^2  \sim \frac{\Gamma\left(\tfrac{d}{2}+l\right)}{n!l!\Gamma(\frac{d}{2}+l+n)}\mu^{2n+l}.
\end{equation}
This simple answer is a large $\Delta$ limit of the general expression for OPE coefficients in MFT \cite{Fitzpatrick:2011dm,Karateev:2018oml}, which we give here only to illustrate how complicated it is:
\begin{align*}\label{eq:MFTOPE}
	f_{l,n}^2 =&\frac{(-1)^{n} \Gamma (\frac{d}{2}-\Delta_{1}) \Gamma (\frac{d}{2}-\Delta_{2}) \Gamma (\frac{d}{2}+l) \Gamma (l+n+\Delta_{1}) \Gamma (l+n+\Delta_{2}) 
}{
\Gamma (\Delta_{1}) \Gamma (\Delta_{2}) \Gamma (l+1) \Gamma (n+1) \Gamma (\frac{d}{2}+l+n) \Gamma (\frac{d}{2}-n-\Delta_{1}) \Gamma (\frac{d}{2}-n-\Delta_{2})}
\\ &\qquad\times \frac{\Gamma (d-2 n-\Delta_{1}-\Delta_{2}) \Gamma (l+2 n+\Delta_{1}+\Delta_{2}-1) \Gamma (-\frac{d}{2}+l+n+\Delta_{1}+\Delta_{2})
}{
 \Gamma (d-n-\Delta_{1}-\Delta_{2}) \Gamma (2 l+2 n+\Delta_{1}+\Delta_{2}-1) \Gamma (-\frac{d}{2}+l+2 n+\Delta_{1}+\Delta_{2})
}.
\end{align*}
It is satisfying to start with this formula and see the dependence on $\Delta_{1,2}$ collapsing to the simple reduced mass combination $\mu\sim \frac{\Delta_1\Delta_2}{\Delta_1+\Delta_2}$ so familiar from undergraduate mechanics!

From this spectrum, we can explicitly construct the correlation function by performing the sum over states \eqref{eq:Glsum}:
\begin{equation}\label{eq:freePartialWaves}
\begin{aligned}
	G_l(\tau) &= \pi^\frac{d}{2} \sum_{n=0}^\infty \frac{2}{n!\Gamma(\frac{d}{2}+l+n)} e^{-(2n+l)\tau }, \\
\end{aligned}
\end{equation}
which indeed evaluates to the Bessel function \eqref{eq:Glfree} giving the partial waves of the free correlator \eqref{eq:Gfree}. 

\subsection{The S-channel resolvent}\label{sec:resolvent}

We have written our correlation function $G(\tau,\theta)$ in \eqref{eq:Schannel} as a sum over physical intermediate states with energies $E=E_{l,n}$. For some purposes (one example of which is given in section \ref{sec:WKBtoG}), an alternate expression for $G$ may be useful, where the discrete sum of energies $E_{l,n}$ in \eqref{eq:Glsum} is replaced by a contour integral over complex $E$:
\begin{equation}\label{eq:Schannel2}
	G_l(\tau) =  \pi^{\frac{d}{2}} \int_\Gamma \frac{dE}{2\pi i} \,  R_l(E) e^{-\tau (E-\frac{d}{2})},
\end{equation}
for an appropriate contour $\Gamma$. The function $R_l(E)$, which we call an `S-channel resolvent', is a meromorphic function of $E$ with simple poles only at the energies $E_{l,n}$ of intermediate states, and corresponding residues proportional to $A_{l,n}^2$:
\begin{equation}\label{eq:Rlres}
	\Res_{E\to E_{l,n}} R_l(E) = -A_{l,n}^2.
\end{equation}
Evaluating this integral as a sum over residues (with $\Gamma$ encircling contours clockwise) recovers the sum over intermediate states labelled by $n$.

We will not make significant use of this representation in the following, so we defer details to appendix \ref{app:resolvent}. There, we discuss non-uniqueness of the resolvent, several methods for constructing functions $R_l$, and the relation to the CFT `Euclidean inversion formula' \cite{Caron-Huot:2017vep}.

\section{Classical dynamics}\label{sec:classical}

We now discuss classical non-relativistic dynamics in AdS. Firstly, we compute the correlation function $G(\tau,\theta)$ in terms of an on-shell action for classical particle motion in Euclidean time. Secondly, we find the spectrum of states and OPE coefficients in the classical limit using the WKB approximation. Finally, we sketch how these are related, with $G(\tau,\theta)$ obtained as a certain saddle-point approximation to the sum over intermediate states.

We note that we are considering a classical limit described by particles, not fields, so our classical solutions extremise a `worldline' action in terms of particle trajectories rather than classical field configurations. This can be thought of as a limit $\hbar \to 0$ where $m$ is held fixed, rather than fixing the Compton wavelength $\frac{\hbar}{m c}$. This is suitable in circumstances where particle number is small and typically conserved such as the non-relativistic limit in which particle production is energetically suppressed. See \cite{Maxfield:2017rkn} for a detailed discussion of the worldline formalism of QFT in AdS.

There is a separate limit of non-relativistic classical fields in AdS ($c\to\infty$, $\hbar\to 0$, $m\to 0$ with $\frac{\hbar}{m \omega}$ fixed), which is interesting but beyond the scope of this paper. The example of a self-gravitating free scalar field was studied in \cite{Bizon:2018frv}, and a field with quartic self-interaction in \cite{Craps:2021xmk}. These works illustrate our philosophy that many interesting problems in AdS have simpler non-relativistic analogues that nonetheless retain the salient features of the full problem.

\subsection{Euclidean dynamics and CFT correlation functions}\label{sec:classcorr}

We now consider in more detail the `Euclidean scattering' of two particles in AdS interacting via a potential $V(r)$, as introduced in section \ref{sec:Escat}. After fixing the centre-of-mass as discussed in section \ref{sec:classSym}, their non-relativistic dynamics is  governed by the Euclidean action
\begin{equation}\label{eq:NRaction}
	S_E = \int dt_E \left[\tfrac{1}{2}\mu(\dot{r}^2 + r^2 \dot{\phi}^2 + r^2) + V(r)\right],
\end{equation}
where $\phi$ is an anglular coordinate in the plane of motion. The dot denotes differentiation with respect to Euclidean time $t_E = it$. The rotational and time-translation symmetries give rise to conserved angular momentum and energy,
\begin{equation}
	J_E = \mu r^2 \dot{\phi}, \qquad E = -\tfrac{1}{2}\mu(\dot{r}^2+r^2\dot{\phi}^2- r^2) + V(r).
\end{equation}
We have defined a `Euclidean' angular momentum $J_E$ which is real for real solutions in Euclidean time; it is related to `Lorentzian' angular momentum $l$ by $l \sim iJ_E$. We have defined $E$ (with `Euclidean kinetic energy' contributing negatively) to match with the usual energy on continuation back to real time. Using these conserved quantities, we obtain a first-order equation of motion,
\begin{equation}\label{eq:NREOM}
	\tfrac{1}{2}\mu \dot{r}^2 = \tfrac{1}{2}\mu r^2+V(r) - \frac{J_E^2}{2\mu r^2}-E = \frac{\kappa(r)^2}{2\mu}.
\end{equation}
The second equality defines the radial Euclidean momentum $\kappa(r) = \mu |\dot{r}|$.

The boundary conditions define a `time delay' $\tau$ and a `scattering angle' $\theta$ by
\begin{gather}
	r(t_E) \sim \frac{1}{\sqrt{\mu}}\exp\left(|t_E|-\tfrac{1}{2} \tau\right), \qquad t_E\to \pm \infty, \label{eq:rBC} \\
	\phi(\infty)-\phi(-\infty) = \theta \; .\qquad\qquad
\end{gather}
From the considerations of section \eqref{sec:Escat}, such a solution will correspond to a saddle-point for the correlation function $G(\tau,\theta)$: if we extrapolate the particle motion into the relativistic regime (where the interactions are neglected so they follow geodesics), they meet the AdS boundary at the locations of operator insertions.

Integrating the first-order equation of motion \eqref{eq:NREOM}, we can express $\tau$ and $\theta$ in terms of the conserved quantities $E$ and $J_E$:
\begin{gather}
	\theta(E,J_E) = 2J_E\int_{r_0}^\infty \frac{dr}{r^2 \kappa(r)} \label{eq:thetaint}\\
	\tau(E,J_E) = \lim_{R\to\infty} \left[2\mu \int_{r_0}^{R} \frac{dr}{\kappa(r)} - \log\left(\mu R^2\right)\right].\label{eq:tauint}
\end{gather}
The lower limit of integration $r_0$ is determined by the minimal separation of the particles, the largest $r_0$ for which $\kappa(r_0)=0$. We may think of $R$ in the expression for $\tau$ as a `matching radius', where we match the non-relativistic regime ($r<R$) to the non-interacting regime ($r>R$): $R$ is sufficiently large that \eqref{eq:rBC} applies, though still much smaller than the AdS length. The factors of two account for both the ingoing ($t_E<0$) and outgoing ($t_E>0$) parts of the evolution.

These expressions identify the classical solutions corresponding to a correlation function $G(\tau,\theta)$ by relating the kinematics $\tau,\theta$ to conserved quantities $E,J_E$. The classical value of the correlator is obtained from the on-shell action of this solution, roughly $G\approx e^{-S_E}$. To properly account for the complete action, in this case we add three different contributions. The most obvious and interesting piece is the non-relativistic Euclidean action \eqref{eq:NRaction}, up to some radius $R$ as above. To this we must add an additional contribution coming from the rest-energy of the particles, their mass times elapsed time. Finally, we must include the action from the early and late stages of the scattering, where the interaction is weak but the non-relativistic limit is inapplicable: this is given by the mass times arclength along the geodesic particle worldlines.

We can write the interesting non-relativistic piece of the action as an integral by making use of conserved quantities, just as we did for $\tau$ and $\theta$:
\begin{equation}\label{eq:S0}
	S_0(E,J_E) = 2\lim_{R\to\infty} \left[\int_{r_0}^R \frac{\mu^2 r^2+2\mu V(r)-\mu E}{\kappa(r)}dr - \frac{1}{2}\mu R^2\right].
\end{equation}
We have subtracted a term which grows as we take $R\to\infty$; this subtraction appears naturally from the other contributions. The limit exists under the very mild fall-off condition that $\frac{V(r)}{r}$ is integrable.
 
To account for the complete action, we first consider the geodesic action for $r>R$. For convenience, we repeat the equation for the geodesics:
\begin{equation}
	r_{1,2} = e^{s_{1,2}},\quad (t_E)_{1,2} = s_{1,2}  - \tfrac{1}{2}\log\left(1+e^{2s_{1,2}}\right) + \tfrac{1}{2}\tau +\tfrac{1}{2}\log\tfrac{m_{1,2}^2}{\mu}.
\end{equation}
 In terms of the individual particle positions, the matching radius $r=R$ corresponds to $r_{1,2}=\frac{\mu}{m_{1,2}}R$, so $s_{1,2} = \log\left(\frac{\mu}{m_{1,2}}R\right)$. The action contributed by a geodesic is the mass times the arclength, which gives
 \begin{equation}
 	S_\mathrm{geodesics} =2(m_1+m_2)\log r_c -2m_1\log\left(\tfrac{\mu}{m_{1}}R\right)-2m_2\log\left(\tfrac{\mu}{m_{2}}R\right),
 \end{equation}
 where $r_c$ is a cutoff radius much larger than the AdS length. We subtract this cutoff term, interpreted as a renormalisation of the operators to have a canonically normalised two-point function.

The last piece comes from the rest-energy, for which we need the time $t_E$ at which we match to the geodesics:
\begin{equation}
	(t_E)_{1,2}=\frac{1}{2}\log(\mu R^2) + \frac{\tau}{2} - \frac{\mu^2}{2m_{1,2}^2}R^2 + \cdots.
\end{equation}
The last term comes from expanding the geodesics as $s\to -\infty$ to subleading order: it is important to keep this term because it is finite in the non-relativistic limit (it contributes an action independent of $c$, while the other included terms give contributions proportional to $c^2$).  This term provides the subtraction of $\tfrac{1}{2}\mu R^2$ in \eqref{eq:S0}.
 
 Adding up these pieces, the total regulated action for the whole scattering is
\begin{equation}
	S = (m_1+m_2)\tau + m_1\log \frac{m_1^2}{\mu} + m_2\log \frac{m_2^2}{\mu}+S_0,
\end{equation}
so the saddle-point for the relevant four-point function contributes $e^{-S}$. The first three terms give the classical limit of $-\log\mathcal{N}$, where $\mathcal{N}$ is the normalising factor in \eqref{eq:Gdef}, so assuming our classical solution provides the dominant contribution to the correlator of interest, we find
\begin{equation}\label{eq:Gclass}
	G(\tau,\theta) \approx e^{-S_0} \,.
\end{equation}

The map from conserved quantities $E,J_E$ to kinematics $\tau,\theta$ is not guaranteed to be one-to-one. There may be several solutions for given $\tau,\theta$, in which case we should sum over saddle-points so the correlation function is dominated by the minimal action $S_0$. First-order phase transitions between different solutions are possible as $\tau,\theta$ are varied. There may also be no classical solutions for some $\tau,\theta$, in which case the quantum theory is required to compute the correlator (or perhaps a complex saddle-point solution).

The integrals for $\tau$, $\theta$ and $S_0$ are not independent functions of $E$ and $J_E$, since they obey Hamilton-Jacobi relations. These relate the variations of the on-shell action with respect to boundary conditions (here the kinematic variables $\tau,\theta$) to the conserved quantities $E,J_E$:
\begin{equation}\label{eq:HJ}
	\frac{\partial S_0}{\partial \tau} = E, \quad \frac{\partial S_0}{\partial \theta} = J_E.
\end{equation}
This provides an alternative method to compute the  non-relativistic action $S_0$ instead of the integral \eqref{eq:S0} (or a consistency check).

For the free case $V=0$, evaluating the integrals gives the following:
\begin{equation}\label{eq:freetauthetaS0}
	\tau = -\log\left(\frac{\sqrt{E^2+J_E^2}}{2}\right), \quad \theta = \tan^{-1}\left(\frac{J_E}{E}\right), \quad S_0 = -E \,,
\end{equation}
where we take a branch of $\tan^{-1}$ so that $\theta$ is continuous at $E=0$. Inverting the relation between kinematics $\tau,\theta$ and conserved quantities $E,J_E$, we find
\begin{equation}
	E = 2e^{-\tau} \cos\theta , \quad J_E =  2e^{-\tau} \sin\theta \implies S_0 = -2e^{-\tau} \cos\theta.
\end{equation}
The Hamilton-Jacobi relations \eqref{eq:HJ} indeed hold, and $e^{-S_0}$ gives the free correlator \eqref{eq:Gfree} once again.

This classical non-relativistic limit is valid when $1 \ll |E| \ll 
\mu$. The lower limit avoids quantum corrections (since the energy is large in units of $\hbar \omega$, or the action is large compared to $\hbar$), and the upper limit avoids relativistic velocities. The lower limit may be modified by a sufficiently strong interaction potential.

\subsection{Spectrum from the WKB approximation}

We now describe the spectrum of states using the WKB approximation, determining the spectrum of energies $E_{l,n}$ and associated coefficients $A_{l,n}$. 

Assume for simplicity of presentation that the classically allowed region $\tfrac{1}{2}\mu r^2 + \tfrac{J^2}{2\mu r^2} + V(r) < E$ is a single interval $r_1<r<r_0$ (generalised with minor modifications). We write $J^2-\frac{1}{4} = (l+\frac{d-3}{2})(l+\frac{d-1}{2})$ for the coefficient of the angular momentum potential, so $J = l +\frac{d-2}{2}$; in the classical limit $J\gg 1$ we may neglect the $\frac{1}{4}$ shift. The WKB solution to the time-independent radial Schr\"odinger equation \eqref{eq:radialSchr} in that region is
\begin{gather}
	\phi(r) \sim \sqrt{\frac{4\mu }{T \,k(r)}} \cos\left(\int_r^{r_0} k(r')dr' - \frac{\pi}{4} \right), \\
	\text{where}\quad k(r) = \sqrt{2\mu\left(E-\tfrac{1}{2}\mu r^2 - \tfrac{J^2}{2\mu r^2} - V(r)\right)},
\end{gather}
with energies determined by the quantisation condition
\begin{equation}
	\int_{r_1}^{r_0} k(r)dr = (n+\tfrac{1}{2})\pi \,.
\end{equation}
This gives a normalised wavefunction when $T$ is the classical period of the motion,
\begin{equation}
	T=2\mu \int_{r_1}^{r_0} \frac{dr}{k(r)}.
\end{equation}

For $r>r_0$, the WKB wavefunction is
\begin{gather}
	\phi(r) \sim \sqrt{\frac{\mu }{T \,\kappa(r)}} \exp\left(-\int_{r_0}^{r} \kappa(r')dr' \right) \,, \\
	\text{where}\quad \kappa(r) = \sqrt{2\mu\left(\tfrac{1}{2}\mu r^2 + \tfrac{J^2}{2\mu r^2} + V(r)-E\right)} \,.
\end{gather}
By expanding this at large $r$, we obtain the fall-off coefficient $A_{l,n}$ defined in \eqref{eq:phiasymp}:
\begin{equation}
\begin{gathered}
	A_{l,n}^2 \sim \frac{e^{-I}}{T}, \\
	\text{where }I(E,J) = \lim_{R\to\infty} \left[2\int_{r_0}^R \kappa(r) dr -\mu R^2 + E \log(\mu R^2)\right]. \label{eq:WKBOPE}
\end{gathered}
\end{equation}

For the free case $V=0$, we have $E=\frac{\mu}{2}(r_0^2+r_1^2)$ and $J = \mu r_0 r_1$. The WKB quantisation condition gives back precisely the MFT spectrum. We have $T=\pi$ (the orbital period of the radial motion is half the full period of the harmonic oscillator) and $I = \left(\frac{E-J}{2}\right)\log\left(\frac{E-J}{2e}\right)+ \left(\frac{E+J}{2}\right)\log\left(\frac{E+J}{2e}\right)$, reproducing the Stirling approximation of \eqref{eq:freeAnl}.

\subsection{From WKB spectrum to classical correlator}\label{sec:WKBtoG}

Now, we expect our result \eqref{eq:Gclass} for the correlator $G(\tau,\theta)$ to be obtained from the sum over states \eqref{eq:Schannel} using the WKB spectrum. This requires us to write the sums over angular momentum and energy ($l$ and $n$) as integrals, which are evaluated in a saddle-point approximation.

To motivate how we might recover the classical correlator \eqref{eq:Gclass} from the classical OPE coefficients \eqref{eq:WKBOPE}, we first note that they are Legendre transforms of one another. Specifically, the on-shell action $S_0(\tau,\theta)$ is the Legendre transform of the WKB integral $I(E,i J_E)$, after continuing to imaginary angular momentum $J = iJ_E$.
This follows from the identities
\begin{gather}
	\frac{\partial }{\partial E}I(E,iJ_E) = -\tau(E,J_E), \label{eq:leg1} \qquad \frac{\partial }{\partial J_E}I(E,iJ_E) = -\theta(E,J_E), \\
	I(E,iJ_E) + E\, \tau(E,J_E) + J_E\, \theta(E,J_E) = S_0(E,J_E).\label{eq:leg2}
\end{gather}
These relations can be thought of as inverses to the Hamilton-Jacobi identities \eqref{eq:HJ}, which give the inverse Legendre transform from $S_0(\tau,\theta)$ to $I(E,iJ_E)$.

Since the Legendre transform is the saddle-point approximation of a Laplace transform, we expect to recover the correlation function from integral of the following form:
\begin{equation}\label{eq:Gsaddle}
	G(\tau,\theta)\approx \int dE \,dJ_E \, e^{-I(E,iJ_E)-E \tau - J_E \theta} \approx e^{-S_0(\tau,\theta)}.
\end{equation}
The relations \eqref{eq:leg1} determine the location of a saddle-point, while \eqref{eq:leg2} gives the value of the exponent at the saddle.

This observation is also useful for practical calculation of the classical correlator under perturbations of the potential. To determine how $S_0(\tau,\theta)$ changes under perturbations of the potential, it appears that we must first find the variations $E$ and $J_E$ that leave $\tau$ and $\theta$ unaltered. But we bypass this step when writing $S_0$ as a Legendre transform, since at the saddle-point, $I-E \tau - J_E \theta$ is stationary to all variations of $E,J_E$. As a result, the variation of $S_0$ at fixed $\tau,\theta$ is equal to the variation of $I$ at fixed $E,J_E$:
\begin{equation}
	\delta S_0(\tau,\theta) = \left. \delta I(E,i J_E)\right|_{\mathrm{saddle}}.
\end{equation}
In particular, by perturbing around $V=0$ we can write the first-order correction to the free correlation function:
\begin{align}\label{eq:deltaS0}
	\delta S_0(\tau,\theta) &= 2\int_{r_0}^\infty dr \frac{r V(r)}{\sqrt{(r^2-\frac{4}{\mu} e^{-\tau } \cos ^2\frac{\theta }{2})(r^2+\frac{4}{\mu} e^{-\tau } \sin^2\frac{\theta }{2})}} \\
	&= \int_{-\infty}^\infty V(r(t_E))dt_E, \qquad r(t_E)  = \sqrt{\tfrac{2 }{\mu}e^{-\tau } (\cos\theta+\cosh (2 t_E))} \,. \nonumber
\end{align}
In the second line we have written the result suggestively as an integral of the potential evaluated on the free particle trajectory in Euclidean time. We will compare this formula to the full quantum perturbation theory and to T-channel conformal blocks in \cite{companion}.

It still remains to find an integral representation of the correlator which admits a saddle-point approximation of the form \eqref{eq:Gsaddle}. Note that this is not a straightforward matter of approximating the discrete sum over $l,n$ as an integral, since the saddle-point values of the conserved quantities (real $J_E$ or imaginary $l$ in particular) do not correspond to physical states (with real $l,E$). We here sketch an approach to the problem.

First, there is a standard trick to write the sum over angular momentum $l\in\NN$ as an integral, namely the Sommerfeld-Watson transform:
\begin{equation}
\begin{aligned}
	G(\tau,\theta) 
	&= \frac{i}{\Omega_{d-1}}\int_\Gamma \frac{ dl}{2\sin(\pi l)} \frac{2l+d-2}{d-2} C_l(-\cos\theta) G_{l}(\tau).\label{eq:SW}
\end{aligned}
\end{equation}
The contour $\Gamma$ initially consists of clockwise circles surrounding each integer $l$, so the residues of poles reproduces the sum over $l$ \eqref{eq:Gl} (using $C_l(-x)=(-1)^l C_l(x)$ for $l\in\NN$). We then hope to be able to deform $\Gamma$ to pass through a saddle point.  To write this, we must choose an analytic extension of the integrand ($C_l$ and $G_l$) from integer $l$ to complex $l$. If we choose to extend $C_l$ to complex $l$ as a hypergeometric function as in \eqref{eq:Cl2F1}, then when we take $l=iJ_E$ with $J_E\gg 1$ it provides the factor $e^{-J_E\theta}$ required for the saddle-point approximation \eqref{eq:Gsaddle}  \cite{durand2019complex}.

To extend $G_l(\tau)$ to complex $l$, we may continue to use the sum \eqref{eq:Schannel} over solutions to the radial Schr\"odinger problem \eqref{eq:radialSchr}, with boundary conditions $\phi(r)\sim r^{l+\frac{d-1}{2}}$ as $r\to 0$ (which defines a unique solution for $\Re l>-\frac{d-2}{2}$) and decay \eqref{eq:phiasymp} as $r\to\infty$. To determine the coefficients $A_{l,n}^2$ we normalise states by $\int \phi^2=1$, \emph{not} using $\int |\phi|^2$ since that would fail to give analyticity in $l$.

Next, we must replace the sum over $n$ by an integral over energies. As for the sum over $l$, this integral must be deformed away from the bound-state values of the energies, since once $l$ is complex $E_{l,n}$ will not be real. This occurs because the Hamiltonian is no longer Hermitian (even if $l+\frac{d-2}{2}$ is imaginary and hence the angular potential is real, since the boundary conditions break Hermiticity). We introduced one way to replace the sum over energies as an integral (including a factor of $e^{-E\tau}$) in section \ref{sec:resolvent}, by packaging the spectrum and OPE coefficients into the poles of a resolvent function $R_l(E)$. If we construct $R_l$ from WKB solutions as explained in appendix \eqref{app:resolvent}, it will be approximated by the WKB integral $I$, $R_l(E)\approx e^{-I(E,J)}$. These ingredients construct an integral of the form \eqref{eq:Gsaddle}.

Several details of this construction remain to be checked and fleshed out. In particular, we must show that the defining contours of the $l,E$ integrals can be deformed to pass through the saddle-points corresponding to the classical solutions of section \ref{sec:classcorr}, and that these saddles are dominant. A detailed analysis would also allow us to extract the prefactor of the classical correlation function (and higher order quantum corrections), which could be compared with a direct one-loop fluctuation determinant about the classical Euclidean solution. This would be an interesting exercise to carry out, since this is a particularly clean and rigorously-defined example of a classical saddle-point approximation to a quantum problem.

\section{Flat spacetime limit}\label{sec:flat}

So far, we have been studying a non-relativistic limit which includes the effects of the AdS curvature through the harmonic gravitational potential $\Phi = \tfrac{1}{2}\omega^2 r^2$. In this section we take an additional flat spacetime limit. This allows us to discuss physics with typical duration much smaller than the cosmological time $1/\omega$  and its realisation in a dual CFT (in an appropriate limit). For this section it is useful to reinstate $\omega$ (previously set to unity), so we can describe the relevant limit as $\omega \to 0$ with the interaction potential $V(r)$ and energies of particles held fixed.

In this flat spacetime limit, the spectrum of the Hamiltonian can be separated into two regimes. For $E<0$ we have bound states, which are not sensitive to the weak harmonic potential. For $E>0$, the $\omega=0$ Hamiltonian has a continuum of scattering states. In the presence of a small nonzero harmonic confining potential, this continuum is resolved into a discrete but closely spaced set of eigenstates, separated by energy gaps $\Delta E\approx 2\omega$. The precise spectrum is determined by the flat space phase shifts. This will give us direct access to flat spacetime scattering amplitudes from a CFT correlation function.

\subsection{Scattering}\label{sec:scatstates}

We begin by considering the spectrum of states of positive energy, which we write in terms of the wavenumber $k$:
\begin{equation}
	E = \frac{k^2}{2\mu} >0 \;.
\end{equation}
In the flat limit $\omega\to 0$ with $k$ fixed, this regime will be controlled by the physics of scattering off the potential $V(r)$.

\subsubsection{Review of scattering in quantum mechanics}

We begin with a brief review of scattering in quantum mechanics (working in arbitrary dimension to generalise from the familiar $d=3$ case).

In the region where we can neglect the potential,  we can write scattering solutions to the Schr\"odinger equation consisting of an  incoming plane wave with momentum $\vec{k}$ (of magnitude $k$) plus a radial outgoing scattered wave depending on the angle $\theta$ relative to $\vec{k}$:
\begin{equation}
	\psi_k(\vec{x}) \sim \frac{1}{(2\pi)^{\frac{d}{2}}}\left[e^{i \vec{k}\cdot \vec{x}} + f_k(\theta) \frac{e^{i k r}}{r^\frac{d-1}{2}}\right] \,.
\end{equation}
This defines the scattering amplitude $f_k(\theta)$, and the differential cross-section $\frac{d\sigma}{d\Omega}=|f_k(\theta)|^2$.

 For a spherically symmetric potential, we may decompose this solution into partial waves $\phi_{l,E}(r)$ solving the radial time-independent Schr\"odinger equation (\eqref{eq:radialSchr} with the quadratic potential neglected), as
\begin{equation}
	\psi_k(\vec{x}) = \frac{1}{\Omega_{d-2}} \sum_{l=0}^\infty \frac{2l+d-2}{d-2}   C_l(\cos\theta) \frac{\phi_{l,E}(r)}{r^{\frac{d-1}{2}} }.
\end{equation}
For the free problem the scattering amplitude vanishes, $f_k(\theta)=0$, so this becomes the partial wave decomposition of a plane wave:
\begin{align}
	\phi^{(\mathrm{free})}_{l,E}(r) &=  \frac{i^l}{k^\frac{d-1}{2}} \sqrt{kr} J_{\frac{d-2}{2}+l}(k r) \\
	& \sim \frac{2i^l}{k^\frac{d-1}{2}} \cos\left(k r-(\tfrac{d-1}{2}+l)\tfrac{\pi}{2} \right).
\end{align}
In the second line we have expanded at $kr\gg 1$. A potential modifies the relation between incoming and outgoing waves $e^{\pm i k r}$ for the asymptotic solution to the radial Schr\"odinger equation, which we can write in terms of the phase shift $\delta_l(E)$:
\begin{equation}\label{eq:phaseShift}
	\phi_{l,E}(r) \sim  \frac{2i^l}{k^\frac{d-1}{2}}  e^{i\delta_l(E)}i^l \cos\left(k r - (l+\tfrac{d-1}{2})\tfrac{\pi}{2}+\delta_l(E)\right).
\end{equation}
From these phase shifts, we can write the scattering amplitude by assembling the partial waves, as
\begin{equation}
	f_k(\theta) = \frac{1}{\Omega_{d-1}} \left(\frac{2\pi}{ik}\right)^\frac{d-1}{2} \sum_{l=0}^\infty \frac{2l+d-2}{d-2} C_l(\cos\theta) \left(e^{2i\delta_l(E)}-1\right).
\end{equation}
We may alternatively express this expansion in terms of the S-matrix partial waves $S_l(E)$, defined in terms of the phase shifts by
\begin{equation}
	S_l(E) = e^{2i\delta_l(E)}.
\end{equation}
For $E>0$, the phase shift  $\delta_l(E)$ is real so $S_l(E)$ has modulus one as demanded by unitarity (since scattering is purely elastic in non-relativistic quantum mechanics). We fix the freedom to shift $\delta_l$ by an integer multiple of $\pi$ by demanding that it is a continuous function of energy, and goes to zero as $E\to \infty$.

Finally, we relate this to the S-matrix $S_{\vec{k}',\vec{k}}$ more familiar from relativistic scattering, defined as the overlap between an in-state with momentum $\vec{k}$ and an out-state with momentum $\vec{k}'$. Separating out the trivial forward scattering and an energy conserving $\delta$-function, we can encode the interesting information in the $T$-matrix $T_{\vec{k}',\vec{k}}$:
\begin{equation}
	S_{\vec{k}',\vec{k}} = \delta^{(d)}(\vec{k}-\vec{k}') -2\pi i \delta(E_{\vec{k}}-E_{\vec{k}'})T_{\vec{k}',\vec{k}} \,.
\end{equation}
With spherical symmetry, we can write $T_{\vec{k}',\vec{k}} = T_k(\theta)$, where $\vec{k}$ and $\vec{k}'$ have equal magnitude $k$ and angle $\theta$ between them. This is related to the scattering amplitude $f_k(\theta)$ by
\begin{equation}\label{eq:scatS}
	f_k(\theta) = e^{-i(d+1)\frac{\pi}{4}}(2\pi)^\frac{d+1}{2}\mu k^\frac{d-3}{2} T_k(\theta).
\end{equation}
We were unable to find this relation for general dimension in the literature, so for completeness have included a derivation in appendix \ref{app:Smatrix}. The partial wave decomposition of the $T$-matrix is given by
\begin{equation}
	T_k(\theta) = - \frac{1}{2\pi i \mu k^{d-2}}  \frac{1}{\Omega_{d-1}}  \sum_{l=0}^\infty \frac{2l+d-2}{d-2} C_l(\cos\theta) \left(e^{2i\delta_l(E)}-1\right).
\end{equation}

We can also write the full S-matrix (including forward scattering) by removing the subtracted $-1$ from the phase shift, giving the precise sense in which $S_l(E)$ is a partial wave decomposition of the S-matrix:
\begin{equation}
	S_{\vec{k}',\vec{k}} = \frac{\delta(E_{\vec{k}}-E_{\vec{k}'})}{\mu k^{d-2}}  \frac{1}{\Omega_{d-1}}  \sum_{l=0}^\infty \frac{2l+d-2}{d-2} C_l(\cos\theta) S_l(E) \,.
\end{equation}
Setting $\delta_l(E)=0$ gives us forward scattering $\delta^{(d)}(\vec{k}-\vec{k}')$ as we expect: the sum over partial waves is a delta-function on $S^{d-1}$, $\frac{1}{\Omega_{d-1}}  \sum_{l=0}^\infty \frac{2l+d-2}{d-2} C_l(\cos\theta) = \delta_{S^{d-1}}(\Omega)$, and the factor $\mu k^{d-2}$ is a Jacobian to change to spherical coordinates $(E,\Omega)$ with magnitude expressed in terms of energy.

\subsubsection{Scattering states in AdS\label{ssec:scatStates}}

With these preliminaries out the way, we now proceed to embed this scattering in AdS. To do this, we take the scattering solutions with the harmonic potential neglected, and extrapolate them to the large $r$ region where the harmonic potential is important, but the interaction potential $V(r)$ (as well as the angular momentum barrier) can be ignored. Since in the flat space limit we are taking $\frac{E}{\omega}$ to be large, we may use the WKB approximation for this extrapolation.\footnote{One could alternatively use the exact solution in terms of hypergeometric $U$ functions in the region where the potential is negligible, which gives the same results in the $\omega\to 0$ limit but is less transparent.}

In the classically allowed region $r<\sqrt{\frac{k}{\omega\mu }}$, the WKB solution which matches to the flat spacetime scattering asymptotics (up to overall normalisation) is
\begin{gather}\label{eq:WKB}
	\phi_{l,E}(r) \sim  \sqrt{\frac{4\mu \omega}{\pi K(r)}}\cos\left(\int_0^r\!\!\! K - (l+\tfrac{d-1}{2})\tfrac{\pi}{2}+\delta_l(E)\right),\\
	\text{where}\quad K(r) = \sqrt{k^2-\mu^2\omega^2 r^2}.
\end{gather}
For $r\ll \frac{k}{\omega\mu}$ (but still large enough to neglect the interaction potential), this becomes $\cos(k r+\delta_l(E))$ to match \eqref{eq:phaseShift}. The coefficient is chosen for normalisation, $\int \phi_{l,E}(r)^2 dr=1$: in the $\omega\to 0$ limit, this integral is dominated by the region of this oscillating WKB solution, and can be straightforwardly evaluated by approximating $\cos^2\sim \frac{1}{2}$ by its average value over a wavelength.  

Now, the spectrum is determined by matching to the decaying WKB solution in the classically forbidden region $r>\frac{k}{\mu\omega}$, which fixes the phase at the turning point. Specifically, the argument of the cosine must be $(n+\frac{1}{4})\pi$ for integer $n$, so the energies $E_{l,n}$ are solutions of
\begin{equation}\label{eq:Elnscatter}
	E_{l,n} = \omega\left(l+\frac{d}{2}+2n -\frac{2}{\pi} \delta_l(E_{l,n})\right).
\end{equation}
In the limit we are considering ($\omega\to0 $ with fixed $V,E,l$), we may treat the phase shifts in \eqref{eq:Elnscatter} as effectively constant, since  $\delta_l(E)$ varies very little over an energy change of order $\omega$. As a result, the phase shift is proportional to the anomalous dimension, the difference $\gamma_{l,n} = \Delta_{l,n}-(\Delta_1+\Delta_2+2n+l)$ between the operator dimensions and the MFT spectrum:
\begin{equation}\label{eq:gammalnFlat}
	\gamma_{l,n} =\tfrac{1}{\omega}E_{l,n}-(\tfrac{d}{2}+l+2n) \sim -\frac{2}{\pi}\delta_l(E), \quad \text{where } E\sim\omega(\tfrac{d}{2}+l+2n).
\end{equation}
Since  $\delta_l(E)\to 0$ as $E\to\infty$ for sufficiently rapidly decaying potentials, the operator spectrum approaches that of MFT at very large $n$.

\begin{figure}
\centering
	\begin{tikzpicture}
  		\node at (-5,-2.5)  {\includegraphics[width=.24\textwidth]{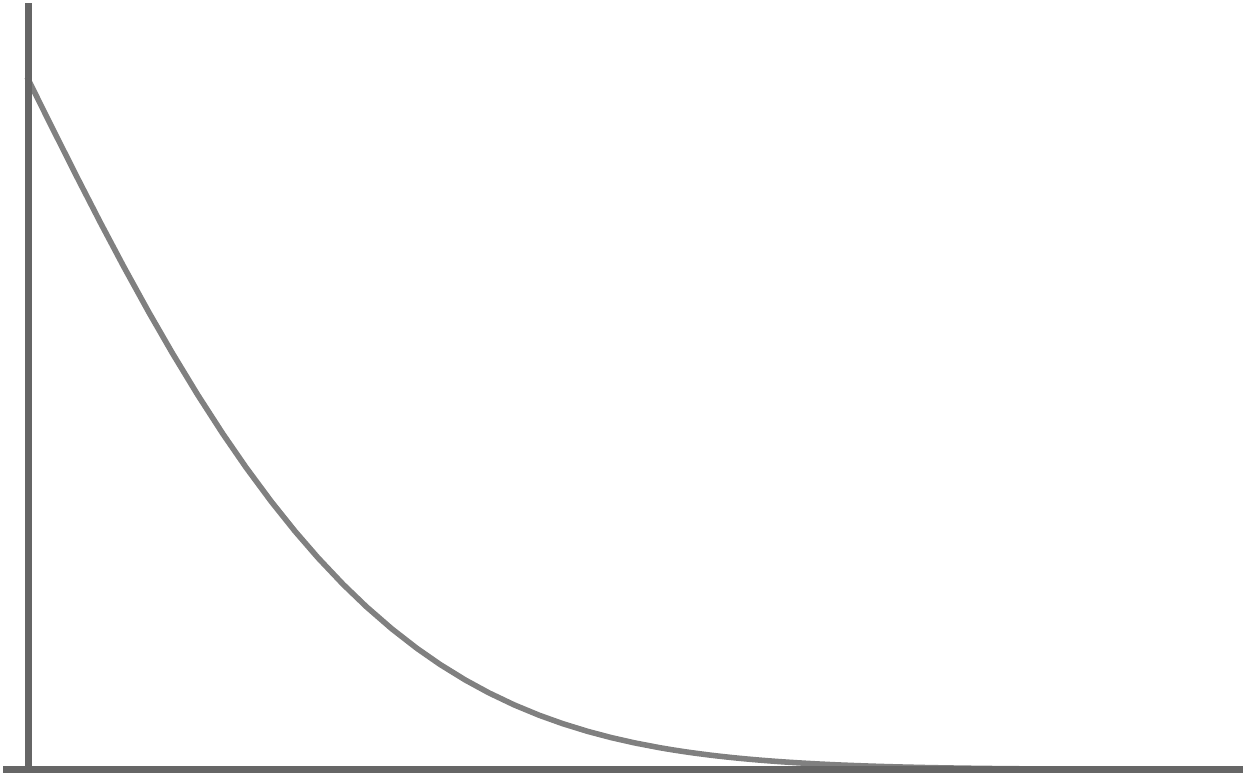}};
  		\node at (-5,0)  {\includegraphics[width=.24\textwidth]{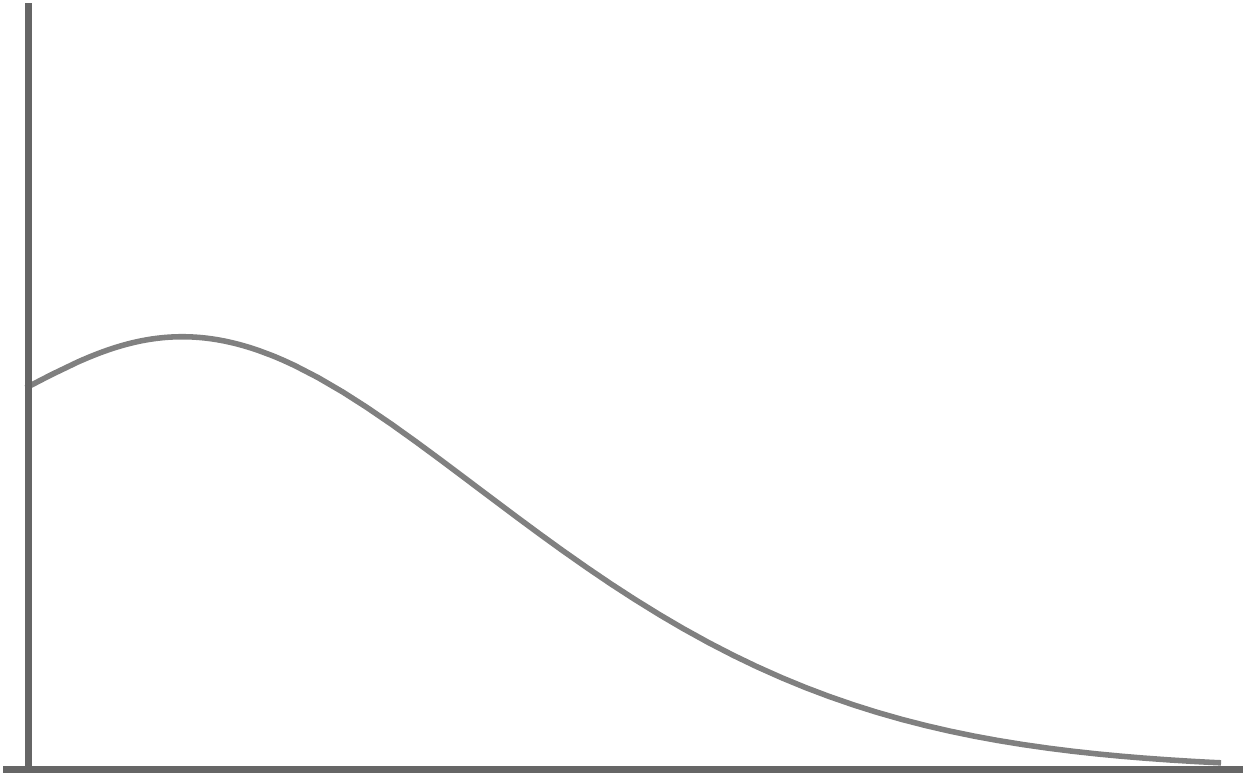}};
  		\node at (-5,2.5)  {\includegraphics[width=.24\textwidth]{deltaPlot2}};
  		\node at (1.8,0)  {\includegraphics[width=.6\textwidth]{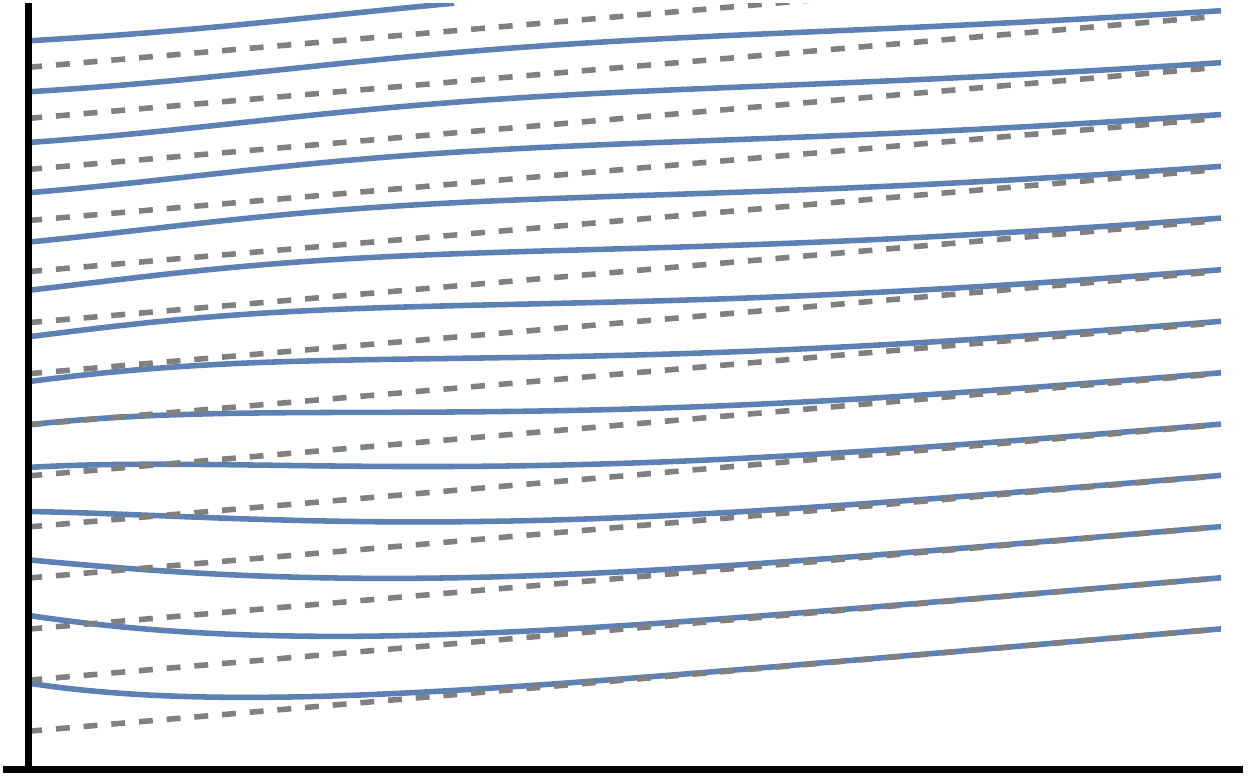}};
  		\node at (-6.9,3.4) {$-\frac{2}{\pi}\delta_l(E_3)$};
  		\node at (-6.9,.9) {$-\frac{2}{\pi}\delta_l(E_2)$};
  		\node at (-6.9,-1.6) {$-\frac{2}{\pi}\delta_l(E_1)$};
  		\node at (-3.6,1.4) {$l$};
  		\node at (-3.6,-1.1) {$l$};
  		\node at (-3.6,-3.6) {$l$};
  		\node at (-2.15,2.3) {\huge$E$};
  		\node at (-2.15,-1.3) {$E_1$};
  		\node at (-2.15,0) {$E_2$};
  		\node at (-2.15,1.3) {$E_3$};
  		\node at (5.6,-2.7) {\huge$l$};
 	\end{tikzpicture}
	\caption{An illustration of the relation between the scattering phase shift $\delta_l(E)$ and the spectrum $E_{l,n}$ in AdS. For an example potential, on the left we plot $-\frac{2}{\pi}\delta_l$ as a function of $l$ for three different energies and on the right we plot the AdS spectrum, with the free spectrum $E_{l,n}=\frac{d}{2}+2n+l$ indicated by the dashed straight lines. The shape of the Regge trajectories on the right follows the plots of the phase shift on the left.
	 \label{fig:scattering}}
\end{figure}

We summarise this result in figure \ref{fig:scattering}, showing how the Regge trajectories of states with $E>0$ depend on the phase shift. These Regge trajectories are approximately evenly spaced (with the dimensions of neighbouring operators differing by two), but not linear; instead, their shape encodes the $l$-dependence of the scattering phase shift. We emphasise that while the anomalous dimensions are linear in the phase shifts, this is a non-perturbative result: the anomalous dimensions need not be small, and may vary rapidly with $l$.

With this constraint on the spectrum, we can match to the decaying WKB solution in the region $r>\frac{k}{\omega\mu}$ and determine the asymptotic decay of the wavefunction as
\begin{equation}
	\phi_{l,E}(r)\sim  \frac{(-1)^n}{\sqrt{\pi r}} e^{-\tfrac{1}{2}\mu\omega r^2}\left(\frac{2\mu\omega r}{k}\right)^\frac{E}{\omega} e^{\frac{E}{2\omega}}.
\end{equation}
From this, we read off the OPE coefficients as given in \eqref{eq:phiasymp}:
\begin{equation}\label{eq:Alnscatter}
	A_{l,n}^2  \sim \frac{1}{\pi} \left(\frac{E_{l,n}}{2 e\omega}\right)^{-\frac{E_{l,n}}{\omega}}.
\end{equation}
We find that the OPE coefficient of a given operator is determined from its energy only. Compared to the MFT OPE coefficient $(f_{l,n}^2)_\mathrm{MFT}$ at the same value of $n$, using \eqref{eq:OPEA} we have
\begin{equation}
	f_{l,n}^2 \sim (f_{l,n}^2)_\mathrm{MFT} \times \left(\frac{E}{2\mu}\right)^{\frac{2}{\pi}\delta(E)}.
\end{equation}

We may also use the results of section \ref{sec:resolvent} and appendix \ref{app:resolvent} to construct a resolvent encoding the spectrum and OPE coefficients of the intermediate bound states. We can compute this from the ratio of the coefficients of the growing and decaying WKB solutions:
\begin{equation}\label{eq:Rlscat}
	R_l(E) = -\frac{1}{2} \left(\frac{E}{2e\omega}\right)^{-\frac{E}{\omega}}  \cot\left(\left(\tfrac{E}{\omega} -l-\tfrac{d}{2}\right)\tfrac{\pi}{2}+\delta_l(E)\right).
\end{equation}
This function has poles at bound state energies \eqref{eq:Elnscatter}, with residues giving the OPE coefficient $-A_{l,n}^2$ from \eqref{eq:Alnscatter} (using $\delta'(E)\ll \omega^{-1}$ to neglect a term coming from the variation of the phase shift).

This analysis can be straightforwardly extended beyond the strict $\omega\to 0$ limit, as long as the flat space scattering regime is cleanly separated from the harmonic potential. In particular, one may scale the strength of the potential while taking $\omega\to 0$ (keeping its width much smaller than $(\mu\omega)^{-1/2}$). In such a case, the wavefunction $\phi_{l,E}$ may have significant support in the scattering regime, and $\delta$ may vary significantly over energy ranges of order $\omega$ or less (these two conditions are equivalent, as we explain in section \ref{sec:resonances}). In such cases, the anomalous dimension is no longer simply proportional to the phase shift (\eqref{eq:Elnscatter} remains valid, but not \eqref{eq:gammalnFlat}), and the OPE coefficient is no longer determined from the energy alone. Long-lived resonances provide an interesting example where this is relevant, as discussed in section \ref{sec:resonances}.

\subsubsection{A correlator for flat space scattering}

Now we have determined the spectrum of states in the flat space limit, we next assemble them into a correlation function. We will introduce a simple correlation function which is proportional to the flat spacetime scattering amplitude (under some mild kinematic restrictions). Essentially the same correlation function was considered (not confined to the non-relativistic limit) in \cite{Hijano:2019qmi,Komatsu:2020sag} and studied further in \cite{Li:2021snj}, though in our non-relativistic context we have the advantage of a full non-perturbative understanding of its validity.

While it will not ultimately be of much interest, we first look at the Euclidean correlation function we studied in section \ref{sec:QM}. For this, we compute the sum over states in \eqref{eq:Glsum}, assuming for now that the relevant intermediate states are the scattering states we found in section \ref{sec:scatstates}:
\begin{equation}
	G_l(\tau) = \pi^{\frac{d}{2}}\sum_n \frac{1}{\pi} \left(\frac{E_{l,n}}{2 e\omega}\right)^{-\frac{E_{l,n}}{\omega}} e^{-\tau (E_{l,n}-\frac{d}{2}\omega)}.
\end{equation}
We have expressed the coefficients in terms of the energies using \eqref{eq:Alnscatter}, and $E_{l,n}$ is given by \eqref{eq:gammalnFlat}. Now, if we take $\tau<0$ and $\omega|\tau|\gg 1$, the function of $E_{l,n}$ appearing in the terms is narrowly peaked around a particular energy. If we write
\begin{equation}\label{eq:Tscatter}
	\tau =-\frac{1}{\omega}\log\left(\frac{E}{2\omega}\right)
\end{equation}
for some $E\gg \omega$, the terms of the sum become
\begin{align}
	A_{l,n}^2 e^{-\tau E_{l,n}} &\sim \frac{1}{\pi} e^{\frac{E_{l,n}}{\omega}}\left(\frac{E_{l,n}}{E}\right)^{-\frac{E_{l,n}}{\omega}} \\
	&\sim \frac{1}{\pi}e^\frac{E}{\omega} e^{-\frac{(E_{l,n}-E)^2}{2\omega E}}, \label{eq:MlscatterTerms}
\end{align}
where in the second line we have expanded around the maximum at $E_{l,n}=E$, taking $E_{l,n}-E$ of order $\sqrt{\omega E}$. This window of energies will dominate the sum in the $\omega\to 0$ limit (unless bound states are important, as discussed in section \ref{sec:BS}).

From \eqref{eq:gammalnFlat}, the spectrum near this maximum consists of evenly spaced energies with gaps $2\omega$ between them, since in the $\omega\to 0$ limit we may treat the phase shift as approximately constant over this range of energies. And the spacing between states is much smaller than the width of the Gaussian, so the sum over states $\sum_n$ is well-approximated by an integral $\frac{1}{2\omega}\int dE $, giving
\begin{equation}
	G_l(\tau=-\tfrac{1}{\omega}\log\left(\tfrac{E}{2\omega}\right))\sim \left(\frac{2\omega\pi}{E}\right)^{\frac{d-1}{2}} e^{\frac{E}{\omega}}.
\end{equation}

This leading order result is not very interesting, since the dependence on the potential through the phase shifts $\delta_l(E)$ has disappeared entirely. We have simply recovered a limit of the free correlator, as can be checked from our earlier result \eqref{eq:freePartialWaves} (it is independent of $l$, giving a delta-function on $S^{d-1}$ supported at $\theta=0$; this is resolved to a narrow Gaussian $\sim e^{-\frac{E}{2\omega}\theta^2}$ by corrections at large $l$). We could go to subleading order in the small $\omega$ limit, but this is technically involved, and mixes corrections from the harmonic potential so does not give a clean signature of the flat spacetime physics. Fortunately, we can instead consider a slightly modified correlator that does give an extremely direct probe of the flat spacetime scattering, and has a very direct physical interpretation as a scattering experiment in AdS.

The new correlator is constructed by inserting an additional real time evolution by time $\frac{\pi}{\omega}$, shifting $\tau\rightarrow \tau+\frac{i\pi}{\omega}$. This means we are considering matrix elements of $e^{-i \frac{\pi}{\omega}H}$ between initial and final states as in \eqref{eq:psiscat}:
\begin{gather}\label{eq:Mscatter}
	G^\mathrm{scat}(E,\theta) = \langle \psi(\tau,\Omega')|e^{-i \frac{\pi}{\omega}(H-\frac{d}{2}\omega)}|\psi(\tau,\Omega)\rangle,\\
	\text{where}\quad \tau = -\tfrac{1}{2\omega}\log\left(\tfrac{E}{2\omega}\right),\quad \Omega\cdot\Omega' = -\cos\theta. \label{eq:Mscattertautheta}
\end{gather}
This is the same as the matrix element giving $G$ from \eqref{eq:scatterM}, with $\tau$ given an imaginary piece $\frac{i\pi}{\omega}$ and $\theta \to \pi-\theta$ as explained momentarily.

We interpret $G^\mathrm{scat}$ in terms of the following process, depicted in figure \ref{fig:scattering}.
\begin{figure}[h]
	\centering
	\begin{tikzpicture}
  		\node at (-1.5,0)  {\includegraphics[width=.26\textwidth]{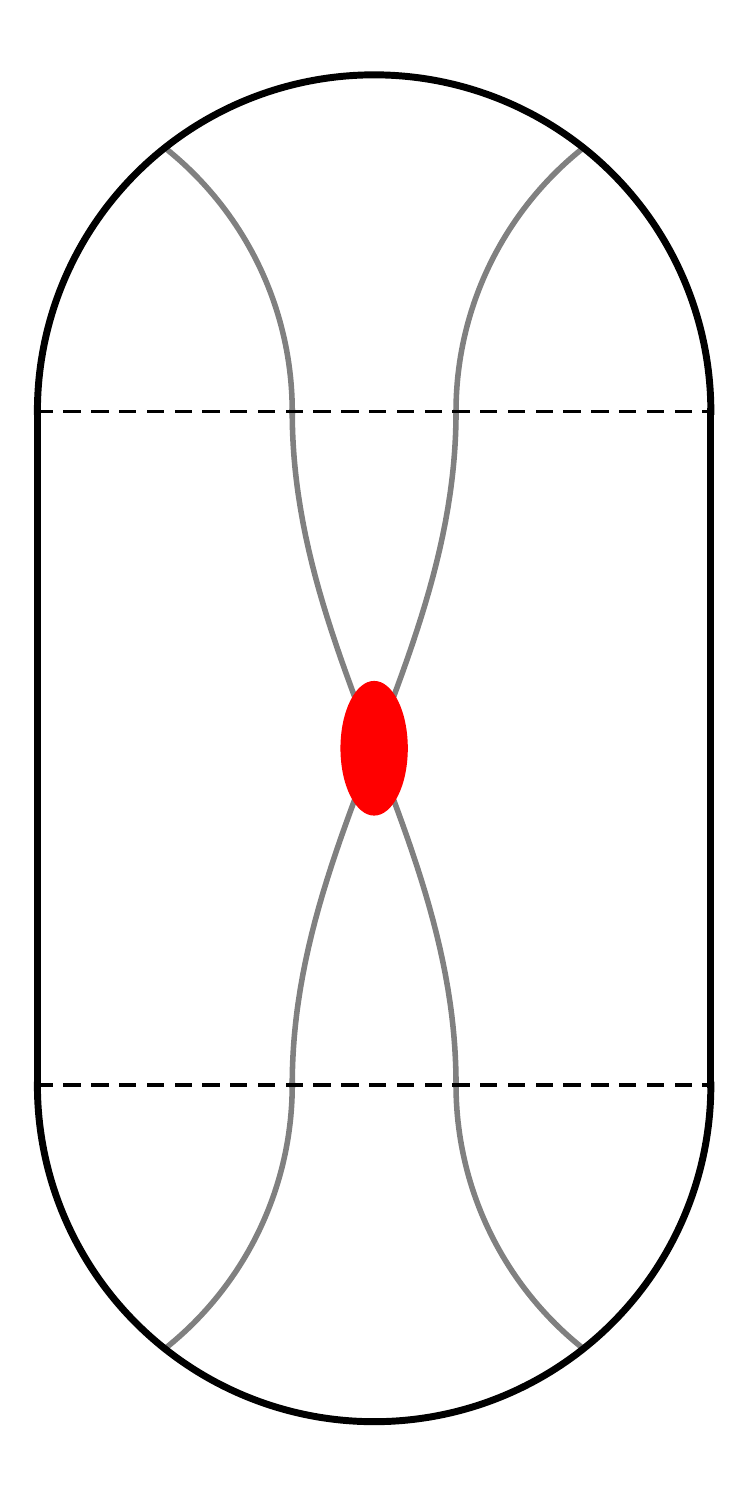}};
  		\node at (1.4,-2.3) {\huge$|\psi(\tau,\Omega)\rangle$};
  		\node at (-4.4,2.7) {\huge$\langle\psi(\tau,\Omega')|$};
  		\node at (1.,.4) {\huge$e^{-i\frac{\pi}{\omega}H}$};
  		\node[font=\color{red}] at (-1.,.2) {\huge$S$};
 	\end{tikzpicture}
	\caption{A cartoon interpreting the scattering correlation function $G^\mathrm{scat}(E,\theta)$ defined in \eqref{eq:Gscat}. The bottom and top pieces indicate Euclidean hemispheres with local operator insertions, preparing initial and final states $|\psi(\tau,\Omega)\rangle$, $|\psi(\tau,\Omega')\rangle$. These states have coherent wavefunctions of zero average momentum on the times indicated by the dashed lines. The middle section denotes evolution by real time $\frac{\pi}{\omega}$. The particles are accelerated by the AdS potential, scatter in the region $S$, and then decelerate back to rest. The resulting correlator is proportional to the scattering amplitude.
	 \label{fig:scattering}}
\end{figure}
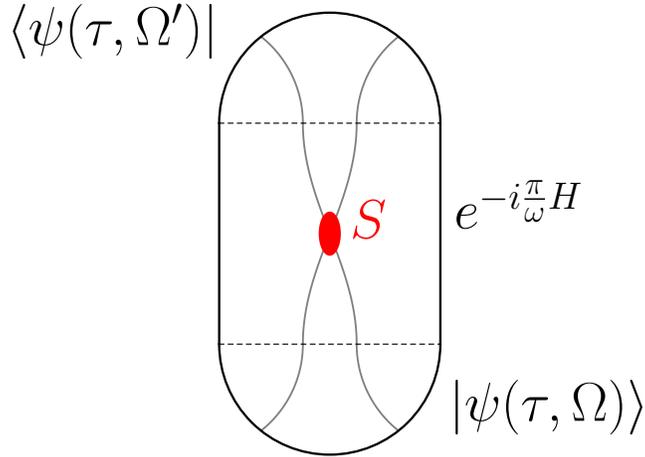
 Our initial state is well-approximated by a coherent state (as in \eqref{eq:scatCoherent}),
\begin{gather}
	|\psi(\tau,\Omega)\rangle\sim e^{\vec{\alpha}\cdot \vec{a}^\dag}|0\rangle, \quad \vec{\alpha}^2 = \tfrac{E}{\omega} \gg 1.
\end{gather}
It's intuitive that this should hold in the $\omega\to 0$ limit since the wavefunction is supported at a distance $r \sim \sqrt{2E}{\omega^2\mu}$ (much greater than its width $\sqrt{\mu\omega}$), far from the potential $V(r)$ (though for low enough energies it can fail by tunnelling into a bound state, see section \ref{sec:BS}). This state describes particles at rest, but elevated in the harmonic potential to have potential energy $E$. We then evolve in real time, so that the particles are accelerated by this potential. After a time of around $\frac{\pi}{2\omega}$, they arrive at the bottom of the harmonic potential with kinetic energy $E$, and scatter with interaction potential $V(r)$. They then travel back out, decelerated by the harmonic potential until they come to rest after total time $\sim \frac{\pi}{\omega}$. We evaluate the overlap of this final state wavefunction with another coherent state to determine the amplitude to scatter through a given angle $\theta$. Note that the angle between initial and final states is $\pi-\theta$, since forward scattering without any interaction would take our initial state to the antipodal point on $S^{d-1}$. From this description, we expect the matrix element \eqref{eq:Mscatter} to be determined by the flat spacetime scattering amplitude $f_k(\theta)$.

Let's see how this expectation is realised by the sum over intermediate states. The real time evolution multiplies the terms by a phase $e^{-\frac{i\pi}{\omega}E_{l,n}}$, where $E_{l,n}$ is given by \eqref{eq:Elnscatter}. Crucially, the equal energy spacing of $2\omega$ ensures that the $n$-dependence drops out so this phase is approximately constant across the energy window dominating the sum over states:
\begin{equation}\label{eq:scatphase}
	e^{-\frac{i\pi}{\omega}(E_{n,l}-\frac{d}{2})} \sim (-1)^l e^{2i\delta_l(E)}.
\end{equation}
We immediately see $S_l(E) = e^{2i\delta_l(E)}$ appearing, which gives the partial wave decomposition of the scattering amplitude:
\begin{equation}\label{eq:Glscat}
	G_l^\mathrm{scat}(E) \sim  \left(\frac{2\omega\pi}{E}\right)^{\frac{d-1}{2}} e^\frac{E}{\omega} S_l(E).
\end{equation}
The factor of $(-1)^l$ in \eqref{eq:scatphase} accounts for the extra sign in the definition of $\cos\theta$ in \eqref{eq:Mscattertautheta}, with $\theta=0$ corresponding to forward scattering (when in and out states are at antipodal points).

Summing the partial waves, we find (for $\theta$ away from zero) a result proportional to the scattering amplitude:
\begin{equation}
	G_\mathrm{scat}(E,\theta) \sim  -2\pi i  \left(4\pi\omega\mu\right)^{\frac{d}{2}} \frac{e^\frac{E}{\omega}}{\sqrt{8\pi\omega E}}  T_k(\theta).
\end{equation}
For $\theta$ close to zero we additionally have a forward scattering piece equal to the free result (a narrowly-supported Gaussian as discussed above).

We can immediately interpret $G_\mathrm{scat}(E,\theta)$ as a CFT four-point function, using \eqref{eq:Gdef} and setting $\omega \tau=i\pi - \log\left(\tfrac{E}{2\omega}\right)$, $\theta \rightarrow \pi-\theta$.
\begin{equation}\label{eq:Gscat}
	G_\mathrm{scat}(E,\theta) = G\left(\tfrac{i \pi}{\omega}-\tfrac{1}{\omega}\log\left(\tfrac{E}{2\omega}\right),\pi-\theta\right)
\end{equation}
Formally, we can think of this as a correlator on $\RR^d$ with operators inserted on a common plane with complex coordinates ($y$,$\bar{y}$), except we allow $y$ and $\bar{y}$ not to be related by complex conjugation due to the mix of Euclidean kinematics (preparing the state) and Lorentzian kinematics (evolution by time $\frac{i\pi}{\omega}$). The insertion points are
\begin{gather}
		y_1=\sqrt{\tfrac{\mu E}{m_1^2}}e^{i(\frac{\theta}{2}-\pi)},\qquad \bar{y}_1=\sqrt{\tfrac{\mu E}{m_1^2}}e^{-i\frac{\theta}{2}},\\
		\qquad y_2 = \sqrt{\tfrac{\mu E}{m_2^2}}e^{i\frac{\theta}{2}}, \quad\qquad\bar{y}_2 = \sqrt{\tfrac{\mu E}{m_2^2}}e^{i(\pi-\frac{\theta}{2})},
\end{gather}
with $y_{3,4}$ related to these by inversion in the unit sphere ($y_1y_4=\bar{y}_1\bar{y}_4=y_2y_3=\bar{y}_2\bar{y}_3=1$). The conformal cross-ratios are given to leading order in the non-relativistic limit by
\begin{equation}
	z\sim e^{-2\pi i}\tfrac{E}{2\mu}e^{i\theta}, \qquad \bar{z}\sim\tfrac{E}{2\mu}e^{-i\theta},
\end{equation}
with the $e^{-2\pi i}$ indicating that we should include phases from encircling the origin anticlockwise in the $z$ coordinate, with no such monodromy in the $\bar{z}$ coordinate.

These kinematics are closely related to the S-channel `double discontinuity' ($\dDisc$) which appears in the Lorentzian inversion formula \cite{Caron-Huot:2017vep,Simmons-Duffin:2017nub}. In terms of cross-ratios, this involves similar configurations where $z$ undergoes a monodromy around the origin while $\bar{z}$ does not, though in that case $z,\bar{z}$ are both real (corresponding to certain Lorentzian kinematics on the plane). Nonetheless, we can similarly define $\dDisc$ from a Euclidean correlator and two Lorentzian correlators in which $z$ encircles the origin in opposite directions:
\begin{align}
	&\dDisc G(E,\theta) = G\left(-\tfrac{1}{\omega}\log\left(\tfrac{E}{2\omega}\right),\theta \right)	\\
	&  -\tfrac{1}{2} G\left(\tfrac{i \pi}{\omega}-\tfrac{1}{\omega}\log\left(\tfrac{E}{2\omega}\right),\pi-\theta\right) -\tfrac{1}{2}  G\left(-\tfrac{i \pi}{\omega}-\tfrac{1}{\omega}\log\left(\tfrac{E}{2\omega}\right),\pi-\theta\right) \nonumber \\
	&\sim -  \left(4\pi\mu\omega\right)^{\frac{d}{2}} \sqrt{\frac{\pi}{2\omega E}} e^\frac{E}{\omega}  \Im T_k(\theta).
\end{align}
Note that in terms of pure four-point functions the second two terms involve extra phases $e^{\pm i(\Delta_1+\Delta_2)}$, which cancel with the normalisation factor in \eqref{eq:Gdef}. The result applies for all angles including $\theta$ close to zero, since the first term cancels the free piece concentrated at $\theta=0$ (the `$1$' in the S-matrix).
The analogy between the double discontinuity and $\Im T$ in a scattering amplitude was noted already in \cite{Caron-Huot:2017vep}: here this relationship is precisely true. From this perspective, there is a physical picture for why double-trace operators with zero anomalous dimension (free two-particle states with $\Delta=\Delta_1+\Delta_2+2n$) give vanishing contribution to the $\dDisc$. They pick up a simple phase $(-1)^l$ upon evolution by time $\frac{\pi}{\omega}$, corresponding to a half-period of free motion in the harmonic AdS potential; they thus return to the same state up to the antipodal map $\theta\mapsto \pi-\theta$.

In some cases, it interesting to somewhat generalise $G_\mathrm{scat}$ to allow Lorentzian evolution by times other than $\frac{\pi}{\omega}$, and to go slightly beyond the strict flat space limit to a context in which the phase shifts $\delta_l(E)$ vary significantly over the Gaussian energy window. We discuss these in section \ref{sec:resonances}.

\subsection{Bound states and low energies}\label{sec:BS}

\subsubsection{Spectrum of bound states}

We now discuss the states with $E<0$. The AdS potential has no important effect on these states, so their spectrum is essentially the same as that of bound states in the absence of the harmonic potential. Indeed, using first order perturbation theory we can estimate the change in energy due the the AdS potential to be $\frac{1}{2}m \omega^2 \langle r^2\rangle$, which goes to zero as $\omega\to 0$.

To determine the OPE coefficient corresponding to a bound state, we extrapolate the flat spacetime bound state wavefunction to large $r$, where the harmonic potential is important. Write
\begin{equation}
	E_{l,n} = -\frac{\kappa^2}{2\mu}.
\end{equation}
For sufficiently large $r$ that the interaction potential is negligible, the wavefunction  can be approximated by a WKB solution with the harmonic potential alone. The WKB solution that decays at infinity is given by
\begin{equation}\label{eq:boundWKB}
	\phi(r) \sim  \frac{ \tilde{A}_{l,n}\kappa}{(\mu^2\omega^2 r^2+\kappa^2)^\frac{1}{4}}\exp\left(-\tfrac{1}{2}r \sqrt{\mu^2\omega^2 r^2+\kappa^2}-\frac{\kappa^2}{2\mu\omega}\sinh ^{-1}\left(\tfrac{\mu \omega}{\kappa}r\right)\right)
\end{equation}
for some dimensionless coefficient $\tilde{A}_{l,n}$.

For $r\ll \frac{\kappa}{\mu\omega}$, we can match this onto the decay of the flat spacetime bound state wavefunction:
\begin{equation}\label{eq:phiasympflat}
	\phi(r) \sim \tilde{A}_{l,n}\sqrt{\kappa} e^{-\kappa r}\, , \qquad r\ll \frac{\kappa}{\mu\omega}.
\end{equation}
The coefficient $\tilde{A}_{l,n}$ is fixed by normalisation of this wavefunction (the change to the normalisation integral from the tails with $r\gtrsim \frac{\kappa}{\mu\omega}$ is negligible), and chosen to be real and positive. In particular, $\tilde{A}_{l,n}$ has a finite non-zero limit in the flat limit $\omega\to 0$. Expanding in the far asymptotic region $r\ll \frac{\kappa_0}{\mu\omega}$, we have
\begin{equation}
	\phi(r) \sim \tilde{A}_{l,n} \frac{\kappa}{\sqrt{\mu\omega r}} e^{-\frac{\kappa^2}{4\mu\omega}} \left(\tfrac{2\mu \omega}{\kappa}r\right)^{-\frac{\kappa^2}{2\mu\omega}} e^{-\frac{1}{2}\mu \omega r^2}.
\end{equation}
Matching this with the asymptotic behaviour \eqref{eq:phiasymp} that we found in the previous section, we find
\begin{equation}\label{eq:Alnbound}
	A_{l,n}^2 \sim \frac{2|E_{l,n}|}{\omega} \left(\frac{|E_{l,n}|}{2 e\omega}\right)^{\frac{|E_{l,n}|}{\omega}} \tilde{A}_{l,n}^2.
\end{equation}
This determines (via \eqref{eq:OPEA}) the OPE coefficients of bound states in the flat space limit.

This data can also be extracted by analytically continuing the S-matrix to complex energy, where bound states give rise to poles at $k=i\kappa$ (since the coefficient of $e^{-i k r}= e^{\kappa r}$ in the asymptotic solution to the time-independent Schr\"odinger equation vanishes). The residues of these poles are proportional to the $|\tilde{A}_{l,n}|^2$,
\begin{equation}
	\Res_{E\to E_{l,n}} S_l(E) = e^{2\pi i (l+\tfrac{d}{2})}\frac{\kappa}{\mu} |\tilde{A}_{l,n}|^2,
\end{equation}
which can be shown with the same methods used in appendix \ref{app:resolvent}. With this result, one could encode the bound states in a resolvent $R_l(E)$ proportional to $S_l(E)$, with an energy-dependent prefactor from \eqref{eq:Alnbound}. Note that this is not what we would obtain from na\"ive continuation of our scattering result \eqref{eq:Rlscat}: analytic continuation from $E>0$ to $E<0$ does not commute with the flat spacetime limit.

\subsubsection{Bound states in correlation functions}\label{sec:BScorr}

Note that the OPE coefficient from \eqref{eq:Alnbound} grows extremely rapidly with the binding energy $|E|$ (faster than exponentially). This means that a sufficiently tightly bound state can dominate the correlation functions we have been considering. A given bound state contributes to the sum over states for the Euclidean correlator \eqref{eq:Glsum} as
\begin{equation}\label{eq:Mlbound}
	G_l(T) \supset \pi^\frac{d}{2}  \frac{2|E_{l,n}|}{\omega} \left(\frac{|E_{l,n}|}{2 e\omega}\right)^{\frac{|E_{l,n}|}{\omega}} |\tilde{A}_{l,n}|^2 e^{-T (E_{l,n}-\frac{d}{2})}.
\end{equation}
If we choose $T = -\frac{1}{\omega}\log\frac{E}{2\omega}$ so that the sum over scattering states is dominated by energy $E>0$ as in \eqref{eq:Tscatter}, the coefficient in \eqref{eq:Mlbound} has a \emph{minimum} as a function of $E_{l,n}$ when $|E_{l,n}|=E$. For sufficiently negative $E_{l,n}$, it becomes larger than the dominant coefficient (of order $e^\frac{E}{\omega}$) for scattering states \eqref{eq:MlscatterTerms}. Thus, if the potential with spin $l$ has a bound state (so $E_{0,l}<0$), the ground state dominates the sum for sufficiently small energy. Specifically,
\begin{equation}\label{eq:Ecrit}
	E< W(e^{-1}) |E_{0,l}|\implies G_l(E) \text{ dominated by ground state.}
\end{equation}
Here, $W(e^{-1})\approx 0.278$ is the solution to $1+w+\log w=0$ ($W$ is the Lambert product log function).

As a result, in the presence of bound states both the Euclidean and Lorentzian correlators undergo a first-order phase transition as a function of the kinematics, where the important intermediate states jump from scattering states of the critical energy \eqref{eq:Ecrit} to the ground state. Our initial and final states $|\psi(\tau,\Omega)\rangle$ are no longer simply coherent states supported far away, since there is a large amplitude for the particles to `tunnel' into the ground state. To extract the properties of states at intermediate energies, one must either subtract the dominant contributions from low-energy states, or perhaps consider different correlation functions.

Essentially the same phenomenon was discussed in \cite{Komatsu:2020sag}. The advantage of the non-relativistic limit is that we know the precise regime of validity \eqref{eq:Ecrit} for our scattering formula non-perturbatively.

\subsubsection{Low energies and Levinson's theorem\label{sec:Lev}}

We have so far considered states with fixed energy in the $\omega\to 0$ limit: our discussion has encompassed both scattering states ($E>0$) and bound states ($E<0$), but was only valid in the regime $|E|\gg \omega$. To complete the picture, we now briefly examine the intermediate regime, where $E$ is of order $\omega$.

Generically, this has a simple answer: for energies $E$ of order $\omega$, the spectrum matches that of MFT, with energies $E_{l,n} = \frac{d}{2}+l+2(n-n_b(l))$. We have shifted the index $n$ by the number of bound states $n_b(l)$ for the potential at a given angular momentum $l$, so that this applies for integers $n\geq n_b(l)$. The exception to this generic situation occurs when the potential $V(r)$ has a marginally bound state.

We argue for this as follows. For small sufficiently small $r$, we may ignore the harmonic potential and the energy, solving the Schr\"odinger equation with potential $V(r)$ at zero energy. At sufficiently large $r$, we may ignore the interaction potential $V(r)$. In the intermediate regime, only the $\frac{1}{r^2}$ potential from the angular momentum barrier is relevant, so the solution will be a combination of powers $\phi(r) \sim c_1 \left(\frac{r}{a}\right)^{l+\frac{d-1}{2}} + c_2 \left(\frac{r}{a}\right)^{-l-\frac{d-3}{2}}$, where $a$ is a characteristic length scale of the potential $V$, and the ratio between coefficients $c_{1,2}$ is generically of order unity. For $r\gg a$, where the harmonic potential becomes relevant, the second term becomes negligible, so we are effectively solving the Schr\"odinger problem for the harmonic oscillator with the usual boundary condition $\phi \propto r^{l+\frac{d-1}{2}}$ as $r\to 0$, and we recover the harmonic oscillator spectrum. This argument fails when $c_1=0$: that is, when the potential $V(r)$ has a marginally bound state with angular momentum $l$.\footnote{Another exception is the case $l=0,d=2$, where the two solutions with the angular momentum potential alone are $\phi\sim \sqrt{r}$ and $\phi\sim \sqrt{r}\log r$: the more singular solution at $r\to 0$ is also the faster growing for large $r$. The argument must be slightly modified, though the result remains true.}

This result smoothly matches up with the scattering spectrum described in section \ref{sec:scatstates}, since the zero energy limit of the phase shift (in the absence of a marginally bound state) is an integer multiple of $\pi$, counting the number of bound states: $\delta_l(0)=\pi n_b(l)$. This is Levinson's theorem \cite{levinson1949uniqueness}. One usually argues for this by placing the system in a box with a `hard wall' at some large radius $R$ and counting states, arguing that states cannot appear or disappear under deformations of the potential (see \S 7.8 of \cite{weinberg2015lectures}, for example). Our context supplies a similar argument, but using the harmonic potential of AdS instead of a hard wall. Since the spectrum approaches that of MFT at large $l$, Levinson's theorem is equivalent to the statement that Regge trajectories are not created or destroyed as we vary $l$ or coupling constants.

\subsection{Resonances and other generalisations}\label{sec:resonances}

It is enlightening to go slightly beyond the strict flat space limit ($\omega\to 0$ with fixed $V$). We continue to insist that we continue to have a separation between the large $r$ region where the harmonic potential is important and the small $r$ region where the interaction potential $V$ is important: this means that the range of the potential $V(r)$ must be much smaller than $\frac{k}{\mu\omega}$. With this condition, much of section \ref{ssec:scatStates} continues to apply. In particular, we have a regime of scattering states with spectrum determined by \eqref{eq:Elnscatter},
\begin{equation}\label{eq:Elnscatter2}
	\frac{E_{l,n}}{\omega} - l -\frac{d}{2} + \frac{2}{\pi} \delta_l(E_{l,n}) = 2n, \qquad n\in \NN.
\end{equation}
However, if we no longer insist that $V$ is fixed in the $\omega\to 0$ limit, the phase shift $\delta_l(E)$ may vary appreciably over narrow windows of energy with width scaling with $\omega$.

\begin{figure}
	\centering
	\begin{tikzpicture}
  		\node (img)  {\includegraphics[width=.7\textwidth]{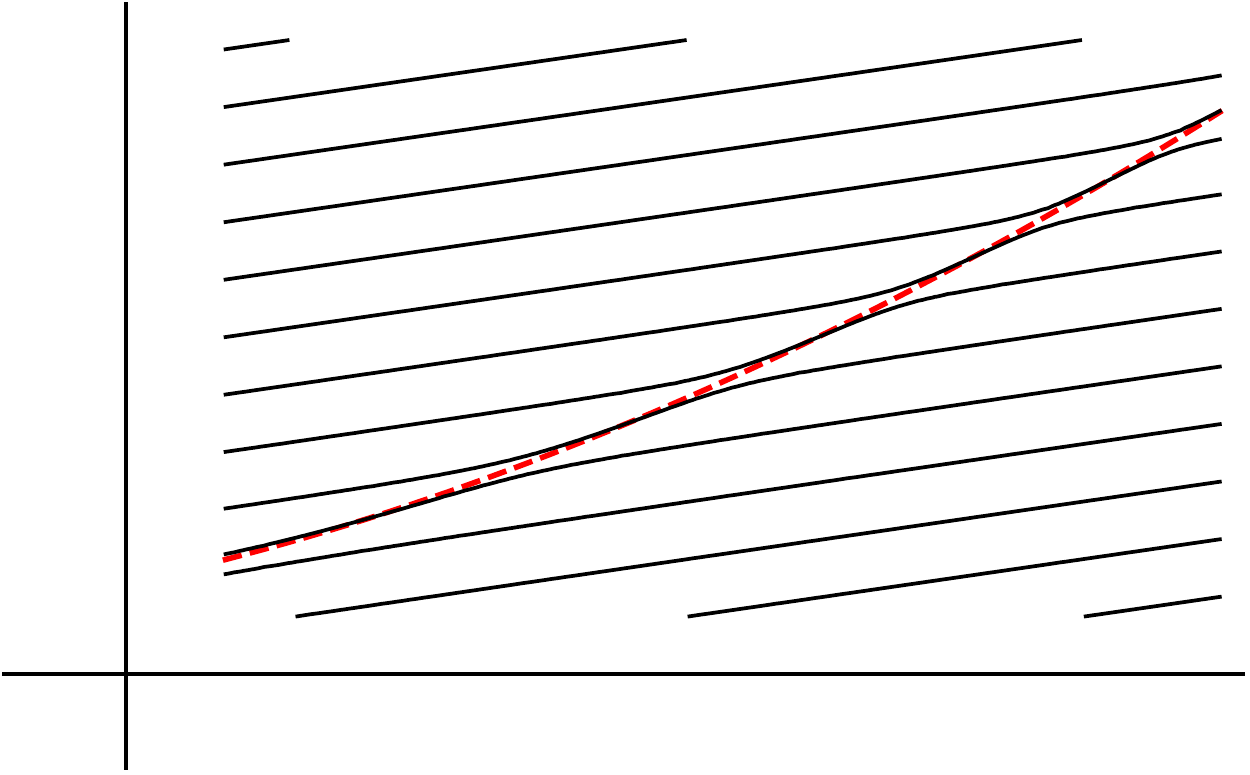}};
  		\node at (4.7,-2.3) {\huge$l$};
  		\node at (-4,3) {\huge$E$};
 	\end{tikzpicture}
	\caption{Regge trajectories of scattering states in the presence of a resonance. The dashed red curve shows the energy $E_R(l)$ of the resonance. Far from the resonance we have evenly-spaced Regge trajectories, determined by the `background' phase shift $\bar{\delta}_l(E)$. There is an additional physical state with energy close to the resonance. Where the resonance crosses a `background' Regge trajectory, the trajectories reconnect to avoid intersecting. Close to such a reconnection (when $E_R$ is close to a background Regge trajectory of energy $\bar{E} = \omega(\frac{d}{2}+l+2n+\frac{2}{\pi}\bar{\delta})$), for narrow resonances $\Gamma\ll \omega $ the Regge trajectories are determined by the solutions to $(E-\bar{E})(E-E_R) = \frac{\omega\Gamma}{\pi}$. These reconnections are the result of strong mixing between the resonance and the background scattering state when $|E_R-\bar{E}|\lesssim \sqrt{\omega\Gamma}$
	 \label{fig:resonance}}
\end{figure}

First, suppose that $\delta_l(E)$ changes significantly over an energy range of order $\omega$ or smaller. In this case, the perturbative solution \eqref{eq:gammalnFlat} to \eqref{eq:Elnscatter2} (giving an anomalous dimension proportional to the phase shift) receives large corrections: the spectrum is sensitive to details of the function $\delta_l(E)$. There is one physical situation where this occurs, namely when the potential $V$ admits a long-lived resonance of spin $l$. Near the energy $E_R$ of the resonance, the phase shift increases rapidly by $\pi$  over an energy range $\Gamma$ (the width of the resonance). The phase shift behaves as
\begin{equation}
\begin{gathered}
	\delta_l(E) = \bar{\delta}_l(E) + \delta^{(R)}_l(E),\\
	\tan\delta^{(R)}_l(E) \sim -\frac{1}{2}\frac{\Gamma}{E-E_R},
\end{gathered}
\end{equation}
where $\bar{\delta}_l(E)$ varies more slowly. This resonant increase in the phase shift gives a corresponding increase by $2$ of the left hand side of \eqref{eq:Elnscatter2}, so there is an extra state (relative to the phase shift $\bar{\delta}_l$) in this energy range. Such a long-lived resonance gives rise to a characteristic signature in the Regge trajectories, illustrated and described in figure \ref{fig:resonance}.

Such a long-lived resonance is essentially the only circumstance in which the phase shift can vary significantly over an energy range of order $\omega$. In particular, rapid decreases of the phase shift are excluded. We can see that such a decrease is forbidden from the AdS scattering state spectrum, since it would allow \eqref{eq:Elnscatter2} to have more than one solution for a given value of $n$, indicating multi-valued Regge trajectories: a discontinuous creation or destruction of states as we vary $l$. And indeed, a lower bound on $\delta_l'(E)$ is implied by causality. To understand this, recall that the derivative of the phase shift can be interpreted as a time delay, $\Delta t = 2\delta_l'(E)$: if we send a wavepacket into the potential $V$, $\Delta t$ is the additional time taken for the wavepacket to enter and leave the scattering region relative to $V=0$. This can be negative, but not too negative: we can have a time advance, but at most by the range of the interaction $a$ divided by the velocity $v=\frac{k}{\mu}$:
\begin{equation}
	-\Delta t = -2\delta_l'(E) \lesssim \frac{\mu}{k} a \ll \frac{1}{\omega}\,.
\end{equation}
By our assumption $a \ll \frac{k\mu}{\omega}$ (ensuring validity of the flat spacetime scattering approximation), the time advance should be much less than the AdS time $\omega^{-1}$. The result is that the left-hand-side of \eqref{eq:Elnscatter2} is an increasing function of $E_{l,n}$, so it has exactly one solution for each $n$ within its regime of validity. The scattering approximation breaks down for sufficiently small energy, where the discussion of section \ref{sec:Lev} takes over and guarantees the same result.

Now assuming $|\delta_l'(E)|\ll \omega^{-1}$, our result \eqref{eq:gammalnFlat} that anomalous dimensions are proportional to the phase shift holds at leading order in the flat spacetime limit. Nonetheless, there are corrections suppressed by a factor of order $\omega \delta_l'(E)$, which are nonlinear in the phase shift. Writing $E_0=\omega(\frac{d}{2}+l+2n)$ for the MFT energy, we have
\begin{equation}
	E-E_0 \sim -\frac{2\omega}{\pi} \delta_l(E_0) + \frac{4\omega^2}{\pi^2} \delta_l(E_0)\delta_l'(E_0) + \cdots.
\end{equation}
Note that there are additional corrections coming from subleading effects in matching the scattering region to the harmonic potential, though we expect the correction presented here to be more important near a resonance.

The OPE coefficients also receive corrections from the variation of the phase shift. This is simple to calculate from the residues of the resolvent \eqref{eq:Rlscat}:
\begin{equation}
	|A_{l,n}|^2 \sim \frac{1}{\pi+2\omega\delta_l'(E_{l,n})} \left(\frac{E_{l,n}}{2 e\omega}\right)^{-\frac{E_{l,n}}{\omega}}.
\end{equation}
We can also understand this result from the normalisation of the scattering wavefunction. The factor $\frac{\pi}{\omega} + 2\delta_l'(E)$ can be interpreted in terms of the time spent by the particles either scattering or moving freely in the harmonic potential, and hence how much probability density is concentrated at different separations. The first term $\frac{\pi}{\omega}$ is a half-period of the harmonic potential, giving the time for the particles to escape from and return to the scattering region. The second term $2\delta_l'(E)$ is the time delay for scattering, telling us how long the particles linger in the scattering region (or if it is negative, how much less time they spend there due to being reflected by the potential). For example, a long-lived resonance ($\Gamma \ll \omega$) will have an OPE coefficient suppressed by a factor of its lifetime $\Gamma^{-1}$, since its wavefunction will be concentrated inside the potential, and the overlap with two-particle scattering states is small, coming from the tail of the wavefunction.

Finally, we reexamine the scattering correlation functions \eqref{eq:Mscatter}. These are determined by a sum over states with energies in a window of width $\sqrt{\omega E}$ around the central energy $E$, so we will have interesting corrections if $\delta_l(E)$ varies significantly over this window. In this context, it is also interesting to slightly generalise, evolving by a real time $\frac{\pi}{\omega}+t$ for some $t\ll \omega^{-1}$.

Writing this amplitude as a sum over intermediate scattering states, we have
\begin{equation}
	\begin{aligned}
		G_l\left(-\tfrac{1}{\omega}\log\left(\tfrac{E}{2\omega}\right)+i \tfrac{\pi}{\omega}+i t\right) &\propto \sum_n \frac{e^{-\frac{(E_{l,n}-E)^2}{2\omega E}}}{\pi + 2\omega \delta'(E_{l,n})} e^{-it (E_{l,n}-E)}e^{2i\delta_l(E_{l,n})} \\
		&\sim  \int \frac{dE_{l,n}}{2\pi \omega} e^{-\frac{(E_{l,n}-E)^2}{2\omega E}} e^{-it (E_{l,n}-E)}e^{2i\delta_l(E_{l,n})}.
	\end{aligned}
\end{equation}
On the second line we assumed that $\delta_l'(E)\ll \omega^{-1}$, so that we may approximate the sum over discrete energies with separation $\omega$ by an integral. The result is somewhat similar to the asymptotic late time wavefunction of a scattering state with Gaussian energy distribution in flat space. The difference is the absence of a radius-dependent phase $e^{ikr}$: all momenta are concentrated by the harmonic potential instead of dispersing.

We mention two examples. First, if $\delta_l(E)$ is approximately linear over the relevant energy window, it gives a Gaussian time dependence $e^{-\frac{1}{2}\omega E(t-2\delta_l'(E))^2}$; the amplitude is maximal when $t$ is equal to the time delay $2\delta_l'(E)$.

Second, a resonance with $E_R\approx E$ and $\omega \ll \Gamma \ll \sqrt{\omega E}$, contributes an exponentially decaying term $\Gamma \Theta(t) e^{-\frac{\Gamma t}{2}}$, convolved with a Gaussian time window $e^{-2\omega E t^2}$ of width $(\omega E)^{-\frac{1}{2}}$. Shorter-lived resonances ($\Gamma \gtrsim \sqrt{\omega E}$) are not clearly resolved, since our coherent state wavepackets are too broad. For longer-lived resonances ($\Gamma \lesssim \omega$), it is not valid to approximate their contribution to the sum over states with an integral. They do not give rise to an exponentially decaying correlator, since their lifetime is longer than the AdS time. Instead, the resonance effectively becomes a single additional stable bound state, as in figure \ref{fig:resonance} and surrounding discussion.

\section{The Coulomb potential}\label{sec:Coulomb}

We now apply these ideas to the example of the Coulomb potential in $d=3$, which is universal since it arises from the exchange of massless (or very light) particles:
\begin{equation}
	V(r) = -\frac{g}{r}.
\end{equation}
Apart from its familiarity and importance for atomic and gravitational physics in the real world, this is also a useful example theoretically because it is exactly solvable in the absence of the harmonic potential. This will allow us to establish a complete picture of the associated AdS spectrum in the strongly-coupled limit dominated by the flat spacetime Coulomb physics.

This potential has a single free parameter, the coupling $g$ with units of velocity (after dividing by $\hbar$). For gravity we have $g = G_N m_1m_2$ \cite{newton}, and for the electrostatic force we have $g=-q_1 q_2 \alpha c$ \cite{de1785premier} where $\alpha$ is the fine structure constant and $q_{1,2}$ are charges. Our full non-relativistic problem with the AdS potential is governed by a single dimensionless parameter
\begin{equation}
	\hat{g} = \sqrt{\frac{\mu}{\omega}} \, g.
\end{equation}
 We can think of this as giving the relative strength of the harmonic and Coulomb forces (for example, $\frac{1}{2}\hat{g}^2$ is the binding energy $\frac{1}{2}\mu g^2$ of the Coulomb ground state in units of the harmonic oscillator energy $\omega$). Alternatively, we can think of it as parameterising `how quantum' the problem is: there are no dimensionless parameters classically, and $\hat{g}$ is proportional to $\hbar^{-\frac{3}{2}}$.

 Here we will be mostly interested in the `strongly coupled' regime $\hat{g}\gg 1$, which is largely determined by flat spacetime physics as in section \ref{sec:flat}. The opposite regime $\hat{g}\ll 1$ is well-described by perturbation theory, discussed more generally in \cite{companion}. 
We will also concentrate on the attractive case $g>0$, since a repulsive potential $g<0$ is simpler: it simply excludes bound states, while the other regimes are essentially the same as the attractive case.

For comparison, we can express the parameters in terms of dimensionless CFT quantities. These are the two-point function $c_J$ of the mediating operator (a spin one current $J$ for electrostatic interaction, and the stress tensor $T$ for gravitational interaction so $c_T$ is the central charge) and the charges $q$ (proportional to the three-point functions $\langle \op \op^\dag J \rangle$). We have $g\propto \frac{q_1 q_2}{c_J} c$, so $\hat{g} \propto \frac{\sqrt{\Delta} q^2}{c_J}$. For gravity\footnote{For our universe taking $\omega$ to be the Hubble parameter, the order of magnitude of $c_T$ is the famous large number $10^{120}$. Large $\hat{g}$ (meaning that local gravitational attraction overwhelms cosmological physics) corresponds to masses bigger than roughly $10^{-20} \mathrm{kg}$.} we have  $g \propto \frac{\Delta^2}{c_T} c$ so $\hat{g} \propto \frac{\Delta^\frac{5}{2}}{c_T}$.

As a final general comment, note that we require $g\ll c$ for the non-relativistic limit to be applicable for the whole spectrum. Nonetheless, it determines much of the spectrum even if $g \gg c$.\footnote{For the Sun-Jupiter system $\frac{g}{c}$ is around $10^{73}$, though that is doubtless non-relativistic!}

\subsection{Scattering}\label{sec:CoulombScat}

First we consider the regime of positive energies $E$ of order $\mu g^2$ and $l$ of order one where we are sensitive to scattering physics. This is essentially an application of  section \ref{sec:scatstates}, though we must modify it slightly because the Coulomb potential does not fall off sufficiently rapidly at large $r$: we do not have a well-defined phase shift as in \ref{eq:phaseShift}, since the phase of the asymptotic partial waves grows logarithmically at large $r$. This is a symptom of the infrared effects of massless particles in $d=3$. But AdS acts as a very natural and symmetric infrared regulator, and our considerations of section \ref{sec:scatstates} require only minor modification.

Like in section \ref{sec:scatstates}, we start from the solution to the Schr\"odinger equation neglecting the harmonic potential, valid for sufficiently small $r$. We then expand for $k r\gg 1$ (analogous to \eqref{eq:phaseShift}), where we will match to a WKB solution in the asymptotic region:
\begin{equation}
\begin{aligned}
	\phi(r) &=  e^{-i k r} (k r)^{l+1} \, _1F_1\left(l+1+\frac{i\mu g}{k};2 l+2;2 i k r\right) \\
	&\sim \frac{2^{-l} \Gamma (2 l+2) e^{-\frac{\pi g \mu}{2 k}}}{\sqrt{\Gamma \left(l-\frac{i g \mu}{k}+1\right) \Gamma \left(l+\frac{i g \mu}{k}+1\right)}}\cos\left(\chi(r)+\tfrac{\pi}{4}\right), \label{eq:Cscatasymp}
\end{aligned}
\end{equation}
where the phase is given by
\begin{equation}
	\chi(r) = kr+\frac{g\mu}{k} \log (2 k r) + \Im \log \Gamma \left(1+l-i\frac{g\mu}{k}\right) -\tfrac{\pi}{2} \left(l+\tfrac{3}{2}\right).
\end{equation}
Note the $\log r$ term: as noted above, this means that we cannot immediately read off the phase shift and apply our earlier formulas.

Nonetheless, we can proceed as before with minor modifications. We can solve in the region $k r\gg 1$ using WKB, but now must include the $\frac{1}{r}$ potential as a small perturbation, giving the contribution to the phase which is linear in $g$ but logarithmic in $r$ in the matching region. We compute the WKB phase relative to the classical turning point at $r=r_0 \sim \frac{k}{\mu\omega}$ as follows:
\begin{align}
	\chi(r_0)-\chi(r)&\sim \int_r^{r_0} \sqrt{k^2-\mu^2\omega^2r^2+\tfrac{2\mu g}{r}}dr \nonumber \\
	 &\sim  \int_r^{\frac{k}{\mu\omega}} \left(\sqrt{k^2-\mu^2\omega^2r^2}+ \frac{\mu g}{r\sqrt{k^2-\mu^2\omega^2r^2}} + \cdots\right) dr \nonumber \\
	 &\sim -r\sqrt{k^2-\mu^2\omega^2r^2} +\frac{k^2}{2\mu\omega}\cos^{-1}\left(\frac{\mu\omega r}{k}\right) +\frac{\mu g}{k}\cosh^{-1}\left(\frac{k}{\mu\omega r}\right) + \cdots \nonumber \\
	 &\sim \frac{\pi E}{2\omega} -k r  -\frac{\mu g}{k}\log\left(\tfrac{\mu\omega r}{2k}\right) \qquad \left(r\ll\tfrac{k}{\mu\omega}\right).
\end{align}
This can now be matched to the asymptotics of the exact solution \eqref{eq:Cscatasymp}, giving us the phase $\chi=\chi(r_0)$ at the turning point as
\begin{equation}
	\chi = \tfrac{\pi}{2} \left(\tfrac{E}{\omega}-l-\tfrac{3}{2}\right) + \Im \log \Gamma \left(1+l-i\frac{g\mu}{k}\right)+\frac{\mu g}{k}\log\left(\tfrac{4k^2}{\mu\omega}\right).
\end{equation}
The $\log r$ term has been regulated by replacing $r$ with the infrared length scale $r_0 = \frac{k}{\mu\omega}$ at which the particle is affected by the confining potential.

At this point we can determine the spectrum and correlation functions using the general considerations of section \ref{sec:scatstates}, with phase shift
\begin{equation}\label{eq:Coulombdelta}
	\delta_l(E) = \Im \log \Gamma \left(1+l-i\frac{g\mu}{k}\right)+\frac{\mu g}{k}\log\left(\tfrac{4k^2}{\mu\omega}\right),
\end{equation}
or S-matrix
\begin{equation}\label{eq:SCoulomb}
	S_l(E) =e^{2i\delta_l(E)} = \left(\tfrac{8E}{\omega}\right)^{\frac{2g\mu i}{k}}\frac{\Gamma \left(l+1-i\frac{g \mu}{k}\right) }{\Gamma \left(l+1+i\frac{g \mu}{k}\right)},
\end{equation}
with the slight novelty that $\omega$ enters in the `flat spacetime' S-matrix to provide an infrared energy scale. The Coulomb S-matrix is usually expressed without the prefactor as a ratio of $\Gamma$-functions only. We can think of our additional prefactor as arising from an energy-dependent infrared cutoff, since larger energies travel further before being affected by the confining harmonic potential.

In particular, we may use this S-matrix to construct a scattering correlation function as in \eqref{eq:Glscat} from the sum of partial waves $S_l(E)$:
\begin{equation}
	G_\mathrm{scat}(E,\theta) =  \frac{\omega e^\frac{E}{\omega}}{2E} \left(\tfrac{8E}{\omega}\right)^{\frac{2g\mu i}{k}} \sum_{l=0}^\infty (2l+1)P_l(\cos\theta) \frac{\Gamma \left(l+1-i\frac{g \mu}{k}\right) }{\Gamma \left(l+1+i\frac{g \mu}{k}\right)}.
\end{equation}
Now this sum over $l$ does not converge in the usual sense since the terms do not go to zero, but it does converge in the sense of distributions on $S^2$, giving
\begin{equation}\label{eq:GscatCoulomb}
	G_\mathrm{scat}(E,\theta) = \frac{\omega e^\frac{E}{\omega}}{E} \left(\tfrac{8E}{\omega}\right)^{\frac{2g\mu i}{k}} i\frac{g\mu}{k} \frac{\Gamma \left(1-i\frac{g \mu}{k}\right) }{\Gamma \left(1+i\frac{g \mu}{k}\right)} \left(\sin^2 \tfrac{\theta}{2}\right)^{\frac{i g \mu}{k}-1} 
\end{equation}
for $\theta$ away from zero.\footnote{To show this, it is easier to check the inverse transform \begin{equation}
	\frac{z}{2}\frac{\Gamma(1-z)}{\Gamma(1+z)}\int_{-1}^1 dx \,  \left(\frac{1-x}{2}\right)^{z-1} P_l(x) = \frac{\Gamma(l+1-z)}{\Gamma(l+1+z)}
\end{equation} which converges for $\Re z>0$, and continue to pure imaginary $z=\frac{ig\mu}{k}$.}

The divergence in the sum over $l$ is associated with the singular behaviour of the correlator as $\theta\to 0$. But in AdS, both the divergence of the sum of partial waves and the singularity at $\theta=0$ are artefacts of the leading order flat space limit. In fact, our partial waves \eqref{eq:SCoulomb} are accurate only for sufficiently small $l\ll \frac{E}{\omega}$, receiving corrections for larger $l$ which regulate the sum. For an explicit illustration of such corrections,  we will calculate the anomalous dimensions for $l\sim \frac{E}{\omega}$ in \eqref{eq:WKBgamma}. In turn, this smooths out the singularity at  $\theta=0$ into a finite (but parametrically high and narrow) peak. This is necessary from the CFT point of view, since the correlation function is guaranteed to be free of singularities for our kinematics.

\subsection{Bound states}

We now describe the regime with negative energies of order $\mu g^2$ and $l$ of order one, which is governed by the bound states of the Coulomb potential. These are familiar as the states of the hydrogen atom, with energies
\begin{equation}
	E_{l,n} = -\frac{\mu g^2}{2(1+l+n)^2},\qquad n=0,1,2,\ldots.
\end{equation}
Like the scattering this will essentially be an application of previous work in section \ref{sec:BS}, though with minor modifications arising from the slow decay of $V$.

We can compute the OPE coefficients corresponding to these states from the decay of the wavefunction. The normalised radial wavefunctions of the hydrogen atom are 
\begin{equation}
\begin{aligned}
	\phi_{l,n}(r) &= (-1)^n\sqrt{\frac{ g \mu  \Gamma(2+2l+n)}{n! \Gamma(2l+2)^2  (l+n+1)^2 }} e^{-\kappa r} \left(2\kappa r\right)^{l+1}  {}_1F_1\left(-n;2 l+2;2 \kappa r\right) \\
	 &\hskip6cm \left(\text{with } \kappa = \tfrac{g\mu}{1+l+n}\right) \\
	&\sim \sqrt{\frac{  g \mu }{ n! (l+n+1)^2 \Gamma (2+2 l+n)}} \left(2\kappa r\right)^{n+l+1} e^{-\kappa r} \qquad (\kappa r\gg 1). \label{eq:Coulombdecay}
\end{aligned}
\end{equation}
As  in section \ref{sec:BS}, we would like to match this to a WKB solution in the region $ r\sim \frac{\kappa}{\mu \omega}$. We cannot directly apply the results of that section, since we do not have purely exponential decay: \eqref{eq:Coulombdecay} contains additional powers of $r$. To account for this, we include an extra term in the WKB tunnelling amplitude \eqref{eq:boundWKB} by expanding to leading order in $g$, as we did for the scattering phase:
\begin{equation}
	\int_r^\infty \frac{\mu g dr}{r\sqrt{\kappa^2+\omega^2\mu^2 r^2}} = \frac{g\mu}{\kappa}\sinh^{-1}\left(\tfrac{\kappa}{\omega\mu r}\right) \sim \frac{g\mu}{\kappa}\log \left(2\tfrac{\kappa}{\omega\mu r}\right) ,
\end{equation}
where we have expanded for $r \ll \frac{\kappa}{\omega\mu}$. Including this extra term we can write the wavefunction in the matching region (modifying \eqref{eq:phiasympflat}) as
\begin{equation}
	\phi(r) \sim \tilde{A}_{l,n}\sqrt{\kappa} \left(\tfrac{8|E|}{\omega }\right)^{-(n+l+1)} (2\kappa r)^{n+l+1} e^{-\kappa r}\, , \qquad r\ll \frac{\kappa}{\mu\omega},
\end{equation}
while leaving the relationship \eqref{eq:Alnbound} between $\tilde{A}_{l,n}$ and $A_{l,n}$ unaltered. By comparing with \eqref{eq:Coulombdecay}, we find our OPE coefficients:
\begin{equation}
	A_{l,n}^2 \sim \frac{  g \mu }{4n! (l+n+1)^2 \Gamma (2+2 l+n)}\left(\tfrac{8|E|}{\omega }\right)^{2n+2l+3} \left(\frac{|E_{l,n}|}{2 e\omega}\right)^{\frac{|E_{l,n}|}{\omega}}  .
\end{equation}

Exactly described in section \ref{sec:BScorr}, the $n=0$ bound states will dominate our correlation functions when we go to sufficiently low energies. The special features of the Coulomb potential are not significant enough to alter the conclusions of that section. In particular, our scattering correlator described in the previous section will be infected by a bound state at sufficiently low energy:
\begin{equation}
	\frac{k}{g\mu} <  \frac{\sqrt{W(e^{-1})}}{l+1} \implies G_l(E) \text{ dominated by bound state.}
\end{equation}

\subsection{Classical limit}\label{sec:classicalCoulomb}

We have so far discussed regimes of energy and angular momentum where the spectrum is determined by familiar Coulomb physics: Rutherford scattering and hydrogen bound states. To give a complete picture of the spectrum, we need to understand what happens when the effects of the Coulomb potential and harmonic potential cannot be so cleanly separated, which occurs in particular for sufficiently large angular momentum. 
Fortunately, in this case the physics is in a classical limit so we may apply the methods of section \ref{sec:classical}, using the WKB approximation to determine the spectrum, and an on-shell action to determine correlation functions.

This becomes clearest if we recast the eigenvalue problem in a dimensionless form, using rescaled radial coordinate $x$ and energy $\hat{E}$ defined as follows:
\begin{equation}
	r= \left(\frac{g}{\mu\omega^2}\right)^\frac{1}{3} x, \quad E = (\mu g^2 \omega^2)^\frac{1}{3} \hat{E}.
\end{equation}
The radial Schr\"odinger equation then becomes
\begin{equation}
	-\frac{1}{2}\frac{d^2 \phi}{dx^2}+ \frac{l(l+1)}{2x^2}\phi + \hat{g}^\frac{4}{3}\left(\frac{x^2}{2} - \frac{1}{x} - \hat{E} \right)\phi=0.
\end{equation}
From this form large $\hat{g}$ is clearly a classical limit, since $\hat{g}^{-\frac{2}{3}}$ plays the role of $\hbar$. The classical approximation can only fail if the classically allowed region includes very small values of $x$. This is avoided if $l$ is not too small (or for any $l$ with a repulsive potential $g<0$).

\subsubsection{Classical spectrum}

The spectrum in the classical limit is given by the WKB integral
\begin{equation}\label{eq:WKBCoulomb}
	\int_{x_0}^{x_1} \sqrt{2\left(\hat{E} + \tfrac{1}{x}-\tfrac{x^2}{2}-\tfrac{\hat{l}^2}{2x^2}\right)} \, dx = (n+\tfrac{1}{2})\pi \hat{g}^{-\frac{2}{3}},
\end{equation}
where $\hat{l} = (l+\tfrac{1}{2}) \hat{g}^{-\frac{2}{3}}$ (with an order one shift in $l$ chosen so that the WKB approximation gives the exact spectrum for the harmonic potential alone), and $x_{0,1}$ are the classical turning points where the integrand vanishes. Since the integrand is a rational function of $x$ and $\sqrt{q(x)}$ for a quartic $q(x)$, with sufficient effort it could be evaluated in closed form in terms of elliptic functions. This does not seem likely to be very enlightening, so we instead simply explain various limits.

For small $n$ the particle will sit near the minimum of the effective potential $\frac{x^2}{2}-\frac{1}{x}+\tfrac{\hat{l}^2}{2x^2}$ at $x=x_\mathrm{min}(\hat{l})$. For these first few Regge trajectories, which correspond classically to circular orbits, we can approximate the potential as a quadratic around this minimum. Since the WKB approximation gives the exact spectrum for quadratic potentials, we may use it to determine energies even for small $n$. This gives us our spectrum implicitly in terms of $x_\mathrm{min}$ as follows:
\begin{equation}\label{eq:CoulombNearCircular}
\begin{aligned}
	E_{l,n} &\sim \hat{g}^\frac{2}{3} \omega \left(x_\mathrm{min}^2- \frac{1}{2x_\mathrm{min}}\right) + \sqrt{4+x_\mathrm{min}^{-3}} (n+\tfrac{1}{2})\omega + \cdots  \\
	l &\sim  \sqrt{x_\mathrm{min}(1+x_\mathrm{min}^3)} \, \hat{g}^\frac{2}{3} - \frac{1}{2}.
\end{aligned}
\end{equation}
For small $x_\mathrm{min}$ this approaches the hydrogen spectrum expanded for fixed $n$ and large $l+\tfrac{1}{2}$, giving $E_{l,n}\sim \mu g^2\left(-\frac{1}{2(l+\frac{1}{2})^2}+\frac{n+\frac{1}{2}}{(l+\frac{1}{2})^3}+\cdots\right)$. For large $x_\mathrm{min}$ it approaches the harmonic oscillator spectrum $E_{l,n} = (2n+l+\frac{3}{2})\omega$, with leading order corrections giving the anomalous dimensions $\gamma_{l,n}\sim  - g\sqrt{\frac{\mu \omega}{l}}$. The same formula can be obtained from perturbation theory at large $l$, discussed in \cite{companion}.

At large angular momentum $\hat{l}\gg 1$, the classically allowed region is pushed to large $x$ and the Coulomb potential can be treated as a perturbation to the free problem. Without the Coulomb potential, the energy and angular momentum are related to the classical turning points $x_{0,1}$ by  $\hat{E} = \frac{x_0^2+x_1^2}{2}$, $\hat{l} = x_0x_1$, and we can expand the WKB integral \eqref{eq:WKBCoulomb} as
\begin{equation}
\begin{gathered}
	\int_{x_0}^{x_1} \frac{\sqrt{(x_1^2-x^2)(x^2-x_0^2)} }{x}\, dx + \int_{x_0}^{x_1} \frac{dx}{\sqrt{(x_1^2-x^2)(x^2-x_0^2)} } + \cdots \\
	= \frac{\pi}{4}(x_1-x_0)^2 + \frac{1}{x_1}K\left(1-\tfrac{x_0^2}{x_1^2}\right)+ \cdots.
\end{gathered}
\end{equation}
Equating this with $(n+\tfrac{1}{2})\pi \hat{g}^{-\frac{2}{3}}$ and solving for $E_{l,n}$, the first term gives us once again the MFT spectrum $E_{l,n}=2n+l+\frac{3}{2}$, and the second term gives anomalous dimensions. After using identities for the elliptic $K$ function to simplify, we obtain
\begin{equation}\label{eq:WKBgamma}
	\gamma_{l,n} \sim-\frac{2}{\pi}\sqrt{\frac{\mu g^2\omega}{l}} K\left(-\frac{n}{l}\right),
\end{equation}
which holds for large $l$ with the ratio $\frac{n}{l}$ held fixed. This result is simply the energy shift in first-order perturbation theory for small $g$, computed using the WKB wavefunction. We will discuss perturbative anomalous dimensions in more detail in the companion paper  \cite{companion}.

 This expression simplifies further in the limits $l\ll n$ and $n\ll l$:
\begin{align}
	\delta E_{l,n} &\sim -g \sqrt{\frac{\mu\omega}{l}}  \qquad \qquad \qquad\qquad (l\gg n),\\
	&\sim -\frac{1}{\pi}\sqrt{\frac{\mu g^2\omega}{n}} \log\left(16\frac{n}{l}\right) \qquad (l\ll n).\label{eq:CoulombWKB2}
\end{align}
The first of these is identical to the result we already found for the first few Regge trajectories, and remains valid for small $n$. The second matches with the large $l$ expansion of the phase shift \eqref{eq:Coulombdelta}, showing that our classical limit has overlapping regime of validity with the scattering regime discussed in section \ref{sec:CoulombScat}.

Together, the three approximations described here (scattering, bound states, and WKB) can be used to determine the entire spectrum for $\hat{g}\gg 1$.\footnote{The only tricky regime is when $\hat{E}$ and $l$ are both of order one or smaller. In this case, the WKB approximation becomes invalid for very small $r$ of order $\frac{1}{\mu g}$, where we have a Coulomb problem with negligible energy. However, while WKB is not valid for this problem, the exact solution nonetheless matches precisely with a WKB solution for $\mu g r\gg 1$ if we shift $l$ by $\frac{1}{2}$ to write the angular potential as $\frac{(l+\frac{1}{2})^2}{2\mu r^2}$. With this small adjustment, the WKB spectrum is parametrically accurate even in this regime.}

\subsubsection{Classical correlation functions}

Finally, we compute our Euclidean correlator $G(\tau,\theta)$ in the classical limit using methods described in section \ref{sec:classical}. We will do this in two regimes: a `perturbative' regime where the Coulomb potential provides a small correction to the harmonic potential, and a `flat spacetime' regime where the scattering is dominated by the Coulomb potential.

First, for the perturbative regime we use \eqref{eq:deltaS0}, which gives us the perturbation to the free classical action:
\begin{equation}\label{eq:deltaS0Coulomb}
	\delta S_0(\tau,\theta) = -\sqrt{\frac{\mu g^2}{\omega}} e^{\frac{\tau}{2}} \sec\tfrac{\theta}{2} \, K(-\tan^2\tfrac{\theta}{2}).
\end{equation}
In \cite{companion} we will find the same expression as the non-relativistic limit of a T-channel conformal block in  $d=3$ from exchange of a conserved current such as the stress tensor (or indeed any sufficiently light operator).

We can also evaluate the action in a limit where scattering occurs in a region where the  Coulomb potential dominates over the harmonic potential, since the pure Coulomb problem is solvable. The orbits are hyperbolas,
\begin{equation}
	r = \frac{J_E^2}{\mu g} \frac{\cos \left(\frac{\theta }{2}\right)}{\cos (\phi )-\cos \left(\frac{\theta }{2}\right)},
\end{equation}
with energy
\begin{equation}
	E = -\frac{g^2 \mu  }{2 J_E^2}\tan ^2\tfrac{\theta }{2}.
\end{equation}

To relate this to the correlation function in AdS we use the methods of section \ref{sec:classcorr}. First, find the kinematic variables $\tau$ and $\theta$ as well as the on-shell action $S_0$ in terms of conserved quantities $E$ and $J_E$. Then, invert the relation to find conserved quantities in terms of kinematics, so that we have an action as a function of $\tau,\theta$. We carry this out in  appendix \ref{app:Coulombscat}, with the final result
\begin{equation}\label{eq:CoulombAction}
	S_0(\tau,\theta) \sim  2 e^{-\omega \tau} -\hat{g} e^{\frac{1}{2}\omega \tau} \left[1+\log\left(\frac{32 \cos\tfrac{\theta}{2}}{\hat{g}e^{\frac{3}{2}\omega\tau}}\right)\right] + \cdots,
\end{equation}
where the relevant small parameter for the expansion is $\hat{g}e^{\frac{3}{2}\omega\tau}$. Our Euclidean correlation function $G(\tau,\theta)$ is approximated by $\exp(-S_0)$ in the regime $\hat{g}^\frac{1}{3}\ll e^{-\frac{1}{2}\omega\tau}\ll \hat{g}$ (for $\theta$ of order one): the lower bound ensures that the AdS potential is negligible in the scattering region, while the upper bound ensures that quantum corrections are small.

This result was obtained for purely Euclidean kinematics, but it is interesting to analytically continue to the Lorentzian scattering kinematics given in \eqref{eq:Gscat} (with $E = \frac{k^2}{2\mu}$):
\begin{equation}
	S_0\left(\tfrac{i\pi}{\omega}-\tfrac{1}{\omega}\log\left(\tfrac{E}{2\omega}\right),\pi-\theta\right) \sim -\frac{E}{\omega} -\frac{2i g\mu }{k} \left[1+\log\left(\frac{4ik^3 \sin\tfrac{\theta}{2}}{ \mu^2\omega g}\right)\right] + \cdots.
\end{equation}
This should be compared to the classical expansion ($k\ll \mu g$) of $-\log G_\mathrm{scat}$, computed in \eqref{eq:GscatCoulomb}.

\section{Discussion and outlook}\label{sec:disc}

\subsection{The T-channel}

In this paper we described a novel limit that we expect to be a universal sector of holographic theories. But we have not made any use of a powerful tool for constraining and understanding CFTs, namely the conformal bootstrap. We discussed how our correlation function can be decomposed into a sum over intermediate `S-channel' states, but crossing symmetry means that the same result can also be written as a sum over `T-channel' exchanges, that is over operators appearing in the OPE of $\op_1\op_1$ and of $\op_2\op_2$.

We can think of the sum over these T-channel exchanges as building the interaction potential $V(r)$, and the precise spectrum of light operators will determine the form of that potential. For example, we expect the exchange of the stress tensor $T$ (and multi-trace operators $[T\cdots T]$) to build the Newtonian Coulomb potential. This is a simple example of building a dynamical spacetime from composite operators, so it is interesting to understand the details as a concrete model of emergent spacetime.

We make the first steps in this direction in the companion paper \cite{companion}. Much of what we are able to say in that paper is constrained to weak interactions, so we also discuss perturbation theory in non-relativistic AdS in that paper, and compare to the results using the T-channel conformal block decomposition. We address both the correlation functions $G$ (comparing perturbative results to expressions for conformal blocks) and the spectrum $E_{l,n}$ (comparing to anomalous dimensions from T-channel exchanges computed using the Lorentzian inversion formula).

\subsection{Relativistic corrections and radiation}

Our entire discussion has been in the strict relativistic limit, so an obvious next direction is to incorporate relativistic corrections. 

Most obviously these corrections must become important for large $l$ or $n$, since for these large orbits the particle velocity becomes large and the spatial curvature of AdS also becomes relevant. Such corrections appear even for very weak interactions, and necessarily depend on the details of AdS spacetime. However, for strong interactions there can also be relativistic corrections that become more important at small $n$ and $l$ and are insensitive to AdS, coming entirely from flat spacetime physics. Such examples are the fine structure of the hydrogen atom, and the analogous post-Newtonian corrections for gravity \cite{Levi:2018nxp}.

At least for perturbative interactions, we should be able to recover these corrections from the T-channel conformal blocks and inversion formula using the methods of \cite{companion} by going to higher order in the non-relativistic expansion. 

In addition, we did not treat the fields mediating the interaction dynamically, including only the interaction potential that they generate and neglecting propagating excitations. Including these means that each of our S-channel states becomes a tower of states containing additional photons, gravitons, or other light particles, with energy separations of order $\omega$. At leading order in the non-relativistic limit these will typically be independent (giving a tensor product of our non-relativistic states with a Hilbert space describing free light excitations), but interactions between the particles and the radiation field will mix these sectors. We expect the resulting corrections to the energy spectrum to typically be suppressed by powers of $c$.

These mixing effects are simply the AdS version of radiative decays, whereby excited states decay by emitting electromagnetic or gravitational waves. Similarly to the resonances discussed in section \ref{sec:resonances}, the AdS potential resolves the spectrum of the full Hamiltonian to a discrete tower, but the exact eigenstates will be a superposition over states containing different numbers of radiation particles. This is similar for energy eigenstates of black holes in AdS (sufficiently large so that their decay time is longer than $\omega^{-1}$), which really describe a superposition over many internal black hole microstates in equilibrium with their Hawking radiation. The non-relativistic limit offers a simple context in which to understand the quantitative details of this feature.

The mixing of states of different particle number will also show up for scattering states in the flat space limit, where it will give rise to inelastic scattering. This will clearly require a modification of the simple relation between anomalous dimensions and phase shifts $\gamma \sim -\frac{2}{\pi}\delta$. The non-relativistic limit offers a straightforward setting to investigate these effects.

\subsection{Hidden symmetries and precession}

Section \ref{sec:Coulomb} was a detailed discussion of the Coulomb potential in $d=3$, chosen partly because this is an exactly solvable problem  when $\omega\to 0$. The underlying reason for this is a hidden symmetry: the conserved Laplace-Runge-Lenz vector enhances the usual $SO(3)$ rotational symmetry to $SO(4)$. We expect Coulomb interactions to be ubiquitous in theories with local four-dimensional bulk descriptions, so such an enhanced symmetry must emerge in the appropriate regime of $d=3$ holographic CFTs (without large compact internal spaces). Can we understand the emergence of such a hidden symmetry from a CFT perspective?

This symmetry shows up in the spectrum as a simple dependence of the spectrum $E_{l,n}$ on the quantum numbers $l,n$, only through the combination $l+n$ (which labels representations of the enhanced symmetry) \cite{pauli1926spectrum}, and hence we have degeneracies between states of different spin. Classically, the symmetry results in closed orbits. These (the quantum degenerate spectrum and closed classical orbits) are equivalent. If we add effects that break the symmetry, we can relate the violation of the spectral degeneracies to the precession of the orbits.

First, we briefly review how the period of classical orbits $T$ is related to the quantum spectrum for a one-dimensional problem. In a classical regime, there are many states with approximately equal energy differences $\Delta E$ between them, and a classical state is a coherent superposition over many such states. Under time evolution each picks up a phase $e^{-i E t}$, so we return to the same state after a time $T=\frac{2\pi}{\Delta E}$ when the relative phases between eigenstates cancels. This is the period of classical motion (as can be verified directly from the WKB approximation).

Now consider a higher-dimensional problem of motion in a central potential. The radial motion for fixed $l$ reduces to a one-dimensional problem with the same period $T=\frac{2\pi}{\Delta_n E}$ given by energy spacings $\Delta_n E=E_{l,n+1}-E_{l,n}$ at fixed spin (the energy gaps between Regge trajectories). But classical states are superpositions over many values of $l$ (for a localised wavefunction in the angular directions), and evolution $e^{-i H T}$ will typically give relative phases between different spins. This is because classical orbits need not be closed: they generically precess, going through an angle $\Phi$ between periapsides. To account for this, we evolve by $e^{-i H T+i L \Phi}$ where $L$ is a rotation in the axis perpendicular to the plane of the orbit, and demand that phases cancel between states with nearby values of $l$ as well as $n$. This requires $T\Delta_l E -\Phi$ to be an integer multiple of $2\pi$, where $\Delta_l E = E_{l+1,n}-E_{l,n}$ (the slope of the Regge trajectories), so
\begin{equation}
	T = \frac{2\pi}{\Delta_n E}, \qquad \Phi = 2\pi \frac{\Delta_l E}{\Delta_n E}.
\end{equation}
As an example, we can look at the near-circular (i.e., small $n$) classical Coulomb orbits with spectrum given in \eqref{eq:CoulombNearCircular}. As a function of the radius $r$ of the orbit, the angle $\Phi$ is
\begin{equation}
	\Phi =  2\pi\sqrt{\frac{\mu \omega^2 r^3+g}{4\mu \omega^2 r^3+g}}.
\end{equation}
This interpolates between $\Phi=2\pi$ for Keplerian orbits at small $r$, where energies depend only on $l+n$, and $\Phi=\pi$ for harmonic oscillator orbits at large $r$, where energies depend on $l+2n$. There is a factor of two because harmonic orbits are ellipses with $r=0$ at the centre (rather than a focus as for Kepler orbits), so they reach their minimum radius twice per orbit.

These considerations allow us to study precession via the quantum spectrum. A historically important example is the precession of the perihelion of Mercury from general relativistic corrections. Since we expect this result to be universal for a local gravitational theory in four dimensions, perhaps one can bootstrap this famous result from a CFT with minimal assumptions? It is also interesting to study examples where the symmetry is not broken by relativistic corrections, such as $\mathcal{N}=8$ supergravity \cite{Caron-Huot:2018ape}, though inevitably it must be broken at large spin by AdS curvature.

\subsection{Top down examples}

Our non-relativistic limit gives a plethora of new predictions for holographic CFTs, so it would be interesting to realise these in known explicit examples such as $\mathcal{N}=4$ Yang-Mills. The main challenge is that we require large dimension operators $\Delta\gg 1$, which excludes typical supergravity fields with AdS scale masses. One candidate is an operator dual to an excited string state, for which $\Delta$ scales as the ratio between AdS and string lengths.

Top down examples are not typically local only in AdS, but have large internal spaces:  AdS$_5\times S^5$ for  $\mathcal{N}=4$, with the radius of $S^5$ the same order as the AdS length. These compact directions must be accounted for, and make a significant qualitative difference since there is no confining potential in these internal directions. Additionally, we expect the interaction potential relevant to the higher dimension (including the compact directions).

\subsection{de Sitter}

There is a very similar description for non-relativistic physics in de Sitter spacetime. The difference is that the de Sitter curvature becomes a negative quadratic Newtonian potential: for each particle we add
\begin{equation}
	V_\mathrm{dS}(r) = -\frac{1}{2} m H^2 r^2,
\end{equation}
where $H$ is the Hubble parameter. Like the AdS case we have been discussing, this is a deformation of ordinary quantum mechanics that retains the Galilean symmetries in a modified form. This limit was recently used for a toy model of static patch holography in \cite{Susskind:2021omt}.

The qualitative difference from AdS is that instead of a completely discrete spectrum of states confined by the potential, we have an entirely continuous spectrum of `scattering states'. For any interaction potential and initial state, the wavefunction eventually disperses exponentially rapidly to large $r$. This non-relativistic limit gives us an extremely simple demonstration of the effects of an exponentially expanding universe in a regime where relativity is unnecessary, which is nonetheless parametrically accurate in the regime of slow relative motion.

In particular, a state which would be bound in flat spacetime with energy $E<0$ becomes unstable in dS for arbitrarily small $H$. The bound state decays by tunnelling to the point at which the energy $E$ matches the de Sitter potential $-\frac{1}{2} \mu H^2 r^2$ (in the limit $\omega H \ll |E|$), so $r = \frac{1}{H}\sqrt{\frac{2|E|}{\mu}}$; for non-relativistic binding energies $|E|\ll \mu c^2$, we have $r\ll \frac{c}{H}$ so the decay takes place well inside the cosmological horizon. When $\omega H \ll |E|$, the lifetime of the bound state can be computed simply from a WKB tunnelling amplitude, scaling as $\exp({\frac{\pi |E|}{H}})$ (a result which previously appeared in \cite{Volovik:2009eb}). Note that this is much smaller (by a square root) than the Boltzmann factor at the de Sitter temperature $T_\mathrm{dS} = \frac{H}{2\pi}$ so this decay is much faster than ionisation by Hawking photons. Nonetheless, for the hydrogen atom in our universe this factor is something like $e^{10^{33}}$, so we shouldn't worry too much about instability of matter due to $\Lambda$.

\paragraph{Acknowledgements}

We would like to thank Simon Caron-Huot for collaboration in the very early stages, and helpful comments and insights. HM is supported by DOE grant DE-SC0021085 and a Bloch fellowship from Q-FARM. HM was also supported by NSF grant PH-1801805, by a DeBenedictis Postdoctoral Fellowship,  and by funds from the University of California. ZZ is funded by Fonds de Recherche du Québec \textemdash \  Nature et Technologies, and the Simons Foundation through the Simons Collaboration on the Nonperturbative Bootstrap.

\appendix

\addtocontents{toc}{\setcounter{tocdepth}{1}}

\section{Spherical harmonics in general dimension}\label{app:SH}

Spherical harmonics are eigenfunctions of the Laplacian on the unit $S^{d-1}$:
\begin{equation}
\nabla^2_{S^{d-1}}Y_{l,m}=-l(l-d+2)Y_{l,m},
\end{equation}
where $l=0,1,2,\ldots$. This value labels the representation of rotations $SO(d)$ under which the functions transform, while the values of $m$  for any given $l$ label a basis of this representation. The representation in question is the $l$-fold symmetric tensor product, which is made manifest by writing $Y_{l,m}$ as harmonic homogeneous polynomial of degree $l$ restricted to the unit $(d-1)$-sphere.

For each $l$, there is a unique spherical harmonic depending only on a latitude $\theta$, called the `zonal spherical harmonics'. These are polynomials of $\cos\theta$,
\begin{equation}
	Z_{l}(\vec {\theta}) = \frac{1}{\sqrt{\mathcal{N}_l}} C_l(\cos\theta),
\end{equation}
where $C_l$ is the Gegenbauer polynomial of order $l$ with paramater $\alpha=\frac{d-2}{2}$, generalising the familiar Legendre polynomials for $d=3$. Choosing  use the Wikipedia/Mathematica normalisation convention $C_l(x)=\mathtt{GegenbauerC[l,\tfrac{d-2}{2},x]}$, we can write them as 
\begin{equation}\label{eq:Cl2F1}
	C_l(x) = \binom{l+d-3}{l} \, _2F_1\left(-l,d+l-2;\tfrac{d-1}{2};\tfrac{1-x}{2}\right)
\end{equation}
 (except for $d=2$, where a different normalisation is required). The constants $\mathcal{N}_l$ are chosen to satisfy orthonormality when integrated on the unit $S^{d-1}$,
\begin{equation}
	\int_{S^{d-1}} d\Omega\, Z_l(\theta)Z_{l'}(\theta) = \frac{\Omega_{d-2}}{\sqrt{\mathcal{N}_l\mathcal{N}_{l'}}} \int_{-1}^1 dx (1-x^2)^{\frac{d-3}{2}} C_l(x)C_{l'}(x) = \delta_{l,l'}.
\end{equation}
Here $\Omega_n = \frac{2\pi^{\frac{n+1}{2}}}{\Gamma(\frac{n+1}{2})}$ is the volume of the unit $S^n$, and the normalisation constant is $\mathcal{N}_l = \frac{d-2}{2l+d-2}\binom{l+d-3}{l}\Omega_{d-1}$.

The zonal harmonics form a complete basis for the functions of latitude on $S^{d-1}$:
\begin{equation}
	f(\theta) = \frac{1}{\Omega_{d-1}}\sum_{l=0}^\infty \frac{2l+d-2}{d-2} a_l \, C_l(\cos\theta)
\end{equation}
for any sufficiently nice function $f$. We call the coefficients $a_{l}$ the partial waves of $f$. We compute them using the orthogonality conditions:
\begin{equation}
\begin{aligned}
	a_l &= \frac{1}{\binom{l+d-3}{l}}\int_{S^{d-1}	}\mkern-18mu d\Omega f(\theta)C_l(\cos\theta) \\
	&	= \frac{\Omega_{d-2}}{\binom{l+d-3}{l}} \int_{-1}^1 f(\theta) C_l(x) \left(1-x^2\right)^\frac{d-3}{2} \, dx \, ,
\end{aligned}
\end{equation}
where $x=\cos\theta$ in the last line. We normalised the coefficients such that if $f(\theta)$ represents a $\delta$-function supported at the North pole $\theta=0$, we have $a_l=1$ for all $l$. 

As an example we look at the partial wave decomposition of the free correlator given in eq.~\eqref{eq:Gfree}:
\begin{equation}
G_\mathrm{free}(\tau,\theta)=e^{2e^{-\tau}\cos\theta}.
\end{equation}
The partial wave decomposition of this correlator can then be written as
\begin{equation}
G_{l}(\tau)=\frac{\Omega_{d-2}}{\binom{l+d-3}{l}} \int_{-1}^1 e^{2e^{-\tau}x} C_l(x) \left(1-x^2\right)^\frac{d-3}{2} \, dx\,.
\end{equation}
This integral can be performed using the Rodrigues formula for $C_l$,
\begin{equation}
C_{l}(x)=\frac{(-1)^{l}}{2^{l}l!}\frac{\Gamma(\frac{d-1}{2})\Gamma(l+d-2)}{\Gamma(d-2)\Gamma(\frac{d-1}{2}+l)}(1-x^2)^{\frac{3-d}{2}}\frac{d^{l}}{dx^{l}}\left[(1-x^2)^{l+\frac{d-3}{2}}\right],
\end{equation}
giving
\begin{equation}
G_{l}(\tau)=\frac{(-1)^{l}2\pi^{\frac{d-1}{2}}}{2^{l}\Gamma(l+\frac{d-1}{2})}\int^1_{-1} e^{2e^{-\tau}x} \frac{d^{l}}{dx^{l}}\left[(1-x^2)^{l+\frac{d-3}{2}}\right]dx \,.
\end{equation}
This integral can be done by using partial integration $l$ times and realizing that the  total derivative term created each time vanishes since the integrand at every step is zero at the endpoints. This leaves an us with
\begin{equation}
G_{l}(\tau)=\frac{2\pi{\frac{d-1}{2}}(e^{-\tau})^{l}}{\Gamma(l+\frac{d-1}{2})}\int^1_{-1}e^{2e^{-\tau}x}(1-x^2)^{l+\frac{d-3}{2}}dx \,,
\end{equation}
which is an integral representation of the Bessel $I$ function:
\begin{equation}
G_{l}(\tau)=2\pi^{d/2}e^{\frac{d-2}{2}\tau}I_{l+\frac{d}{2}-1}(2e^{-\tau}) \, .
\end{equation}
This result is the partial amplitude quoted in \eqref{eq:Glfree}

\section{S-channel conformal blocks}\label{app:Sblocks}

In this appendix we determine the S-channel conformal blocks in the non-relativistic limit.

First, the conformal blocks admit simple closed-form expressions in even dimensions. These are all in terms of the `lightcone block'
\begin{equation}
\begin{aligned}
	k_\beta(x) &= x^\frac{\beta}{2} {}_2F_1\left(\tfrac{\beta+\Delta_2-\Delta_1}{2},\tfrac{\beta+\Delta_2-\Delta_1}{2};\beta;x\right),
\end{aligned}	
\end{equation}
which also appears as the lightcone ($z\to 0$) limit of blocks in general dimension. For $d=2,4$, these expressions are the following:
\begin{align}
g_{\Delta,l}(z,\bar{z}) &= \frac{k_{\Delta-l}(z)k_{\Delta+l}(\bar{z})+k_{\Delta+l}(z)k_{\Delta-l}(\bar{z})}{1+\delta_{l,0}} \qquad (d=2) \\
	g_{\Delta,l}(z,\bar{z}) &= \frac{z\bar{z}}{\bar{z}-z}(k_{\Delta-l-2}(z)k_{\Delta+l}(\bar{z})-k_{\Delta+l}(z)k_{\Delta-l-2}(\bar{z}))  \quad (d=4)
\end{align}

For the non-relativistic limit, we are interested in small cross-ratios and large dimension, which requires the limit of small $x$ and large $\beta$ and $\Delta_{1,2}$ of the same order as $x^{-1}$. This is simple to compute by taking the limit term by term in the hypergeometric series, which becomes an exponential:
\begin{equation}
	k_\beta(x) \sim  x^\frac{\beta}{2} \exp\left(\tfrac{(\beta+\Delta_2-\Delta_1)^2}{4\beta}x\right).
\end{equation}
This holds as long as $\beta$  doesn't approach a negative integer as it becomes large.

Using this in $d=2$, we find
\begin{equation}
	g_{\Delta,l}(z,\bar{z}) \sim (z\bar{z})^\frac{\Delta}{2}\frac{\left(\frac{\bar{z}}{z}\right)^\frac{l}{2} + \left(\frac{z}{\bar{z}}\right)^\frac{l}{2}}{1+\delta_{l,0}} \exp\left(\frac{\Delta_2^2}{\Delta_1+\Delta_2}(z+\bar{z})\right).
\end{equation}
In $d=2$, the normalisation $\mathcal{N}_{d,l}$ blow up and the Gegenbauer polynomials $C_l$ go to zero, but their combination is well-defined, giving $\mathcal{N}_{d,l}C_l(\cos\theta) = 2 \,{}_2F_1(-l,l;\frac{1}{2};\frac{1-\cos\theta}{2})=2T_l(\cos\theta)$ where $T_l$ are Chebyshev polynomials (except at $l=0$, where we get unity). This is precisely the spin-dependence of the above, so we can write this as
\begin{equation}
	g_{\Delta,l}(z,\bar{z}) \sim 2T_l(\cos \theta) (z\bar{z})^\frac{\Delta}{2}\exp\left(\frac{\Delta_2^2}{\Delta_1+\Delta_2}(z+\bar{z})\right),
\end{equation}
where $\cos \theta=\frac{z+\bar{z}}{2\sqrt{z\bar{z}}}$.

For $d=4$, we first note that $\mathcal{N}_{d,l}=1$, and the Gegenbauer polynomial $C_l$ becomes a Chebyshev polynomial of the second kind,
\begin{equation}
	C^{(d=4)}_l(\cos\theta) = U_l(\cos\theta) = \frac{\sin((l+1)\theta)}{\sin(\theta)} = \frac{\left(\frac{\bar{z}}{z}\right)^\frac{l+1}{2} - \left(\frac{z}{\bar{z}}\right)^\frac{l+1}{2}}{\left(\frac{\bar{z}}{z}\right)^\frac{1}{2} - \left(\frac{z}{\bar{z}}\right)^\frac{1}{2}}.
\end{equation}
Using this, the above formula for the blocks becomes
\begin{equation}
	g_{\Delta,l}(z,\bar{z}) \sim (z\bar{z})^\frac{\Delta}{2}U_l(x) \exp\left(\frac{\Delta_2^2}{\Delta_1+\Delta_2}(z+\bar{z})\right).
\end{equation}

These two examples lead to suggest a formula for general dimension $d$:
\begin{equation}
	g_{\Delta,l}(z,\bar{z}) \sim \mathcal{N}_{d,l}C_l(\cos\theta) (z\bar{z})^\frac{\Delta}{2}\exp\left(\frac{\Delta_2^2}{\Delta_1+\Delta_2}(z+\bar{z})\right).
\end{equation}
This has the correct expansion at small $z,\bar{z}$. To confirm that this indeed holds in general dimension we check that it obeys the Casimir differential equation in the appropriate limit. This equation can be found in \cite{Poland:2018epd}. We make an ansatz $g_{\Delta,l}(z,\bar{z}) = C_l(\cos\theta) (z\bar{z})^\frac{\Delta}{2} \tilde{g}(z,\bar{z})$, where $\tilde{g}$ is of order one in the appropriate limit (with corrections given by powers of the small parameter). To leading order, $\tilde{g}$ obeys a first order equation with general solution $\exp\left(\frac{\Delta_2^2}{\Delta_1+\Delta_2}(z+\bar{z})\right)$ times any function of $\theta$ (the spin does not appear at this order in the equation). The proposed formula is fixed uniquely by the $z,\bar{z}\to 0$ expansion.

\section{The S-channel resolvent}\label{app:resolvent}

In this appendix we expand on the resolvent $R_l(E)$ discussed in section \ref{sec:resolvent}. This is a meromorphic function with poles at $E=E_{l,n}$ with residues $-A_{l,n}^2$, so that we have
\begin{equation}\label{eq:Glresolvent}
	G_l(\tau) =  \pi^{\frac{d}{2}} \int_\Gamma \frac{dE}{2\pi i} \,  R_l(E) e^{-\tau (E-\frac{d}{2})}
\end{equation}
for an appropriate contour $\Gamma$ encircling the poles clockwise. From the residue theorem we recover \eqref{eq:Glsum}. Depending on the asymptotic behaviour of $R_l$, we may take $\Gamma$ to be a contour running from the lower to upper half-plane, passing to the left of all the poles, and going to infinity in an appropriate direction.

Importantly, the poles and residues do not uniquely define $R_l(E)$, since we have the freedom to add any entire analytic function of $E$. We will mention several ways to construct functions $R_l(E)$, but these will have ambiguities and will not always give rise to the same function. It is not clear whether there is any particular `preferred' choice for $R_l(E)$.

\subsection{Constructions of $R_l(E)$}

\subsubsection{From the Schr\"odinger equation at general $E\in\CC$}

One way to construct $R_l(E)$ is from the solutions to the time-independent radial Schr\"odinger equation \eqref{eq:radialSchr} for general $E\in\CC$. Up to scaling, there is a unique solution $\phi_{l,E}(r)$ with regular behaviour ($\phi_{l,E}(r)\sim r^l$) at $r\to 0$. We can write this as a linear combination of solutions $\phi_{l,E}^{(\pm)}(r)$ with specified behaviour as $r\to\infty$:
\begin{equation}\label{eq:pmasymp}
\begin{gathered}
	\phi_{l,E}(r) = A_{l,E}^{(+)} \phi_{l,E}^{(+)}(r) + A_{l,E}^{(-)} \phi_{l,E}^{(-)}(r),\\
	\phi_{l,E}^{(\pm)}(r) \sim \frac{1}{\sqrt{r}}(\mu r^2)^{\mp \frac{E}{2}} e^{\pm \frac{1}{2}\mu r^2} \qquad (r\to\infty).
\end{gathered}
\end{equation}
This defines $\phi_{l,E}^{(-)}(r)$ uniquely, but not $\phi_{l,E}^{(+)}$ (we can add any multiple of the decaying solution $\phi_{l,E}^{(-)}$ without changing it). We insist only that our choice depends analytically on $E$.\footnote{This is always possible; for example we can define $\phi_{l,E}^{(+)}$ by its behaviour as $r\to \pm i\infty$.} From this, define
\begin{equation}\label{eq:Cldef}
	R_l(E) =  \frac{A_{l,E}^{(-)}}{A_{l,E}^{(+)}}.
\end{equation}
 The ambiguity in the definition of $\phi_{l,E}^{(+)}$ corresponds precisely to the freedom to add an entire function to $R_l(E)$.
 
This function will have the desired poles and residues. The presence of poles for physical states is clear from \eqref{eq:pmasymp}, since the denominator $A_{l,E}^{(+)}$ vanishes precisely when the solution $\phi_{l,E}$ with good boundary conditions at $r=0$ also decays at $r\to \infty$. The residues require a calculation.

Suppose we have a solution $\phi_{l,E}$ at energy $E\in\CC$, and would like to construct its variation under a small change in the energy $\delta E$. The linear variation $\delta\phi = \phi_{l,E+\delta E} -\phi_{l,E}$ satisfies
\begin{equation}
	(H_l-E)\delta\phi = \delta E \phi_{l,E},
\end{equation}
which can be solved in terms of the two independent unperturbed solutions $\phi_{l,E}^{(\pm)}$ by the method of variation of parameters. We can write the general solution as
\begin{equation}
\begin{gathered}
	\delta\phi(r) = a_+(r) \phi_{l,E}^{(+)}(r)+ a_-(r) \phi_{l,E}^{(-)}(r), \\
	\text{where } a_\pm'(r) = \mp \delta E \phi_{l,E}^{(\mp)}(r)\phi_{l,E}(r).
\end{gathered}
\end{equation}
This solves the perturbed equation, since $\phi_{l,E}^{(\pm)}$ are linearly independent solutions to the unperturbed equation. In particular, we use the fact that the Wronskian $\phi_{l,E}^{(-)}(\phi_{l,E}^{(+)})'- \phi_{l,E}^{(+)}(\phi_{l,E}^{(-)})'$ is a constant independent of $r$; we can evaluate this constant to be $2\mu$ using the asymptotics \eqref{eq:pmasymp}.

Now, if the unperturbed energy corresponds to an eigenstate $E=E_{l,n}$, we have $A^{(+)}_{l,E}=0$, and $\phi_{l,E} = A^{(-)}_{l,E}\phi_{l,E}^{(-)}$. We are interested in computing the variation $\delta A^{(+)} = A^{(+)}_{l,E+\delta E}$, which is given by the asymptotic value $a_+(\infty)$ of $a_+$. Since $\phi_{l,E}^{(+)}$ is a linearly independent solution to the unperturbed Schr\"odinger equation, it will not have the required behaviour as $r\to 0$, so the boundary conditions for $\delta\phi$ at $r\to 0$ require that $a_+(0)=0$. We can thus compute $\delta A^{(+)}$ from an integral,
\begin{equation}
	\delta A^{(+)} = -\delta E \,  \int_0^\infty \phi_{l,E}(r) \phi_{l,E}^{(-)}(r) dr = -\delta E A^{(-)}\int_0^\infty \phi_{l,E}^{(-)}(r)^2 dr.
\end{equation}
From this we read off the residue:
\begin{equation}
	\Res_{E\to E_{l,n}}R_l(E) = - \left(\int_0^\infty \phi_{l,E}^{(-)}(r)^2 dr\right)^{-1} = -A_{l,n}^2.
\end{equation}
The second equality follows from the definition of the coefficients $A_{l,n}$, which gives us $\phi_{l,n} = A_{l,n}\phi_{l,E}^{(-)}$, and the normalisation $\int\phi_{l,n}^2=1$.

Thus,  we find that $R_l(E)$ has simple poles at the energies $E=E_{l,n}$ with the claimed residues.

\subsubsection{As matrix elements of the resolvent of $H$}

An alternative construction of $R_l(E)$ begins with the matrix elements of the resolvent $(H-E)^{-1}$ of the non-relativistic Hamiltonian between states $|\psi(\tau_\mathrm{in},\Omega_\mathrm{in})\rangle$ (after decomposing into partial waves). By inserting a complete set of states, we can express this as a weighted sum over the poles,
\begin{equation}\label{eq:Rlsum}
	R_l^{(\tau_0)}(E) = \sum_{n=0}^\infty A_{l,n}^2 \frac{e^{-(E_{l,n}-E)\tau_0}}{E_{l,n}-E},
\end{equation}
where $\tau_0$ is a parameter depending on the choice of $\tau_\mathrm{in}$ in the states $|\psi(\tau_\mathrm{in},\Omega_\mathrm{in})\rangle$. This sum converges absolutely for all $E$ (away from poles), since the tail of the sum is determined by large energies where the potential becomes negligible and we may use the free result, with coefficients decaying as $(n!)^{-2}$.

This is equivalent to an expression for $R_l(E)$ as a Laplace transform of $G_l(\tau)$, noticing that \eqref{eq:Glresolvent} takes the form of a Bromwich integral for an inverse Laplace transform:
\begin{equation}\label{eq:resInt}
	R_l^{(\tau_0)}(E) = \pi^{-\frac{d}{2}}\int_{\tau_0}^\infty e^{\tau(E-\frac{d}{2})}G_l(\tau)d\tau
\end{equation}
for $\Re E< E_{l,0}$, and extended analytically.

In this form, there is only a one-parameter ambiguity coming from the choice of $\tau_0$, and no choice is obviously more natural than any other ($\tau_0=0$ might look natural, but it depends on the precise conventions used in our definitions). The poles and residues are manifest from the expression as a sum over eigenstates.

\subsection{Example: free particles}

We illustrate the constructions above with the example of the free problem ($V=0$). 

First, we use the Schr\"odinger equation method. The solution to the Schr\"odinger equation with regular behaviour at $r=0$ is given by
\begin{equation}
\phi_{l,E}(r)  =  \frac{e^{-\frac{1}{2}\mu r^2}}{\sqrt{r}}(\mu r^2)^{\frac{l}{2}+\frac{d}{4}}{}_1F_1(\tfrac{d}{4}+\tfrac{l}{2}-\tfrac{E}{2};l+\tfrac{d}{2};\mu r^2),
\end{equation}
and the solution decaying at infinity is
\begin{equation}
	\phi^{(-)}_{l,E}(r) = \frac{e^{-\frac{1}{2}\mu r^2}}{\sqrt{r}}(\mu r^2)^{\frac{l}{2}+\frac{d}{4}}U(\tfrac{d}{4}+\tfrac{l}{2}-\tfrac{E}{2};l+\tfrac{d}{2};\mu r^2),
\end{equation}
where ${}_1F_1$ and $U$ are confluent hypergeometric functions. Now, there is not an immediately obvious choice for the growing solution. One possibility is to take a solution that decays as $r\to i \infty$ or as $r\to -i \infty$. We will take the average of these (to give a real solution for real $E$),  which can be written as
\begin{equation}
\begin{aligned}
	\phi^{(+)}_{l,E}(r) &= \tfrac{1}{2}e^{-\frac{i\pi}{2}\left(\frac{d}{2}+l+E\right)} \frac{e^{\frac{1}{2}\mu r^2}}{\sqrt{r}}(\mu r^2)^{\frac{l}{2}+\frac{d}{4}}U(\tfrac{d}{4}+\tfrac{l}{2}+\tfrac{E}{2};l+\tfrac{d}{2}; e^{-i\pi}\mu r^2) \\
	&+ \tfrac{1}{2}e^{+\frac{i\pi}{2}\left(\frac{d}{2}+l+E\right)} \frac{e^{\frac{1}{2}\mu r^2}}{\sqrt{r}}(\mu r^2)^{\frac{l}{2}+\frac{d}{4}}U(\tfrac{d}{4}+\tfrac{l}{2}+\tfrac{E}{2};l+\tfrac{d}{2}; e^{+i\pi}\mu r^2).
\end{aligned}
\end{equation}
The confluent hypergeometric $U$ function is defined with a branch cut along the negative real axis; we have written the arguments with factors of $e^{\pm i\pi}$ to indicate whether it is to be evaluated above or below the branch cut.

Now, our solution $\phi_{l,E}$ is a linear combination of $\phi^{(\pm)}_{l,E}$ with the appropriate behaviour $\phi_{l,E}(r)\sim \frac{1}{\sqrt{r}}(\mu r^2)^{\frac{l}{2}+\frac{d}{4}}$ as $r\to 0$. The ratio $R_l(E)$ of the coefficients of $\phi^{(\pm)}_{l,E}$ is fixed by setting the coefficient of the more singular solution to zero, using the expansion of the confluent hypergeometric $U$ around $z=0$,
\begin{equation}
	U(a,b,z) \sim  \frac{\Gamma(b-1)}{\Gamma(a)} z^{1-b}  \qquad (\Re b>1).
\end{equation}
From this, we obtain
\begin{equation}\label{eq:freeRes1}
	R_l(E) = \cos\left(\tfrac{\pi}{2}\left(E-l-\tfrac{d}{2}\right)\right)\frac{\Gamma(\frac{d}{4}+\frac{l}{2}-\frac{E}{2})}{\Gamma(\frac{d}{4}+\frac{l}{2}+\frac{E}{2})}.
\end{equation}
As expected, this is a meromorphic function of $E$, with simple poles at $E=E_{l,n}=\frac{d}{2}+l+2n$ for $n=0,1,2,\ldots$, and the residues as required from \eqref{eq:freeAnl}.

An alternative expression is given by the sum \eqref{eq:Rlsum} with MFT spectrum and OPE coefficients, or the integral \eqref{eq:resInt} using the partial wave amplitudes \eqref{eq:Glfree}. Either way, we obtain
\begin{equation}
	R_l^{(\tau_0)}(E) = \frac{2 e^{\tau_0 \left(E-\frac{d}{2}-l\right)}}{(\frac{d}{2}+l-E) \Gamma \left(\frac{d}{2}+l\right)}\, _1F_2\left(\tfrac{d}{4}+\tfrac{l}{2}-\tfrac{E}{2};1+\tfrac{d}{4}+\tfrac{l}{2}-\tfrac{E}{2},\tfrac{d}{2}+l;e^{-2 \tau_0}\right).\nonumber
\end{equation}
Note that (for any $\tau_0$) this is different from our previous expression \eqref{eq:freeRes1}.

\subsection{The Euclidean inversion formula}

The idea of the resolvent $R_l(E)$ may sound familiar to CFT experts from the conformal partial wave decomposition of correlation functions, used in both the `Euclidean' and  `Lorentzian' inversion formulas \cite{Caron-Huot:2017vep}. These formulas uniquely define a similar function $c(\Delta,l)$  with poles at the locations of S-channel operators and residues giving OPE coefficients. We might hope that $c(\Delta,l)$ obtained from the inversion formulas becomes a resolvent $R_l(E)$ in the non-relativistic limit. Unfortunately, there does not appear to be a nice non-relativistic limit for $c(\Delta,l)$ or the Euclidean inversion formula.

We can see this first by taking a non-relativistic limit of the Euclidean inversion formula, simply expanding the integrand in the non-relativistic regime. The conformal partial wave is dominated by the `shadow block', which is evaluated using the same methods as appendix \ref{app:Sblocks}. The result is a formula like \eqref{eq:resInt} giving a Laplace transform of the partial waves $G_l$, but with $\tau_0\to -\infty$ (since it is an integral over all Euclidean cross-ratios). The $\tau\to-\infty$ region introduces dependence beyond the non-relativistic limit (though we expect this to be independent of interactions). This indicates that we may not recover a nice non-relativistic limit.

Indeed, we see this explicitly from the example of MFT. The function $c(\Delta,l)$ is given by
\begin{align}
		c(\Delta,l) = &\frac{ \Gamma (\Delta -1) \Gamma (\frac{d}{2}-\Delta_{1}) \Gamma (\frac{d}{2}-\Delta_{2}) \Gamma (\frac{d}{2}+l) \Gamma (d+l-\Delta )
		}{
		2\Gamma (\Delta_{1}) \Gamma (\Delta_{2}) \Gamma (l+1) \Gamma (\Delta -\frac{d}{2}) \Gamma (\Delta+l -1)
		}
		\\
		\times &
		\frac{
		\Gamma (\frac{\Delta+l +\Delta_{1}-\Delta_{2}}{2})
		\Gamma (\frac{\Delta+l -\Delta_{1}+\Delta_{2}}{2})
		\Gamma (\frac{l-\Delta +\Delta_{1}+\Delta_{2}}{2})
		\Gamma (\frac{\Delta+l-d +\Delta_{1}+\Delta_{2}}{2})
		}{
		\Gamma (\frac{d-\Delta+l +\Delta_{1}-\Delta_{2}}{2})
		\Gamma (\frac{d-\Delta+l -\Delta_{1}+\Delta_{2}}{2})
		\Gamma (\frac{2 d+l-\Delta -\Delta_{1}-\Delta_{2}}{2})
		\Gamma (\frac{\Delta+l+d -\Delta_{1}-\Delta_{2}}{2})
		}\nonumber,
\end{align}
from equation (3.118) of \cite{Karateev:2018oml} (up to a factor of $2^J$ since we use a different normalisation convention). It is relatively straightforward to take a non-relativistic limit (writing $\Delta = \Delta_1+\Delta_2-\frac{d}{2}+E$, and taking $\Delta_{1,2}\to\infty$ with fixed $l,E$), and whenever the argument of a $\Gamma$-function is going to negative infinity, we rewrite it using the reflection formula $\Gamma (z)=\frac{\pi }{\sin (\pi  z) \Gamma (1-z)}$ and thereafter use Stirling's formula. This gives us a candidate $R_l(E)$ by stripping off  a factor of $\frac{1}{2l!}\Gamma\left(\frac{d}{2}+l\right)\mu^{E-\frac{d}{2}}$ relating $f_{l,n}^2$ to $A_{l,n}^2$ as in \eqref{eq:OPEA}, with the following result:
\begin{align}
R_l&(E=\tfrac{d}{2}+l+2n) \sim \nonumber \\
&\frac{\sin \left(\pi  \left(\Delta_1 -\frac{d}{2}+ n\right)\right)}{\sin \left(\pi  \left(\Delta_1 -\frac{d}{2}\right)\right)}\frac{\sin \left(\pi  \left(\Delta_2 -\frac{d}{2}+ n\right)\right)}{\sin \left(\pi  \left(\Delta_2 -\frac{d}{2}\right)\right)} \frac{\sin \left(\pi  \left(\Delta_1 + \Delta_2 -\frac{d}{2}+ n\right)\right)}{\sin \left(\pi  \left(\Delta_1 + \Delta_2 -\frac{d}{2}+2n\right)\right)}\nonumber \\
&\qquad\qquad\qquad\qquad\times \frac{\Gamma(\frac{d}{4}+\frac{l}{2}-\frac{E}{2})}{\Gamma(\frac{d}{4}+\frac{l}{2}+\frac{E}{2})} \;.
\end{align}
We have written the top line in terms of $n$ instead of $E$, partly to simplify the expression, and partly to emphasise that these factors simply give $(-1)^n$ when evaluated on integers $n$ where we have physical operators.

This result is similar to \eqref{eq:freeRes1} above, and in particular has the correct residues at the poles $E=E_{l,n}$. But it has oscillatory factors that depend on the fractional parts of $\Delta_{1,2}$, and additional spurious poles when $\Delta-l-\frac{d}{2}$ is an integer. Presumably, these features arise from the region $\tau\to -\infty$ in the inversion integral which goes beyond the non-relativistic limit as discussed above.

\section{Scattering amplitudes and the S-matrix}\label{app:Smatrix}

Non-relativistic potential scattering is usually expressed in terms of the scattering amplitude $f_{\vec{k}}(\Omega)$, defined by the asymptotic wavefunction of the scattering in-states $\psi_{\vec{k}}^\mathrm{in}(x)$:
\begin{equation}
	\psi_{\vec{k}}^\mathrm{in}(\vec{x}) \sim \frac{1}{(2\pi)^\frac{d}{2}}\left[e^{i\vec{k}\cdot\vec{x}} + f_{\vec{k}}(\Omega) \frac{e^{i k r}}{r^{\frac{d-1}{2}}}\right],
\end{equation}
where $r=|\vec{x}|$ and $\Omega$ denotes the direction on $S^{d-1}$. But in most other situations (in particular for relativistic scattering in QFT) we use the S-matrix, defined as the overlap between in- and out-states, here labelled by incoming and outgoing momenta:
\begin{equation}
	S_{\vec{q},\vec{p}} = \langle \psi_{\vec{q}}^\mathrm{out}|\psi_{\vec{p}}^\mathrm{in}\rangle.
\end{equation}
This is a formal relation giving a distribution, understood to apply after superposing in- and out-states, weighted by appropriate test functions.

In this appendix we explain how to relate these, in particular the result \eqref{eq:scatS}. This uses standard methods in scattering theory, though we were not able to find the result in standard texts (particularly for general dimension) so we include it here for convenience. We most closely follow \cite{weinberg2015lectures}.

The key tool we make use of is the Lippmann-Schwinger equation for the in-state:
\begin{equation}
	|\psi_{\vec{p}}^\mathrm{in}\rangle = |\phi_{\vec{p}}\rangle+ \frac{1}{E_{\vec{p}}-H_0+i\epsilon} V |\psi_{\vec{p}}^\mathrm{in}\rangle.
\end{equation}
Here, $|\phi_{\vec{p}}\rangle$ a plane wave with wavefunction $\phi_{\vec{p}}(x)=\frac{1}{(2\pi)^\frac{d}{2}} e^{i\vec{p}\cdot\vec{x}}$, an eigenstate  of momentum and of the free Hamiltonian $H_0$ with eigenvalue $E_{\vec{p}}=\frac{p^2}{2\mu}$. The full Hamiltonian is $H=H_0+V$.

First, we relate this to the S-matrix by inserting a complete set of plane waves, before considering the wavefunction of a superposition of in-states at late times. The Lippmann-Schwinger relation becomes
\begin{equation}
	|\psi_{\vec{p}}^\mathrm{in}\rangle = |\phi_{\vec{p}}\rangle+ \int d^dq\frac{T_{\vec{q},\vec{p}}}{E_{\vec{p}}-E_{\vec{q}}+i\epsilon} |\phi_{\vec{q}}\rangle,
\end{equation}
where
\begin{equation}
	T_{\vec{q},\vec{p}} = \langle \phi_{\vec{q}}|V|\psi_{\vec{p}}^\mathrm{in}\rangle.
\end{equation}
Smearing with a test function $g(\vec{p})$ and evolving by time $t$, we have
\begin{align}
	|\psi_{g}^\mathrm{in}(t)\rangle &:= \int d^dp  \,g(\vec{p}) e^{-i E_{\vec{p}} t}|\psi_{\vec{p}}^\mathrm{in}\rangle \\
	&= \int d^dq \,e^{-i E_{\vec{q}} t}|\phi_{\vec{q}}\rangle \left[g(\vec{q}) + \int d^dp \frac{T_{\vec{q},\vec{p}}\, g(\vec{p}) e^{-i (E_{\vec{p}}-E_{\vec{q}}) t}}{E_{\vec{p}}-E_{\vec{q}}+i\epsilon} \right] \\
	&\sim \int d^dq \,e^{-i E_{\vec{q}} t}|\phi_{\vec{q}}\rangle \left[g(\vec{q}) -2\pi i \int d^dp\, \delta(E_{\vec{p}}-E_{\vec{q}})T_{\vec{q},\vec{p}}\, g(\vec{p})  \right].
\end{align}
In the last line we have taken a limit of $t\to \infty$, where the integral over energy $E_{\vec{p}}$ is dominated by the residue of the pole at $E_{\vec{q}}-i\epsilon$.\footnote{A more careful argument: we can write 
\begin{equation}
	\int dE f(E)\frac{e^{-iEt}}{E+i\epsilon} = \int dE \frac{f(E)-f(0)e^{-i k |E|}}{E}e^{-iEt} +f(0)\int dE \frac{e^{-iEt-ik|E|}}{E+i\epsilon}
\end{equation} Under weak conditions on $f$, the Riemann-Lebesgue lemma ensures that the first integral goes to zero as $t\to\infty$, while the second integral gives $-2\pi i f(0)$.}

Now, for very large $t$ the out states $|\psi_{\vec{q}}^\mathrm{out}\rangle$ approach plane waves $|\phi_{\vec{q}}\rangle$ (by definition), so we can interpret the last line as an expression for an in-state in terms of out-states. This means that the contents of the square brackets equals $\int d^d p S_{\vec{q},\vec{p}} g(\vec{p})$, so we have
\begin{equation}
	S_{\vec{q},\vec{p}} = \delta^{(d)}(\vec{p}-\vec{q}) - 2\pi i \delta(E_{\vec{p}}-E_{\vec{q}}) T_{\vec{q},\vec{p}}\, .
\end{equation}

The last step is to relate $T_{\vec{q},\vec{p}}$ to the asymptotics of the in-state wavefunction. For this, we insert a complete set of position eigenstates in the Lippmann-Schwinger equation:
\begin{equation}
	\psi_{\vec{p}}^\mathrm{in}(\vec{x}) = \phi_{\vec{p}}(\vec{x})+ \int d^d y \,G^+_{E_{\vec{p}}}(x,y) V(y)  \psi_{\vec{p}}^\mathrm{in}(\vec{y}),
\end{equation}
where $G^+$ is the Green's function for the free Schr\"odinger operator,
\begin{equation}
	G^+_{E}(x,y) = \langle x|\frac{1}{E-H_0+i\epsilon}|y\rangle.
\end{equation}
This is the solution to the free Schr\"odinger equation with delta-function source $\frac{1}{2\mu}(k^2-\nabla^2)G(x,y)=\delta^{(d)}(x-y)$ (where $E=\frac{k^2}{2\mu}$), and purely outgoing boundary conditions (going as $e^{ik r}$ at large $r$ without terms $e^{-ik r}$). It is a function of $r=|x-y|$ only, which we can write in terms of a Hankel function, fixing the coefficient by integrating over a small ball:
\begin{align}
	G^+_{E}(x,y) &= \frac{\mu}{2i} \left(\frac{k}{2\pi r}\right)^\frac{d-2}{2} H_{\frac{d-2}{2}}^{(1)}(kr)  \\
	&\sim \frac{1}{(2\pi)^\frac{d}{2}} e^{-i(d+1)\frac{\pi}{4}}\sqrt{2\pi}\mu k^\frac{d-3}{2} \frac{e^{ikr}}{r^\frac{d-1}{2}} \qquad (r\to \infty).
\end{align}
For sufficiently rapidly decaying $V$, we can use the asymptotic form of $G^+$, along with $|\vec{x}-\vec{y}| \sim r - \vec{\Omega}\cdot \vec{y} + O(r^{-1})$ where $r=|x|$ and $\vec{\Omega}$ is a unit vector pointing in the direction of $\vec{x}$. From this we recover the asymptotic in-state wavefunction
\begin{gather}
	\psi_{\vec{p}}^\mathrm{in}(\vec{x}) \sim \frac{1}{(2\pi)^\frac{d}{2}}\left[e^{i\vec{p}\cdot\vec{x}}+ f_p(\Omega) \frac{e^{ipr}}{r^\frac{d-1}{2}}\right], \qquad \text{with} \\
	f_p(\Omega) = e^{-i(d+1)\frac{\pi}{4}}\sqrt{2\pi}\mu p^\frac{d-3}{2} \int d^d y e^{i p \hat{\Omega}\cdot y} V(y)  \psi_{\vec{p}}^\mathrm{in}(\vec{y}).
\end{gather}
But the integral is simply the position space representation of $T_{\vec{q},\vec{p}}$, with $\vec{q}$ pointing in the direction of $\hat{x}$ with magnitude $p$:
\begin{equation}
	f_p(\Omega) = e^{-i(d+1)\frac{\pi}{4}}(2\pi)^\frac{d+1}{2}\mu p^\frac{d-3}{2} T_{\vec{q},\vec{p}}, \qquad \vec{q}=p\vec{\Omega}.
\end{equation}
In particular, for $d=3$ we have
\begin{equation}
	f_p(\Omega) = -(2\pi)^2\mu  T_{\vec{q},\vec{p}}\,.
\end{equation}

\section{The classical Coulomb correlation function}\label{app:Coulombscat}

In this appendix we describe the evaluation of the on-shell classical action for the Coulomb correlation function in section \ref{sec:classicalCoulomb}, giving the result for the action in \eqref{eq:GscatCoulomb}.

First, we rewrite the problem in terms of dimensionless parameters:
\begin{equation}
	r = \frac{J_E^2}{\mu g} x, \quad E = -\frac{\mu g^2}{2J_E^2} u^2, \quad \epsilon = \frac{J_E^3 \omega}{\mu g^2}\ll 1,
\end{equation}
We are interested in the limit where $\epsilon$ is small with $u$ held fixed. Then the on-shell Euclidean momentum becomes
\begin{equation}
	\kappa = \frac{g \mu}{J_E}\sqrt{u^2 - \frac{2}{x} - \frac{1}{x^2}+\epsilon^2 x^2}.
\end{equation}

To evaluate the integrals \eqref{eq:thetaint}, \eqref{eq:tauint} for kinematics $\theta$, $\tau$ and \eqref{eq:S0} for action $S_0$, we must must split them up into two pieces at $x=\xi$, with $1\ll \xi\ll \epsilon^{-1}$. For the `inner' piece $x<\xi$ we expand the integrand in powers of $\epsilon$ with $x$ of order one (so the harmonic potential term $\epsilon^2 x^2$ is taken to be small), and integrate term by term. For  the `outer' piece $x>\xi$ we take $y=\epsilon x$ to be of order one (neglecting Coulomb and angular momentum terms relative to energy and harmonic potential) before expanding the integrand and integrating. We can then combine the two, checking that the splitting location $\xi$ drops out of the result. In particular, a term in the expanded integrands which goes as $\frac{1}{x}$ in the overlap region gives rise to a $\log\epsilon$ in the result. It is most convenient for the inner integral to take the limit fixing the turning point $x=x_0$ where $\kappa=0$ rather then $u$, and then substitute back for $u$.

Explicitly, for the $\theta$ integral \eqref{eq:thetaint}, the inner contribution is
\begin{equation}
\begin{aligned}
	\theta_\mathrm{in} &= \int_{x_0}^\xi \frac{2dx}{x^2\sqrt{u^2 - \frac{2}{x} - \frac{1}{x^2}+\epsilon^2 x^2}} \\
	& =4 \cot ^{-1}\left(\sqrt{2 x_0+1}\right)-\frac{2x_0}{\sqrt{2 x_0+1}}\xi^{-1} +O(\epsilon^2\xi,\xi^{-2}) \\
	&= 2\tan^{-1}u - \frac{2}{u}\xi^{-1} +O(\epsilon^2\xi,\xi^{-2}),
\end{aligned}
\end{equation}
while the outer piece gives
\begin{equation}
\begin{aligned}
	\theta_\mathrm{out} &= \int_{\epsilon\xi}^\infty \frac{2\epsilon dy}{y^2\sqrt{u^2+y^2 - \frac{2\epsilon}{y} - \frac{\epsilon^2}{y^2}}} \\
	&=-2\frac{\epsilon}{u^2} + \frac{2}{u}\xi^{-1} +O(\epsilon^2\xi,\xi^{-2}).
\end{aligned}
\end{equation}
Combining these, we get
\begin{equation}
	\theta = \theta_\mathrm{in}+\theta_\mathrm{out} = 2\tan^{-1} u -\frac{2}{u^2} \epsilon + \cdots.
\end{equation}

We similarly split the $\tau$ integral \eqref{eq:tauint},
\begin{equation}
\begin{aligned}
	\omega\tau_\mathrm{in} &= \frac{2\epsilon\xi}{u} + \frac{2\epsilon}{u^3} \left(\log \left(\frac{2 \xi  u^2}{\sqrt{u^2+1}}\right)-1\right) + \cdots \\
	\omega\tau_\mathrm{out} &= 2\log\left(\tfrac{2y_c}{u}\right) - \frac{2\epsilon\xi}{u}+\frac{2\epsilon}{u^3}\left(\log\left(\tfrac{2u}{\epsilon\xi}\right)-1\right) + \cdots
\end{aligned}
\end{equation}
where we have integrated up to the cutoff $y=y_c\gg 1$. Combining these and subtracting $\mu R^2 = \log(\epsilon^{-1}J_E y_c^2)$, this gives
\begin{equation}
	\omega \tau =-\log\left(\frac{u^2J_E}{4\epsilon}\right) +   \frac{2\epsilon}{u^3} \left(\log \left(\frac{4  u^3}{\epsilon\sqrt{u^2+1}}\right)-2\right) +\cdots.
\end{equation}

Finally, for the action integral \eqref{eq:S0} we get
\begin{equation}
\begin{aligned}
	S_\mathrm{in} &= J_E\xi u- \frac{J_E}{u}\left(1+3\log\frac{2 \xi  u^2}{\sqrt{u^2+1}}\right) \cdots \\
	S_\mathrm{out} &= \frac{J_E}{\epsilon}y_c^2+\frac{J_E}{2\epsilon}u^2 - J_E u \xi+ \frac{J_E}{u}\left(1-3\log\frac{2u}{\epsilon\xi}\right)  + \cdots,
\end{aligned}
\end{equation}
giving us
\begin{equation}
	S_0 = \frac{J_E}{2\epsilon}u^2- \frac{3J_E}{u}\log\left(\frac{4   u^3}{\epsilon\sqrt{u^2+1}}\right)  + \cdots.
\end{equation}

The next step is to invert these relations to express conserved quantities $E,J_E$ (or equivalently $u$,$\epsilon$) in terms of kinematics $\tau,\theta$. This is made slightly trickier by the explicit appearence of $J_E\propto\epsilon^{\frac{1}{3}}$ in the expression for $\tau$. To eliminate this problem, exponentiate and eliminate $J_E$ in favour of $\epsilon$ to write
\begin{equation}
	\hat{g}e^{\frac{3}{2}\omega\tau} = \frac{8\epsilon}{u^3 } +   \frac{24\epsilon^2}{u^6} \left(\log \left(\frac{4  u^3}{\epsilon\sqrt{u^2+1}}\right)-2\right) +\cdots,
\end{equation}
so $\hat{g}e^{\frac{3}{2}\omega\tau}$ will become the small parameter we're expanding in. Now it's simple to see the leading order results $u=\tan\frac{\theta}{2}$ and $\epsilon = \frac{1}{8}\hat{g}e^{\frac{3}{2}\omega\tau}\tan^3\frac{\theta}{2}$, and we can go to the next order in the expansion:
\begin{align}
	u &=\tan\tfrac{\theta}{2}\left(1+\frac{\hat{g}e^{\frac{3}{2}\omega\tau}}{8\cos^2\frac{\theta}{2}} +\cdots\right) \\
	\epsilon &= \frac{1}{8}\hat{g}e^{\frac{3}{2}\omega\tau}\tan^3\tfrac{\theta}{2}\left[1+\tfrac{3}{8}\hat{g}e^{\frac{3}{2}\omega\tau}\left(2+\sec ^2\tfrac{\theta }{2}+\log \left(\tfrac{\hat{g}e^{\frac{3}{2}\omega\tau}}{32 \cos\frac{\theta }{2}}\right)\right) +\cdots\right]\nonumber
\end{align}

Finally, we can substitute these to find the action $S_0$ (after writing $J_E = \hat{g}^\frac{2}{3}\epsilon^\frac{1}{3}$) in terms of $\tau,\theta$:
\begin{equation}
	S_0 = \frac{2}{\hbar} e^{-\omega \tau} -\hat{g} e^{\frac{1}{2}\omega \tau} \left[1+\log\left(\frac{32 \cos\tfrac{\theta}{2}}{\hat{g}e^{\frac{3}{2}\omega\tau}}\right)\right] + \cdots
\end{equation}
as given in the main text as equation \eqref{eq:CoulombAction}.

\bibliographystyle{JHEP}
\bibliography{NRAdSbib}

\providecommand{\href}[2]{#2}\begingroup\raggedright\begin{thebibliography}{10}

\bibitem{Heemskerk:2009pn}
I.~Heemskerk, J.~Penedones, J.~Polchinski and J.~Sully, \emph{{Holography from
  Conformal Field Theory}},
  \href{http://dx.doi.org/10.1088/1126-6708/2009/10/079}{\emph{JHEP} {\bfseries
  10} (2009) 079}, [\href{https://arxiv.org/abs/0907.0151}{{\ttfamily
  0907.0151}}].

\bibitem{El-Showk:2011yvt}
S.~El-Showk and K.~Papadodimas, \emph{{Emergent Spacetime and Holographic
  CFTs}}, \href{http://dx.doi.org/10.1007/JHEP10(2012)106}{\emph{JHEP}
  {\bfseries 10} (2012) 106},
  [\href{https://arxiv.org/abs/1101.4163}{{\ttfamily 1101.4163}}].

\bibitem{Afkhami-Jeddi:2016ntf}
N.~Afkhami-Jeddi, T.~Hartman, S.~Kundu and A.~Tajdini, \emph{{Einstein gravity
  3-point functions from conformal field theory}},
  \href{http://dx.doi.org/10.1007/JHEP12(2017)049}{\emph{JHEP} {\bfseries 12}
  (2017) 049}, [\href{https://arxiv.org/abs/1610.09378}{{\ttfamily
  1610.09378}}].

\bibitem{Penedones:2016voo}
J.~Penedones, \emph{{TASI lectures on AdS/CFT}},  in \emph{{Theoretical
  Advanced Study Institute in Elementary Particle Physics}: {New Frontiers in
  Fields and Strings}}, pp.~75--136, 2017.
\newblock \href{https://arxiv.org/abs/1608.04948}{{\ttfamily 1608.04948}}.
\newblock \href{http://dx.doi.org/10.1142/9789813149441_0002}{DOI}.

\bibitem{Hawking:1982dh}
S.~W. Hawking and D.~N. Page, \emph{{Thermodynamics of Black Holes in anti-De
  Sitter Space}}, \href{http://dx.doi.org/10.1007/BF01208266}{\emph{Commun.
  Math. Phys.} {\bfseries 87} (1983) 577}.

\bibitem{Simmons-Duffin:2016gjk}
D.~Simmons-Duffin, \emph{{The Conformal Bootstrap}},  in \emph{{Theoretical
  Advanced Study Institute in Elementary Particle Physics}: {New Frontiers in
  Fields and Strings}}, pp.~1--74, 2017.
\newblock \href{https://arxiv.org/abs/1602.07982}{{\ttfamily 1602.07982}}.
\newblock \href{http://dx.doi.org/10.1142/9789813149441_0001}{DOI}.

\bibitem{Poland:2018epd}
D.~Poland, S.~Rychkov and A.~Vichi, \emph{{The Conformal Bootstrap: Theory,
  Numerical Techniques, and Applications}},
  \href{http://dx.doi.org/10.1103/RevModPhys.91.015002}{\emph{Rev. Mod. Phys.}
  {\bfseries 91} (2019) 015002},
  [\href{https://arxiv.org/abs/1805.04405}{{\ttfamily 1805.04405}}].

\bibitem{companion}
H.~Maxfield and Z.~Zahraee, \emph{{To appear}}, .

\bibitem{Paulos:2016fap}
M.~F. Paulos, J.~Penedones, J.~Toledo, B.~C. van Rees and P.~Vieira, \emph{{The
  S-matrix bootstrap. Part I: QFT in AdS}},
  \href{http://dx.doi.org/10.1007/JHEP11(2017)133}{\emph{JHEP} {\bfseries 11}
  (2017) 133}, [\href{https://arxiv.org/abs/1607.06109}{{\ttfamily
  1607.06109}}].

\bibitem{Cornalba:2007zb}
L.~Cornalba, M.~S. Costa and J.~Penedones, \emph{{Eikonal approximation in
  AdS/CFT: Resumming the gravitational loop expansion}},
  \href{http://dx.doi.org/10.1088/1126-6708/2007/09/037}{\emph{JHEP} {\bfseries
  09} (2007) 037}, [\href{https://arxiv.org/abs/0707.0120}{{\ttfamily
  0707.0120}}].

\bibitem{Kulaxizi:2019tkd}
M.~Kulaxizi, G.~S. Ng and A.~Parnachev, \emph{{Subleading Eikonal, AdS/CFT and
  Double Stress Tensors}},
  \href{http://dx.doi.org/10.1007/JHEP10(2019)107}{\emph{JHEP} {\bfseries 10}
  (2019) 107}, [\href{https://arxiv.org/abs/1907.00867}{{\ttfamily
  1907.00867}}].

\bibitem{Hijano:2019qmi}
E.~Hijano, \emph{{Flat space physics from AdS/CFT}},
  \href{http://dx.doi.org/10.1007/JHEP07(2019)132}{\emph{JHEP} {\bfseries 07}
  (2019) 132}, [\href{https://arxiv.org/abs/1905.02729}{{\ttfamily
  1905.02729}}].

\bibitem{Komatsu:2020sag}
S.~Komatsu, M.~F. Paulos, B.~C. Van~Rees and X.~Zhao, \emph{{Landau diagrams in
  AdS and S-matrices from conformal correlators}},
  \href{http://dx.doi.org/10.1007/JHEP11(2020)046}{\emph{JHEP} {\bfseries 11}
  (2020) 046}, [\href{https://arxiv.org/abs/2007.13745}{{\ttfamily
  2007.13745}}].

\bibitem{Hogervorst:2013sma}
M.~Hogervorst and S.~Rychkov, \emph{{Radial Coordinates for Conformal Blocks}},
  \href{http://dx.doi.org/10.1103/PhysRevD.87.106004}{\emph{Phys. Rev. D}
  {\bfseries 87} (2013) 106004},
  [\href{https://arxiv.org/abs/1303.1111}{{\ttfamily 1303.1111}}].

\bibitem{Bacry:1968zf}
H.~Bacry and J.~Levy-Leblond, \emph{{Possible kinematics}},
  \href{http://dx.doi.org/10.1063/1.1664490}{\emph{J. Math. Phys.} {\bfseries
  9} (1968) 1605--1614}.

\bibitem{derome1972hooke}
J.-R. Derome and J.-G. Dubois, \emph{Hooke’s symmetries and nonrelativistic
  cosmological kinematics.—i}, {\emph{Il Nuovo Cimento B (1971-1996)}
  {\bfseries 9} (1972) 351--376}.

\bibitem{dubois1973hooke}
J.~Dubois, \emph{Hooke’s symmetries and nonrelativistic cosmological
  kinematics. ii: Irreducible projective representations}, {\emph{Il Nuovo
  Cimento B (1971-1996)} {\bfseries 15} (1973) 1--17}.

\bibitem{shakespeare}
W.~Shakespeare, \emph{Macbeth}.
\newblock 1623.

\bibitem{Sleight:2017krf}
C.~Sleight, \emph{{Metric-like Methods in Higher Spin Holography}},
  \href{http://dx.doi.org/10.22323/1.296.0003}{\emph{PoS} {\bfseries
  Modave2016} (2017) 003}, [\href{https://arxiv.org/abs/1701.08360}{{\ttfamily
  1701.08360}}].

\bibitem{Caron-Huot:2017vep}
S.~Caron-Huot, \emph{{Analyticity in Spin in Conformal Theories}},
  \href{http://dx.doi.org/10.1007/JHEP09(2017)078}{\emph{JHEP} {\bfseries 09}
  (2017) 078}, [\href{https://arxiv.org/abs/1703.00278}{{\ttfamily
  1703.00278}}].

\bibitem{Fitzpatrick:2010zm}
A.~L. Fitzpatrick, E.~Katz, D.~Poland and D.~Simmons-Duffin, \emph{{Effective
  Conformal Theory and the Flat-Space Limit of AdS}},
  \href{http://dx.doi.org/10.1007/JHEP07(2011)023}{\emph{JHEP} {\bfseries 07}
  (2011) 023}, [\href{https://arxiv.org/abs/1007.2412}{{\ttfamily 1007.2412}}].

\bibitem{Fitzpatrick:2011dm}
A.~L. Fitzpatrick and J.~Kaplan, \emph{{Unitarity and the Holographic
  S-Matrix}}, \href{http://dx.doi.org/10.1007/JHEP10(2012)032}{\emph{JHEP}
  {\bfseries 10} (2012) 032},
  [\href{https://arxiv.org/abs/1112.4845}{{\ttfamily 1112.4845}}].

\bibitem{Karateev:2018oml}
D.~Karateev, P.~Kravchuk and D.~Simmons-Duffin, \emph{{Harmonic Analysis and
  Mean Field Theory}},
  \href{http://dx.doi.org/10.1007/JHEP10(2019)217}{\emph{JHEP} {\bfseries 10}
  (2019) 217}, [\href{https://arxiv.org/abs/1809.05111}{{\ttfamily
  1809.05111}}].

\bibitem{Maxfield:2017rkn}
H.~Maxfield, \emph{{A view of the bulk from the worldline}},
  \href{https://arxiv.org/abs/1712.00885}{{\ttfamily 1712.00885}}.

\bibitem{Bizon:2018frv}
P.~Bizon, O.~Evnin and F.~Ficek, \emph{{A nonrelativistic limit for AdS
  perturbations}}, \href{http://dx.doi.org/10.1007/JHEP12(2018)113}{\emph{JHEP}
  {\bfseries 12} (2018) 113},
  [\href{https://arxiv.org/abs/1810.10574}{{\ttfamily 1810.10574}}].

\bibitem{Craps:2021xmk}
B.~Craps, M.~De~Clerck and O.~Evnin, \emph{{Time-periodicities in holographic
  CFTs}}, \href{http://dx.doi.org/10.1007/JHEP09(2021)030}{\emph{JHEP}
  {\bfseries 09} (2021) 030},
  [\href{https://arxiv.org/abs/2103.12798}{{\ttfamily 2103.12798}}].

\bibitem{durand2019complex}
L.~Durand, \emph{Complex asymptotics in $\lambda$ for the gegenbauer functions
  c $\lambda$ $\alpha$ (z) and d $\lambda$ $\alpha$ (z) with z$\in$(- 1, 1)},
  {\emph{Symmetry} {\bfseries 11} (2019) 1465}.

\bibitem{Li:2021snj}
Y.-Z. Li, \emph{{Notes on flat-space limit of AdS/CFT}},
  \href{http://dx.doi.org/10.1007/JHEP09(2021)027}{\emph{JHEP} {\bfseries 09}
  (2021) 027}, [\href{https://arxiv.org/abs/2106.04606}{{\ttfamily
  2106.04606}}].

\bibitem{Simmons-Duffin:2017nub}
D.~Simmons-Duffin, D.~Stanford and E.~Witten, \emph{{A spacetime derivation of
  the Lorentzian OPE inversion formula}},
  \href{http://dx.doi.org/10.1007/JHEP07(2018)085}{\emph{JHEP} {\bfseries 07}
  (2018) 085}, [\href{https://arxiv.org/abs/1711.03816}{{\ttfamily
  1711.03816}}].

\bibitem{levinson1949uniqueness}
N.~Levinson, \emph{On the uniqueness of tne potential in a schr{\"o}dinger
  equation for a given asymptotic phase}, {\emph{Danske Vid. Selsk. Mat.-Fys.
  Medd.} {\bfseries 25} (1949) 29}.

\bibitem{weinberg2015lectures}
S.~Weinberg, \emph{Lectures on quantum mechanics}.
\newblock Cambridge University Press, 2015.

\bibitem{newton}
I.~Newton, \emph{Philosophiae naturalis principia mathematica}.
\newblock 1687.

\bibitem{de1785premier}
C.~A. de~Coulomb, \emph{Premier m{\'e}moire sur l’electricit{\'e} et le
  magn{\'e}tisme}, {\emph{Histoire de l’Acad{\'e}mie Royale des Sciences}
  {\bfseries 569} (1785) }.

\bibitem{Levi:2018nxp}
M.~Levi, \emph{{Effective Field Theories of Post-Newtonian Gravity: A
  comprehensive review}},
  \href{http://dx.doi.org/10.1088/1361-6633/ab12bc}{\emph{Rept. Prog. Phys.}
  {\bfseries 83} (2020) 075901},
  [\href{https://arxiv.org/abs/1807.01699}{{\ttfamily 1807.01699}}].

\bibitem{pauli1926spectrum}
W.~Pauli, \emph{On the spectrum of the hydrogen from the standpoint of the new
  quantum mechanics}, {\emph{Z. Phys} {\bfseries 36} (1926) 336}.

\bibitem{Caron-Huot:2018ape}
S.~Caron-Huot and Z.~Zahraee, \emph{{Integrability of Black Hole Orbits in
  Maximal Supergravity}},
  \href{http://dx.doi.org/10.1007/JHEP07(2019)179}{\emph{JHEP} {\bfseries 07}
  (2019) 179}, [\href{https://arxiv.org/abs/1810.04694}{{\ttfamily
  1810.04694}}].

\bibitem{Susskind:2021omt}
L.~Susskind, \emph{{De Sitter Holography: Fluctuations, Anomalous Symmetry, and
  Wormholes}}, \href{http://dx.doi.org/10.3390/universe7120464}{\emph{Universe}
  {\bfseries 7} (2021) 464},
  [\href{https://arxiv.org/abs/2106.03964}{{\ttfamily 2106.03964}}].

\bibitem{Volovik:2009eb}
G.~E. Volovik, \emph{{Particle decay in de Sitter spacetime via quantum
  tunneling}}, \href{http://dx.doi.org/10.1134/S0021364009130013}{\emph{JETP
  Lett.} {\bfseries 90} (2009) 1--4},
  [\href{https://arxiv.org/abs/0905.4639}{{\ttfamily 0905.4639}}].

\end{thebibliography}\endgroup

\end{document}